\documentclass[reprint,
showpacs,
nofootinbib,
longbibliography,
aps,
rmp,
]{revtex4-1}

\usepackage{amsmath,amssymb,amsfonts,amsthm}
\usepackage[english]{babel}
\usepackage{url}
\usepackage{bm}
\usepackage[latin1]{inputenc}
\usepackage{graphicx,rotating}
\usepackage[colorlinks=true,linkcolor=black, citecolor=blue]{hyperref}
\usepackage{csvsimple}
\usepackage{longtable}
\usepackage{booktabs}
\usepackage{multirow, array}

\usepackage{dcolumn}

\usepackage[perpage,bottom]{footmisc}

\allowdisplaybreaks[1]

\newcommand{\eqtext}[1]{\ensuremath{\stackrel{\text{#1}}{=}}}

\begin{document}

\title{High precision analytical description of the allowed $\beta$ spectrum shape}

\author{Leendert Hayen}
\email[Corresponding author: ]{leendert.hayen@kuleuven.be}

\author{Nathal Severijns}
\affiliation{Instituut voor Kern- en Stralingsfysica, KU Leuven, Celestijnenlaan 200D, B-3001 Leuven, Belgium}

\author{Kazimierz Bodek}

\affiliation{Marian Smoluchowski Institute of Physics, Jagiellonian University, 30-348 Cracow, Poland}

\author{Xavier Mougeot}
\affiliation{CEA, LIST, Laboratoire National Henri Becquerel, F-91191 Gif-sur-Yvette, France}

\author{Dagmara Rozpedzik}

\affiliation{Marian Smoluchowski Institute of Physics, Jagiellonian University, 30-348 Cracow, Poland}

\date{\today}
\begin{abstract}
A fully analytical description of the allowed $\beta$ spectrum shape is given in view of ongoing and planned measurements. Its study forms an invaluable tool in the search for physics beyond the standard electroweak model and the weak magnetism recoil term. Contributions stemming from finite size corrections, mass effects, and radiative corrections are reviewed. A particular focus is placed on atomic and chemical effects, where the existing description is extended and analytically provided. The effects of QCD-induced recoil terms are discussed, and cross-checks were performed for different theoretical formalisms. Special attention was given to a comparison of the treatment of nuclear structure effects in different formalisms. Corrections were derived for both Fermi and Gamow-Teller transitions, and methods of analytical evaluation thoroughly discussed. In its integrated form, calculated $f$ values were in agreement with the most precise numerical results within the aimed for precision. We stress the need for an accurate evaluation of weak magnetism contributions, and note the possible significance of the oft-neglected induced pseudoscalar interaction. Together with improved atomic corrections, we then present an analytical description of the allowed $\beta$ spectrum shape accurate to a few parts in $10^{-4}$ down to 1\,keV for low to medium $Z$ nuclei, thereby extending the work by previous authors by nearly an order of magnitude.

\end{abstract}

\pacs{13.40.Ks, 14.60.Cd, 23.40.-s, 24.80.+y, 31.15.-p}

\maketitle

\tableofcontents

\section{Introduction}
The study of $\beta$ decay played a pivotal role in uncovering the nature of the left-handed `$V$-$A$' weak interaction - and by extension the electroweak sector of the Standard Model (SM) \cite{Weinberg2009} - half a century ago. Over the following decades, the continuous development of new experimental techniques has kept research on $\beta$ decay at the forefront of revealing the structure of the SM and its inner workings \cite{Severijns2006, Vos2015}. Probing an array of different observables, e.g., correlation coefficients or the $\mathcal{F}t$ value, allows for a sensitive investigation of several Beyond Standard Model (BSM) influences. Deviations from pure $V$-$A$ theory can occur as manifestations of exotic interactions, such as e.g., right-handed currents involving new heavy particles. Through investigation of sidereal variations of experimental observables it additionally allows for a study of Lorentz invariance violation \cite{Noordmans2013}.

In the LHC era, competitive results can be extracted from $pp \rightarrow e + MET + X$ reaction channels \cite{Khachatryan2015}. In the past decade, the development of so-called Effective Field Theories (EFT) has seen a tremendous amount of work and interest where, through inclusion of a new physics scale $\Lambda_{\text{BSM}} \gg \Lambda_{\text{LHC}}$, both high- and low energy experimental results can be interpreted in the same theoretical framework \cite{Bhattacharya2012, Cirigliano2013}. The Lee-Yang Hamiltonian \citep{PhysRev.104.254} is generalized to a form where couplings with higher dimensional combinations of SM field operators are categorized according to their transformation properties; see e.g. the work by \textcite{Bhattacharya2012}. Several comparative reviews have been presented in the past few years \cite{Naviliat-Cuncic2013, Cirigliano2013a, Holstein2014, Holstein2014a, Severijns2014}.

Recently, renewed interest has arisen in the beta spectrum (e.g., \cite{Kuzminov2000, Huyan2016, Severijns2014, Towner2005, George2014, Severin2014, Dawson2009, Mougeot2014, Loidl2010, Mougeot2015}) as a tool for precision measurements searching for exotic currents beyond the standard electroweak model and investigating QCD-induced form factors. The latter are related to the fact that the quark involved in beta decay is not a free particle but is embedded in a nucleon. Beyond Standard Model scalar and tensor coupling constants appear in the mathematical description of the beta spectrum shape via the so-called Fierz interference term \cite{Jackson1957}
\begin{align}
\label{bFierzapprox}
b_{\text{Fierz}}  \simeq \pm \frac{1}{ 1 + \rho ^2} &\left[ \text{Re} \left( \frac{ C_S +
C^\prime _S }{ C_V } \right) \right. \nonumber \\ &\left. + \rho ^2 \text{Re} \left( \frac{ C_T + C^\prime _T }{ C_A } \right) \right],
\end{align}
with the upper(lower) sign for electron(positron) decay, respectively, and $C_{S, T}^{(\prime)}$ are the coupling constants for possible scalar or tensor type weak interactions. Further, $C_{V, A}^{(\prime)}$ are the coupling constants for the SM vector and axial-vector interactions and $\rho = \frac{C_A M_{GT}}{C_V M_F}$ is the ratio of the Gamow-Teller and Fermi fractions in the $\beta$ decay, with $M_{GT}$ and $M_F$ being the respective nuclear matrix elements. For pure Fermi and Gamow-Teller transitions this reduces to, respectively,
\begin{subequations}
\begin{align}
\label{bFierz-F-GT}
b_{\text{Fierz}, F}  &\simeq  \pm \text{Re} \left( \frac{ C_S + C^\prime _S }{ C_V } \right), \\ \quad b_{\text{Fierz}, GT} &\simeq  \pm \text{Re} \left( \frac{ C_T + C^\prime _T }{ C_A } \right) ,
\end{align}
\end{subequations}
i.e., searches for non-Standard Model currents become independent (to first order) of the nuclear matrix elements. In order to improve existing limits on these scalar or tensor coupling constants \cite{Severijns2011, Vos2015, Holstein2014} a precision of typically 0.5$\%$ is required when determining $b_{\text{Fierz}}$ \cite{Severijns2013}.

As the required experimental precision increases, QCD-induced form factors have to be taken into account in order not to limit the sensitivity to BSM physics \cite{Severijns2013, Pitcairn2009, Wauters2009a, Wauters2010, Soti2014, Sternberg2015, Fenker2016}. These form factors are part of the SM and are dominated by the so-called weak magnetism term, $b_{\text{wm}}$. Its inclusion is typically sufficient to be accurate below the $10^{-3}$ level, while higher order terms can be incorporated should the need arise. Conversely, similar experimental precision has to be reached in order to test these corrections.

The Fierz and weak magnetism terms modify the shape of the $\beta$ spectrum as follows
\begin{align}
N(W)dW &\propto p W (W_0 - W)^2 \nonumber \\ 
& \times \left(1 + \frac{\gamma m_e}{W} b_{\text{Fierz}} \pm \frac{4}{3} \frac{W}{M} b_{\text{wm}} \right) dW,
\end{align}
with $p, W$ and $W_0$ the $\beta$ particle momentum, its total energy and total energy at the spectrum endpoint, respectively. Additionally, $\gamma = \sqrt{1-(\alpha Z)^2}$ with $\alpha$ the fine structure constant, $Z$ the atomic number of the daughter nucleus, and $m_e$ and $M$ the rest mass of the electron and the average mass of the mother and daughter nuclei, respectively. It is clear that, in order to search for new physics or when trying to establish the effect of weak magnetism, the $\beta$ spectrum has to be sufficiently well understood theoretically in order to avoid other SM effects mimicking a non-zero Fierz term or behaving like a weak magnetism contribution. A description of the $\beta$ spectrum reliable at the 10$^{-4}$ level is then required.

In the past a detailed description of the $\beta$ spectrum has been provided by Behrens and collaborators \cite{Buhring1963, Buhring1965a, Buhring1965b, Behrens1969, Behrens1971, Behrens1978, Behrens1982} - based on the formalism by Stech and Sch\"ulke \cite{Stech1964, Schulke1964} - by Holstein \cite{Holstein1971, Holstein1971a, Holstein1971b, Holstein1972a, Holstein1972b, Holstein1972c, Holstein1974a, Holstein1974b} and more recently by \textcite{Wilkinson1982, Wilkinson1989a, Wilkinson1990, Wilkinson1993a, Wilkinson1993b, Wilkinson1993c, Wilkinson1995, Wilkinson1995b, Wilkinson1997}. The work by Behrens \emph{et al.}, while rigorous and complete, relies on numerical solutions of the Dirac equation for the outgoing leptons, whereas that by Wilkinson provided analytical parametrisations of the dominant effects.

In this work special attention has been given to improve the analytical description of atomic effects. The theoretical work performed by B\"uhring concerning screening effects was combined with high precision atomic potentials to guarantee good behavior for all $Z$ values. An analytical fit is proposed for the atomic exchange effect, based on state-of-the-art numerical calculations. Its contribution can easily exceed 20\% in the lowest energy regions for higher $Z$, and so is not to be neglected. The effects of shake-up and shake-off have been reviewed, as well as their influence on the aforementioned atomic effects. Due to the aimed for precision in this work, molecular effects have been explored and discussed. Limited analytical work is presented which can act as a guideline for estimating the error associated with a neglect of these influences.

The effects of nuclear structure and spatial variations of the final state wave functions inside the nuclear volume have been reviewed and we propose a new correction combining the transparency of the Holstein formalism with the rigor of the Behrens-B\"uhring formalism. Bundling all this information, the analytical atomic $\beta$ spectrum shape presented here is expected to be accurate at the few parts in $10^{4}$ level down to 1\,keV for low to medium $Z$ nuclei.

In the following, the $\beta$ spectrum shape is discussed by first providing an overview of the full description in section Sec.~\ref{sec:complete_expression}. Several sections treat the different electromagnetic and kinematic corrections, discussing the Fermi function (Sec.~\ref{Fermi function}), followed by the effects of the finite size and mass of the nucleus (Sec.~\ref{size-and-mass}), radiative corrections (Sec.~\ref{sec:radiative_corr}) and finally different atomic and chemical effects (Sec.~\ref{atomic_effects}). Another important part deals with corrections related to nuclear structure (Sec. (\ref{sec:nuclear_structure})), where considerable attention has been given to its correct evaluation. To this end a comparison was made between different approaches. It is here the weak magnetism contribution is reviewed and complete expressions are given for both Fermi and Gamow-Teller decays. Significant attention was given to the correct evaluation of the matrix elements in the nuclear structure dependent terms. Finally, the precision required to search for scalar and tensor weak interactions, and to study the effect of weak magnetism will be discussed. In the appendices a comparison is given between the electromagnetic corrections in the Behrens-B\"uhring and Holstein formalisms, and further elaborated on the correct evaluation of nuclear structure dependent effects.

\section{Complete expression}
\label{sec:complete_expression}

Apart from the electromagnetic corrections to the $\beta$ spectrum shape, several other smaller corrections are to be included when a precision at the 10$^{-4}$ level is required. The detailed description of the allowed $\beta$ spectrum shape, including these smaller corrections, is given by
\begin{widetext}
\begin{align}
N(W)dW & = \frac{G_V^2 V_{ud}^2}{2\pi^3} ~ F_{0}(Z, W) ~ L_0(Z, W) ~ U(Z, W) ~ D_\text{FS}(Z, W, \beta_2) ~ R(W, W_0) ~ R_N(W,W_0, M)  \nonumber \\
 & ~~~~ \times ~ Q(Z, W) ~ S(Z, W) ~ X(Z, W) ~ r(Z, W) ~ C(Z, W) ~ D_C(Z, W, \beta_2) ~ p W (W_0 - W)^2 ~ dW  \nonumber \\
 & \equiv \frac{G_V^2 V_{ud}^2}{2\pi^3} ~ K(Z, W, W_0, M) ~ A(Z, W) ~ C'(Z, W) ~ p W (W_0 - W)^2 ~ dW .
\label{full-expression}
\end{align}
\end{widetext}
\noindent Here, $Z$ is the proton number of the daughter nucleus, $W = E/m_e c^2 + 1$ is the total $\beta$ particle energy in units of the electron rest mass, $W_0$ is the total energy at the spectrum endpoint, $p = \sqrt{W^2 - 1}$ the $\beta$ particle momentum in units of $m_e c$, $G_V$ the vector coupling strength in nuclei, and $V_{ud}^2 = \cos^2\theta_C$, with $\theta_C$ the Cabibbo-angle, is the square of the $up$-$down$ matrix element of the Cabibbo-Kobayashi-Maskawa quark-mixing matrix.

The factor $F_0(Z, W)$ is the point charge Fermi function that takes into account the Coulomb interaction between the $\beta$ particle and the daughter nucleus. The product $L_0(Z, W) ~ U(Z, W) ~ D_{\text{FS}}(Z, W, \beta_2))$ describes the required corrections to this Fermi function after evaluation at the origin, which depend on the size and shape of the daughter nucleus (Sec.~\ref{size-and-mass}). Whereas previous effects are electrostatic in origin, $R(W, W_0)$ takes into account radiative corrections calculated using QED (Sec.~\ref{sec:radiative_corr}). Moving from an infinitely massive nucleus to one of finite mass introduces further kinematical corrections described by $R_N$ and $Q$. All these factors are combined into the factor $K(Z, W, W_0, M)$. The nuclear decay occurs in an atomic environment, meaning additional atomic corrections have to be taken into account. Here, $S(Z, W)$ is the screening correction (Sec.~\ref{screening}), $X(Z, W)$ takes into account the so-called atomic exchange effect (Sec.~\ref{exchange}) while $r(Z, W)$ accounts for the atomic mismatch (Sec.~\ref{sec:atomic_mismatch}). These effects are combined into $A(Z, W)$. Finally, the nuclear structure sensitive effects are written as $C(Z, W)$, with $D_C$ its corresponding nuclear deformation correction. These are extensively discussed in Sec. (\ref{sec:nuclear_structure}).

We comment here on the different effects encompassed by the name `finite size effects' used by different authors. For this, we must first realize the Fermi function comes about by extracting the electron amplitude at either the origin or the nuclear radius from the transition amplitude. We will perform the former in this work. As the nucleus is an object of finite size and the electron wave function is not a constant within this surface, this extraction requires corrections from convoluting its wave function with that of initial and final states. As the extracted Fermi function is typically written down in analytical form for a point charge through $F_0$, this too requires corrections stemming from the finite size and shape of the daughter nucleus. We will call these effects `electrostatic finite size' corrections in order to clearly distinguish their origin, and describe them mathematically through $L_0$, $U$, and $D_{\text{FS}}$. This amounts simply to the extraction of a more correct electron wave function evaluated at the origin. We still require a convolution of the correct wave function through the nuclear volume via initial and final nuclear states contributing to the decay. This involves a convolution with all relevant operators contributing to the decay, which we do not artificially separate but write completely as $C$. As this depends on the electron wave function behavior inside the nucleus, Coulomb effects are present in the calculation thereof. In the approach by \textcite{Calaprice1976}, \textcite{Holstein1974} and others these are artificially separated into nuclear structure and Coulomb sensitive factors when describing the spectral functions. Together with the `electrostatic finite size' effects defined above, these are collectively called `finite size' corrections. In the works inspired by \textcite{Behrens1982}, on the other hand, only the part involving the leptonic convolution is typically referred to as the `finite size' correction. Others still refer to only our `electrostatic finite size' effects. By specifiying the electrostatic origin of these corrections, we hope to put these confusions to rest. As the nuclear structure sensitive correction, $C$, is obviously non-zero even for point nuclei, we refrain from calling these `finite size' effects altogether even though we recognize the finite nuclear wave function clearly influences these results. Appendix \ref{app:em_comparison} in particular aims to further discuss the overlap and differences in the different formalisms commonly found in the literature.

In the entire treatment natural units suited for $\beta$ decay are used, i.e., $c=\hbar=m_e=1$. In all formulae presented below we define $Z$ as a positive quantity, and distinguish between $\beta^-$ and $\beta^+$ explicitly unless mentioned otherwise. When estimates of the magnitude of different corrections are given, it represents the relative change in the phase space integral after inclusion of the effect unless mentioned otherwise. As the detailed description of $\beta$ decay was developed over a period of several decades with several different formalisms being used, we also attempt here to relate different theoretical results and trace their origin, in order to avoid double counting issues. The appendices further elaborate on this idea.

Finally, all formulae presented in the work below have been implemented in a custom C++ program \cite{HayenTBP}. Based on simple configuration files describing the decay parameters and basic nuclear properties, the allowed $\beta$ spectrum shape and corresponding (anti)neutrino spectrum can be automatically calculated. This includes the nuclear structure-sensitive terms, allowing for its evaluation in the extreme single particle approximation as well as through connections with shell model of mean field software packages. The former approach is shown here to work very well once one moves to a properly deformed potential rather than a spherical harmonic oscillator potential. Results specific to nuclear structure related parameters utilizing this code are discussed elsewhere \cite{SeverijnsTBP}.

\section{Fermi function}
\label{Fermi function}
The Hamiltonian that governs all types of $\beta$ decay must include not only the weak interaction responsible for the actual decay, but also the electromagnetic interaction of the $\beta$ particle with its surroundings. As the latter is much stronger than the former, it cannot be treated in perturbation theory \cite{Halpern1970, Holstein1979}. Writing down the transition matrix element to first order in the weak Hamiltonian we find
\begin{align}
M_{fi} = &-2\pi i \delta(E_f-E_i) \langle f|T\left[\exp\left(-i\int_0^{\infty}dt\mathcal{H}^{Z}(t)\right)\right] \nonumber \\
& \times \mathcal{H}_{\beta}(0)T\left[\exp \left(-i\int_{-\infty}^0dt\mathcal{H}^{Z'}(t)\right)\right]|i\rangle
\label{eq:transition_general}
\end{align}
with $T$ ensuring a time-ordered product and $Z'$ $(Z)$ the charge of the mother (daughter). The typical approximation made in dealing with $\beta$ decay equates initial and final Coulomb interactions, replacing it instead with only the final. This is then corrected for by including radiative corrections and atomic final state influences discussed below. We choose to follow this approach, such that Eq. (\ref{eq:transition_general}) corresponds to using the solution of the Dirac equation with an electrostatic potential rather than a plane wave for the electron \cite{Roman1965}. It is this change which requires the introduction of the so-called Fermi function. Denoted by $F(Z, W)$, it takes into account the distortion of the electron radial wave function by the nuclear charge, i.e., the Coulomb interaction between the $\beta$ particle and the daughter nucleus. Many authors (e.g. \cite{Konopinski1941, Behrens1969}) start from the point charge Fermi function introduced by \textcite{Fermi1934}\footnote{An English summary of the Fermi theory was presented by, e.g., \textcite{Konopinski1935, Konopinski1943}.}
\begin{equation}
F_0(Z, W) = 4 (2pR)^{2(\gamma - 1)} e^{\pi y} \frac{|\Gamma(\gamma + i y)|^2}{\left( \Gamma(1 + 2\gamma) \right)^2},
\label{eq:F_0}
\end{equation}
\noindent with
\begin{equation}
\gamma = \sqrt{1 - (\alpha Z)^2}, \quad y = \pm \alpha Z W / p  .
\end{equation}
Here, $R$ is the cut-off radius in the evaluation of the Dirac equation for the electron or positron necessitated by its slight divergence at the origin for a fictitious point charge. This cut-off radius is taken to represent the radius of the daughter nucleus, although it has in fact no real physical significance.

Throughout the vast literature of $\beta$ decay many different definitions of the Fermi function have appeared. We attempt to provide a short overview of the most frequently used, and stress that no formulation is in itself superior. Each requires additional corrections stemming from, e.g., the finite size of the nucleus and its precise shape. These have been estimated to various orders of precision and are discussed in the next sections. This is not to say the Fermi function as written, for example, in Eq. (\ref{eq:F_0}) carries no merit. It absorbs most of the $Z$ and $W$ dependence of the tree-order Coulomb interaction, whereas additional corrections are typically limited to at most a few percent.

Before we summarize the different Fermi functions, we briefly discuss the relativistic solution of an electron inside a spherical Coulomb potential. The seminal work by \textcite{Rose1961} elaborates on a fully analytical solution of the Dirac equation for an electron in a hydrogenic approximation. In general the solution can be written as
\begin{equation}
\Psi_{\kappa} = \left(
\begin{array}{c}
\text{sign}(\kappa)f_{\kappa}(r) \sum_{\mu}\chi_{-\kappa}^{\mu} \\
g_{\kappa}(r) \sum_{\mu}\chi_{\kappa}^{\mu}
\end{array}
\right),
\label{eq:sol_dirac}
\end{equation}
with $\chi_{\kappa}^{\mu}$ the traditional spin-angular functions, and $\kappa$ the eigenvalue of the operator\footnote{Here $1$ is a 4$\times$4 unit matrix, $\vec{\sigma}$ stands for the Pauli matrices in four dimensions, $\vec{L}$ is the orbital angular momentum operator, and $\beta$ is the equal to the $\gamma_0$ Dirac matrix. For $s$ orbitals this evaluates to $\kappa = -1$.} $K = \beta(\bm{\sigma} \bm{L}+1)$. For allowed decay all states with $j=1/2$ contribute, meaning $\kappa = -1$ for $s_{1/2}$ and $\kappa=1$ for $p_{1/2}$ orbitals. All Coulombic information is then encoded in the radial wave functions $f_{\kappa}(r)$ and $g_{\kappa}(r)$. When calculating the transition matrix element, one typically extracts the dominant terms of the square of the electron wave function throughout the nucleus, evaluated at some radius $r$, to define\footnote{Additional corrections stemming from the `small' terms $f_{-1}$ and $g_1$ can be written as a multiplication of Eq. (\ref{eq:F_general}) and some power expansion in $(pr)$ and $\alpha Z$, and are included in the convolution finite size correction discussed in Sec. \ref{sec:nuclear_structure}.}
\begin{equation}
\label{eq:F_general}
F(Z, W, r) = \frac{f_1^2(r) + g_{-1}^2(r)}{2 p^2}.
\end{equation}
Historically, this has been evaluated either at the origin or at the nuclear radius. Using the analytical solutions of the Dirac equation in a central Coulomb potential, one finds the factor $4$ in Eq. (\ref{eq:F_0}) replaced by $(1+\gamma)/2$. This was noted by \textcite{Konopinski1941}, and has since been adopted by several authors \textcite{Fano1952, Blatt1952, Konopinski1966, Wilkinson1989a}. Others \cite{Greuling1942, Dismuke1952}, including the seminal text books of \textcite{Schopper1966} and \textcite{Behrens1982}, adhere to the formulation of Eq. (\ref{eq:F_0}). The factor $(1+\gamma)/2$ is then absorbed into an $L_0$ correction, which is also responsible for modifications due to the finite nuclear size. To remain consistent with the latter works, we have opted to do the same and use Eq. (\ref{eq:F_0}) as our Fermi function, and discuss, among others, the $L_0$ correction factor in the following sections.

Additional definitions of the Fermi function have attempted to extend Eq. (\ref{eq:F_0}) by either including finite size corrections, through evaluation at the origin or both\footnote{Tabulations by \textcite{Bhalla1961, Bhalla1964} have been shown to contain errors for $\beta^+$ decay and have issues with double counting \cite{Huffaker1967}, thereby requiring a large correction \cite{Calaprice1976}. We therefore do not consider the Bhalla-Rose Fermi function.}. A frequent formulation that performs both of these is that by \textcite{Behrens1969}
\begin{equation}
F_{BJ}(Z, W) = ~ F_0(Z, W) ~ \left( 1 \mp \frac{13}{15} \alpha Z W R + ... \right),
\label{eq:FBJ_F0}
\end{equation}
discussed, for example, by \textcite{Calaprice1976}. The additional term can easily be traced from the power expansion of $f_1$ and $g_{-1}$ inside the nucleus by \textcite{Huffaker1967}
\begin{align}
\left\{\begin{array}{c}
g_{-1}(r) \\
f_1(r)
\end{array} \right\} &\propto [p^2F_0(Z, W)]^{1/2}\left\{1-\left[\frac{13}{30} + \frac{1}{2}\left(\frac{r}{R}\right)^2 \right] \right.\nonumber \\
&\left.\times \alpha Z W R - \frac{1}{6}(pr)^2\right\}
\label{eq:expansion_huffaker}
\end{align}
and understood as a rough electrostatic finite size correction. The additional correction stemming from the convolution of the leptonic wave functions with the weak charge distribution then brings about nuclear structure-sensitive $\langle r^2 \rangle$ terms. \textcite{Calaprice1976} has treated this with slightly higher accuracy. Since the expansion performed by \textcite{Huffaker1967} neglects terms of order $(\alpha Z)^2$, the aforementioned factor $(1+\gamma)/2$ is absent. The error introduced by this is of the order of 0.5\% for $Z=20$.

As mentioned above, we split up all corrections in the current work. The `finite size' corrections are split up into an \emph{electrostatic} finite size correction, stemming from the electric potential difference, and a \emph{convolution} finite size correction from an integration over the nuclear volume, and sensitive to nuclear structure. A discussion of the latter is postponed until Sec. \ref{sec:nuclear_structure}.

In order to facilitate the discussion of the correction terms on Eq. (\ref{eq:F_0}), we take Eq. (\ref{eq:F_general}) one step further. The results obtained by \textcite{Fermi1934} and \textcite{Konopinski1941} used analytical solutions of the Dirac equation in a simple Coulomb potential for a point charge. Once we move away from this simple picture to include the finite nuclear size or atomic screening, no such analytical solutions are available. Even when this is impossible, however, the behavior of the electron radial wave functions can be developed in a power expansion close to the origin \cite{Behrens1971}
\begin{align}
\left\{\begin{array}{c}
f_{\kappa}(r) \\
g_{\kappa}(r)
\end{array} \right\}
&=
\,\alpha_{\kappa}\{(2|\kappa|-1)!!\}^{-1}(pr)^{|\kappa|-1} \nonumber \\
&~~~\times \sum_{n=0}^{\infty}
\left\{ \begin{array}{c}
a_{\kappa n} \\
b_{\kappa n}
\end{array} \right\}
r^n,
\label{eq:erwf_extensive}
\end{align}
where the electrostatic information is now contained in the so-called `Coulomb amplitudes' $\alpha_{\kappa}$, and $a_{\kappa n}, b_{\kappa n}$ are iteratively defined parameters. In this picture, we are not any more hampered in evaluating the electron wave function at the origin and write instead \cite{Behrens1982}
\begin{equation}
F_0L_0 = \frac{\alpha_{-1}^2 + \alpha_1^2}{2p^2}.
\label{eq:fermi_buhring}
\end{equation}
This result is valid, regardless of the charge distribution of the nucleus even though only a point charge solution is analytically solvable.

\section{Finite mass and electrostatic finite size effects}
\label{size-and-mass}

Attributing a finite mass to the nuclear opens up the decay kinematics from a two-body to three-body process. This has consequences for both the outgoing energy of the lepton fields, as well as the electromagnetic corrections where one considered the nuclear Coulomb potential as static. Giving the nucleus now also a finite size introduces several more corrections on the $S$ matrix as its nuclear volume is not any more a simple delta function. As discussed in the previous section the traditional Fermi function is calculated based on an infinitely heavy point-charge model of the nucleus. As a consequence the electron radial wave functions diverge slightly at the origin and are instead evaluated at the nuclear radius $R$. This requires several corrections, stemming from the finite size of the nucleus. We stress again the ambiguity in available literature when discussing the so-called finite size corrections. For this reason, we have explicitly separated this into an \emph{electrostatic} and \emph{convolution} part. The former originates from a difference in the electrostatic potential when moving to a more realistic nuclear shape, whereas the latter takes into account the integration of the leptonic wave functions throughout the nuclear volume and includes the `small' Coulomb terms $f_{-1}$, $g_{1}$. We focus here on the electrostatic finite size corrections, and leave the convolution part for Sec. \ref{sec:nuclear_structure}.

\subsection{$L_0(Z, W)$, $U(Z, W)$ and $D_\text{FS}(Z, W, \beta_2)$: Electrostatic finite nuclear size corrections}
The deviation from a point charge distribution inherently changes the electron density inside the nucleus, and as such introduces additional electromagnetic corrections to the Fermi function. We discuss the consequence on Eq. (\ref{eq:fermi_buhring}) from moving to a uniformly charged sphere through $L_0$, and go even further to a Fermi distribution in $U$. \textcite{Wilkinson1990, Wilkinson1993b} has discussed the former effects in detail and provided analytical expressions for each of them. Finally, we discuss the electrostatic effect of nuclear deformation, $D_{\text{FS}}$. 

\subsubsection{$L_0(Z, W)$}

When considering instead of a point nucleus one of finite size, the electron and positron wave functions become finite at the center of the nucleus. Any other description but a point charge is, however, not analytically solvable, and so an approximate correction factor, $L_0(Z, W)$, is introduced to be used in combination with the analytical Fermi function. In order to remain consistent with the formulation of the original Fermi function in Eq. (\ref{eq:F_0}), the nucleus is presented by a uniformly charged sphere of radius $R$ that is adjusted to give the experimental $\langle r^2 \rangle^{1/2}$ of the daughter nucleus, i.e.
\begin{equation}
R = \sqrt{5/3} \langle r^2 \rangle^{1/2} .
\end{equation}
According to Eq. (\ref{eq:fermi_buhring}), $L_0$ is defined as embracing all information that is not contained in the Fermi function of Eq. (\ref{eq:F_0}) when evaluating the dominant electron components at the origin. As discussed in the previous section, this entails that even when considering a point charge, $L_0$ differs from unity since $L_0 \sim (1+\gamma)/2$. Analogous to the expansion of $f_{1}$ and $g_{-1}$ in Eq. (\ref{eq:expansion_huffaker}), \textcite{Behrens1982} performed an expansion to higher order (see Eqs. (\ref{eq:erwf_extensive}) and (\ref{eq:fermi_buhring})) and find after a straightforward calculation
\begin{align}
L_0 &\simeq \frac{1+\gamma}{2}\left[1\mp \alpha Z W R + \frac{7}{15}(\alpha Z^2)-\frac{1}{2}\gamma \frac{\alpha Z R}{W} \right] \nonumber \\
&\simeq 1 \mp \alpha Z W R + \frac{13}{60}(\alpha Z^2)-\frac{1}{2}\frac{\alpha Z R}{W}
\label{eq:L0_approx_BB}
\end{align}
for $Z$ sufficiently low. Here $1-\gamma$ was approximated as $\frac{1}{2}(\alpha Z^2)$ and $\gamma$ as unity. More precise numerical calculations were tabulated by \textcite{Behrens1969} using $R=1.20 A^{1/2}\,$fm close to the valley of stability. This was extended by \textcite{Wilkinson1990} to cover all isotopes between the proton and neutron drip lines for $Z \leq 60$. Together with a slight generalization of the prefactors in Eq. (\ref{eq:L0_approx_BB}) (keeping terms proportional to $\gamma$ alive rather than approximate them to unity), numerical results were fitted after inclusion of simple power expansions. The result was presented in analytical form as
\begin{align}
L_0(Z, W) = & ~ 1 ~ + ~ \frac{13}{60} \left( \alpha Z \right)^2 \mp ~ \frac{ \alpha Z W R (41 - 26\gamma)}{[15(2\gamma -1)]}   \nonumber \\
& \mp ~ \frac{\alpha Z R \gamma (17-2\gamma)}{[30 W (2\gamma -1)]}   \nonumber \\
& + ~ a_{-1} \frac{R}{W} + \sum_{n=0}^{5} a_n (WR)^n   \nonumber \\
& + ~ 0.41 (R - 0.0164)(\alpha Z)^{4.5} ~ ,
\label{L0}
\end{align}
for electrons, while for positrons the 0.41 in the last term of this equation is to be replaced by 0.22. The $a_n$-values ($n = -1, 0, 1 ... 5$) are given by the parametrisation
\begin{equation}
a_n  = \sum_{x=1}^{6}b_{x, n} (\alpha Z)^x ~ ,
\end{equation}
with $b_{x, n}$ ($x = 1 ... 6$) being listed in the Tables 1 and 2 of Ref.~\textcite{Wilkinson1990} for electrons and positrons, respectively, and reproduced here in Tables~\ref{Table 5 in Huber} and \ref{Table 2 in Wilkinson509}\footnote{Note that the signs of odd powers of $Z$ have been flipped for positrons to remain consistent with our initial notation. Specifically, $Z$ is always defined as a positive quantity, with the upper (lower) sign for $\beta^-$ ($\beta^+$) decay.} (Note that in Table \ref{Table 2 in Wilkinson509} the number 0.066483 under $b_2$ was replaced by 0.066463 and the number 2.83606 under $b_4$ was replaced by 2.63606 \cite{Wilkinson1993b}). These parametrizations yield $L_0(Z, W)$ accurately to 1 part in 10$^4$ for $p \leq$~45 and for $|Z| \leq$~60. The effect of $L_0(Z, W)$ ranges from a few 0.1\% up to several \% and so is everywhere highly significant.
\begin{table*}
\caption{\label{Table 5 in Huber} Coefficients for the parametrization of $L_0(Z, W)$ for electrons. Reproduction of Table I in \cite{Wilkinson1990}.}
\begin{ruledtabular}
\begin{tabular}{c c c c c c c}
 & {$b_1$} & {$b_2$} & {$b_3$} & {$b_4$} & {$b_5$} & {$b_6$} \\
\hline
$a_{-1}$ & 0.115 & -1.8123 & 8.2498 & -11.223 & -14.854 & 32.086 \\
$a_0$ & -0.00062 & 0.007165 & 0.01841 & -0.53736 & 1.2691 & -1.5467 \\
$a_1$ & 0.02482 & -0.5975 & 4.84199 & -15.3374 & 23.9774 & -12.6534 \\
$a_2$ & -0.14038 & 3.64953 & -38.8143 & 172.1368 & -346.708 & 288.7873 \\
$a_3$ & 0.008152 & -1.15664 & 49.9663 & -273.711 & 657.6292 & -603.7033 \\
$a_4$ & 1.2145 & -23.9931 & 149.9718 & -471.2985 & 662.1909 & -305.6804 \\
$a_5$ & -1.5632 & 33.4192 & -255.1333 & 938.5297 & -1641.2845 & 1095.358 \\
\end{tabular}
\end{ruledtabular}
\end{table*}
\begin{table*}
\caption{\label{Table 2 in Wilkinson509} Coefficients for the parametrization of $L_0(Z, W)$ for positrons. Reproduction of Table I in \cite{Wilkinson1990}, with small modifications as discussed in the text. The signs used in odd powers of $Z$ have been flipped to agree with the convention used in this work.}
\begin{ruledtabular}
\begin{tabular}{c c c c c c c}
 & {$b_1$} & {$b_2$} & {$b_3$} & {$b_4$} & {$b_5$} & {$b_6$} \\
\hline
$a_{-1}$ & 0.0701 & -2.572 & 27.5971 & -128.658 & 272.264 & -214.925 \\
$a_0$ & -0.002308 & 0.066463 & -0.6407 & 2.63606 & -5.6317 & 4.0011 \\
$a_1$ & 0.07936 & -2.09284 & 18.45462 & -80.9375 & 160.8384 & -124.8927 \\
$a_2$ & -0.93832 & 22.02513 & -197.00221 & 807.1878 & -1566.6077 & 1156.3287 \\
$a_3$ & 4.276181 & -96.82411 & 835.26505 & -3355.8441 & 6411.3255 & -4681.573 \\
$a_4$ & -8.2135 & 179.0862 & -1492.1295 & 5872.5362 & -11038.7299 & 7963.4701 \\
$a_5$ & 5.4583 & -115.8922 & 940.8305 & -3633.9181 & 6727.6296 & -4795.0481 \\
\end{tabular}
\end{ruledtabular}
\end{table*}

\subsubsection{$U(Z, W)$}
\label{sec:U}
In the previous section we have reported on the finite size correction when the nuclear shape was assumed to be uniformly spherical with a fixed radius $R$. In reality the nuclear charge distribution is smeared out over a certain distance.
This in turn introduces an additional, smaller, correction term from replacing the uniform spherical charge distribution that defines $L_0(Z, W)$ with a more realistic one having the same $\langle r^2 \rangle^{1/2}$ \cite{Behrens1970, Behrens1972}. Throughout the times several different distributions have been proposed, ranging from a simple Yukawa form to Fermi functions and even Gudermannian combinations \cite{Asai1974, Ogata1974}. An extensive overview of oft-encountered charge distributions in nuclear physics is given by \textcite{Andrae2000}, together with some basic properties. We will explicitly consider here the modified Gaussian and Fermi potentials. The first is given by
\begin{equation}
\rho_{MG}(r) = N_0\left\{1+A\left(\frac{r}{2}\right)^2\right\}e^{-(r/a)^2},
\label{eq:mod_gauss}
\end{equation}
where
\begin{equation}
N_0 = \frac{8}{2+3A}a^{-3}\pi^{-1/2}
\end{equation}
is a normalization constant and $a$ is constrained by $\langle r^2 \rangle = 3/5R^2$ to find
\begin{equation}
a = R\left[\frac{5}{2}(2+5A)/(2+3A)\right]^{-1/2}.
\end{equation}
Here $A$ is a free parameter. The second is a basic Fermi distribution
\begin{equation}
\rho_F(r) \sim \left\{ 1 + \exp\left[ (r-c)/a \right] \right\}^{-1} .
\label{eq:fermi_distribution}
\end{equation}
The former agrees well with charge distributions of low $p$-shell and $sd$-shell nuclei, while the latter more accurately resembles the shape of heavier nuclei. 

We can now provide an analytical approximation of the induced effect by calculating the $L_0$ correction for different charge distributions for low $Z$. The calculation is analogous to that required to obtain Eq. (\ref{eq:L0_approx_BB}). Cancelling common normalization factors and constants, we find 
\begin{align}
\frac{L_0'}{L_0} &= \frac{\alpha_{1}'^2+\alpha_{-1}'^2}{\alpha_1^2+\alpha_{-1}^2} \\
&= \frac{B_1'^{-2}+B_{-1}'^{-2}}{B_{1}^{-2}+B_{-1}^{-2}},
\end{align}
where we have neglected terms of order $\mathcal{O}(WR^2)$, and primes stand for the more advanced charge distribution. Now $B_{\pm 1}^{(')}$ is given by
\begin{equation}
B_{\pm 1}^{(')} = \pm f^I_{\pm 1}(R)g_{02}(R)-g^I_{\pm 1}(R)f_{02}(R).
\end{equation}
Here the superscript $I$ denotes the wave functions inside the nucleus, and $f_{02}$ and $g_{02}$ are radial Coulomb functions that depend on the sign of $\kappa$. The functions inside the nucleus can be found through a series expansion of the solution according to Eq. (\ref{eq:erwf_extensive}) and matching powers in the Dirac equation. This gives a recursive relationship for $a_n$ and $b_n$. The last quantity needed depends on the order of the potential. In the case of a uniformly charged sphere, i.e. a second order dependence, the large wave functions must be expanded up to $n=6$. In general this is $n=2(m+1)$, where $m$ is the largest, relevant radial order of the potential. The radial expansion of the Coulomb functions $f_{02}$ and $g_{02}$ is known and can be plugged in. Expanding results up to order $\mathcal{O}(\alpha ZR)$, $\mathcal{O}((WR)^2)$ and $\mathcal{O}((\alpha Z)^2)$, we find after a tedious but straightforward calculation
\begin{align}
U(Z, W) &\equiv \frac{L_0'}{L_0} \nonumber \\
&\approx 1 +\alpha Z W R \Delta_1 + \frac{\gamma}{W}\alpha Z R \Delta_2 \nonumber \\
&+(\alpha Z)^2\Delta_3 - (WR)^2\Delta_4
\label{eq:U_analytical}
\end{align}
where
\begin{subequations}
\begin{align}
\Delta_1 &= \frac{4}{3}\Delta v_0 + \frac{17}{30}\Delta v_2 + \frac{25}{63}\Delta v_4, \\
\Delta_2 &= \frac{2}{3}\Delta v_0 + \frac{7}{12}\Delta v_2 + \frac{11}{63}\Delta v_4, \\
\Delta_3 &= \frac{1}{3}\Delta v_0^2 + \frac{1}{15}\Delta v_2^2 + \frac{1}{35}\Delta v_4^2 + \frac{1}{6}\Delta v_2v_0  \nonumber \\
&+ \frac{1}{9}\Delta v_4v_0 +\frac{1}{20}\Delta v_4v_2 + \frac{1}{5}\Delta v_2 + \frac{1}{7}\Delta v_4, \\
\Delta_4 &= \frac{4}{3}\Delta v_0 +\frac{4}{5}\Delta v_2 + \frac{4}{7}\Delta v_4,
\end{align}
\end{subequations}
taking into account even-$r$ potentials up to fourth order, and defining $\Delta v_n = v_n'-v_n$. Here we define $v_n$ through
\begin{equation}
V(r) = \sum_{n=0}^{\infty}V_nr^n = \sum_{n=0}^{\infty}\left(-\frac{\alpha Z}{R^{n+1}}\right)v_n r^n.
\end{equation}
As an example, the electrostatic potential of the modified Gaussian distribution of Eq. (\ref{eq:mod_gauss}) is
\begin{equation*}
V(r) = -\alpha Z\left[\frac{\text{erf}(r/a)}{r}-\frac{2A}{(2+3A)a\sqrt{\pi}}\exp(-(r/a)^2) \right]
\end{equation*}
where erf is the error function. Expanding all terms we find
\begin{subequations}
\begin{align}
v_0 &= \sqrt{\frac{5}{2}}\frac{4(1+A)(2+5A)^{1/2}}{\sqrt{\pi}(2+3A)^{3/2}}\\
v_2 &= -\frac{4}{3(3A+2)\sqrt{\pi}}\left(\frac{5(2+5A)}{2(2+3A)}\right)^{3/2} \\
v_4 &= \frac{2-7A}{5(3A+2)\sqrt{\pi}}\left(\frac{5(2+5A)}{2(2+3A)}\right)^{5/2}
\end{align}
\end{subequations}
compared to $v_0=3/2$, $v_2=-1/2$ and $v_4=0$ for the uniformly charged sphere. For sufficiently low $Z$, Eq. (\ref{eq:U_analytical}) can be employed for any charge distribution.

For higher $Z$ the contributions from higher order terms neglected in Eq. (\ref{eq:U_analytical}) cannot be any more ignored, and computational methods must be employed. \textcite{Wilkinson1993b} has done this for the Fermi distribution of Eq. (\ref{eq:fermi_distribution}) with $a \simeq 0.55\,$fm as it agrees quite well with Hartree-Fock calculations and experimental data. The correction is then written as
\begin{equation}
U(Z, W) = 1 + \sum_{n=0}^{2} a_n p^n ~ ,
\label{eq:nuclear_diffuse}
\end{equation}
\noindent where
\begin{align}
a_0 &= -5.6 \times 10^{-5} \mp 4.94 \times 10^{-5}Z +6.23 \times 10^{-8}Z^2, \nonumber \\
a_1 &= 5.17 \times 10^{-6} \pm 2.517 \times 10^{-6}Z +2.00 \times 10^{-8}Z^2, \nonumber \\
a_2 &= -9.17 \times 10^{-8} \pm 5.53 \times 10^{-9}Z +1.25 \times 10^{-10}Z^2
\end{align}
in natural units. The effect of $U(Z, W)$ amounts to some 0.1\% for medium-high masses and so cannot be neglected. As the precise shape of the nucleus remains model-dependent, some uncertainty remains. Experimental electron scattering data have, however, been fitted and tabulated by \textcite{DeVries1987}, allowing one to make a very good estimate. Much work in mean-field theories has also been presented over the years \cite{Anni1995, Anni1995a, Moller2016}. 

Interesting to note is the applicability of Eq. (\ref{eq:U_analytical}) when used in combination with Eq. (\ref{eq:nuclear_diffuse}), for example, in describing the correction due to the oft-used three-parameter Fermi function, also known as the `wine-bottle' distribution \cite{Andrae2000, Towner2015}
\begin{equation}
\rho_{}(r) = \rho_0(1+w(r/c)^2)\left\{1+\exp((r-c)/a)\right\}^{-1}
\end{equation}
where $w$ describes the central depression. As this distribution closely resembles that of the normal Fermi distribution for small $w$, Eq. (\ref{eq:nuclear_diffuse}) can be used as a first approximation, after which Eq. (\ref{eq:U_analytical}) describes the difference in two Fermi distributions. As the difference in $v_n$ will be small, Eq. (\ref{eq:U_analytical}) remains relevant even for higher $Z$.

\subsubsection{$D_\text{FS}(Z, W, \beta_2)$}
\label{sec:D_FS}

\begin{figure}[h!]
\centering
\includegraphics[width=0.45\textwidth]{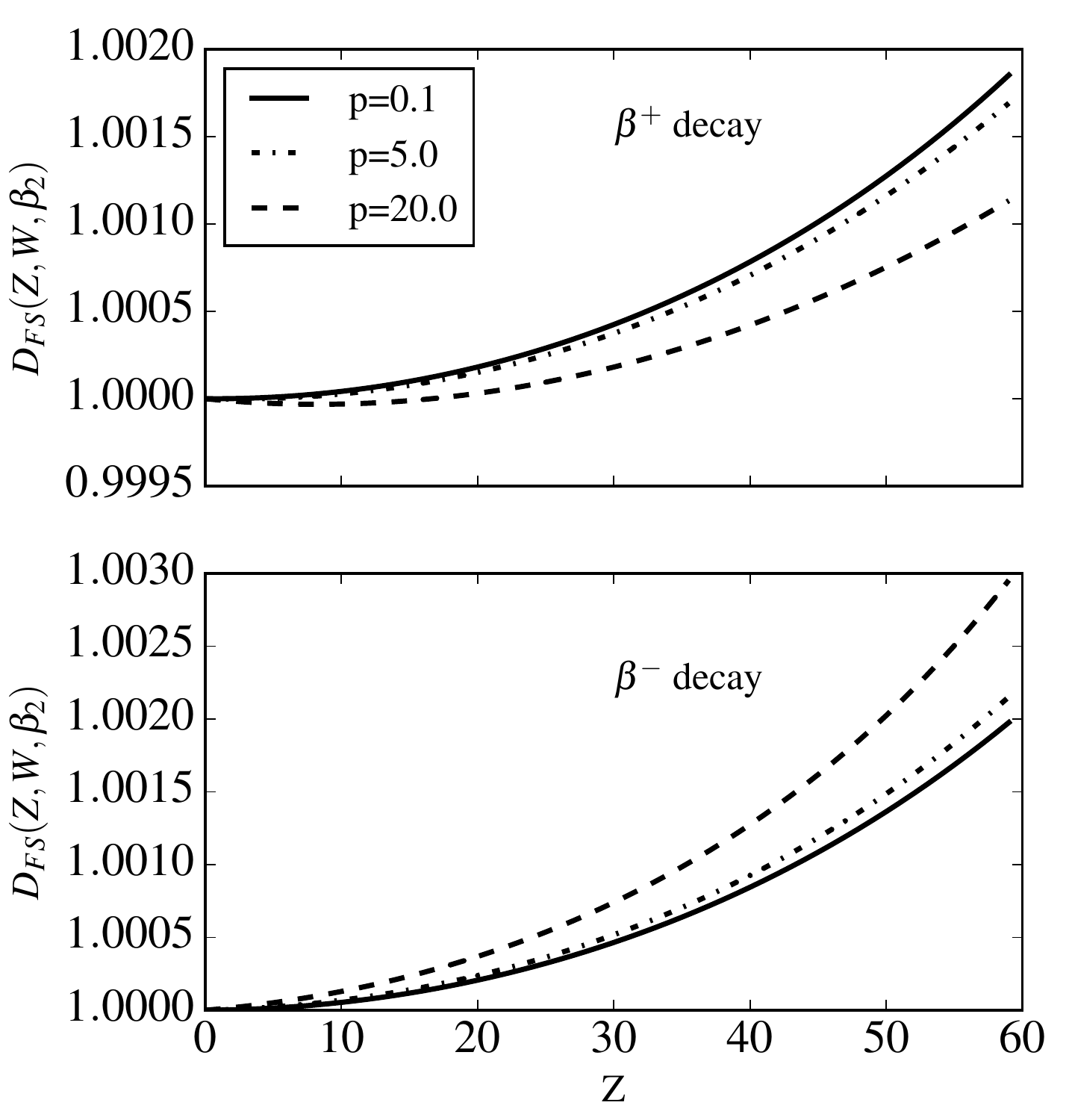}
\caption{Shown are several examples of $D_\text{FS}$ for different momenta, using $\beta_2 = 0.2$. Interesting to note is the reversal of the energy dependence when switching between $\beta^-$ (positive) and $\beta^+$ (negative) decay. Important to note is that even though $D_\text{FS}$ can grow to several parts per $10^3$, it has to be combined with $D_C$ from Sec. \ref{sec:isospin_breakdown}, which decreases the overall effect.}
\label{fig:D_FS}
\end{figure}

Just as in the previous section wherein we considered the change in the Fermi function due to a more realistic, though still spherically symmetric, charge distribution, we must account for a possible nuclear deformation. We limit ourselves here to axially symmetric deformations, writing the surface of our ellipsoid as\footnote{This is merely an approximation, as our factor unity should be replaced by $a_0 = 1-\beta_2^2/4\pi$ to conserve the total volume of the undistorted sphere to second order. This effect is negligible, however, and we continue with Eq. (\ref{eq:R_deformed}).}
\begin{equation}
R(\theta, \phi) = R_0\left(1+\beta_2 Y_{2}^0(\theta, \phi)\right)
\label{eq:R_deformed}
\end{equation}
with $\beta_2$ the traditional measure of quadrupole deformation \cite{Davidson1968, Eisenberg1975}. In the case $\beta_2 >0$, the nucleus has a prolate shape while it is oblate for $\beta_2 < 0$. This deformation gives rise to a non-zero intrinsic electric quadrupole moment typically parametrized as
\begin{equation}
Q_0 = 3\sqrt{\frac{\alpha}{5\pi}}R_0^2Z\beta_2(1+0.16\beta_2)
\label{eq:EQM_deformed}
\end{equation}
in our natural units. This implies experimental $\beta_2$ values can be obtained from measured electric quadrupole measurements, or from theoretical models such as the work by \textcite{Moeller2015}. Following the approach of \textcite{Wilkinson1994}, we define $a$ and $b$ to be the axes perpendicular and along the symmetry axis, respectively, such that $b/a > 1$ for prolate deformations. Combining Eq. (\ref{eq:EQM_deformed}) with Eq. (25) from \textcite{Wilkinson1994} we find
\begin{equation}
\frac{b}{a} = \sqrt{\frac{1+C}{1-C/2}}
\end{equation}
with
\begin{equation}
C = \frac{5}{3}\sqrt{\frac{5}{\pi}}\beta_2(1+0.16\beta_2).
\end{equation}
Together with the requirement that
\begin{equation}
\langle r^2 \rangle = \frac{1}{5}(b^2+2a^2),
\end{equation}
$a$ and $b$ are uniquely determined.
Defining an angle-averaged charge distribution as
\begin{equation}
\rho(r) = \left\{\begin{array}{lcl}
\rho_0 & \text{for} & 0 \leq r \leq a \\
\rho_0\left(1-\frac{b}{r}\left[\frac{r^2-a^2}{b^2-a^2} \right]^{1/2} \right) & \text{for} & a < r \leq b
\end{array} \right.
\label{eq:charge_dist_deformed}
\end{equation}
where
\begin{equation}
\rho_0 = \frac{3}{4\pi}\frac{Z}{a^2b},
\end{equation}
we have transitioned from a deformed to a spherically symmetric charge distribution with which we can continue\footnote{Equation (\ref{eq:charge_dist_deformed}) is valid for prolate deformations. For an oblate nucleus, one simply interchanges $a$ and $b$ in the $r$ ranges, and reverses the signs in the square brackets. We assume prolate deformations unless otherwise specified.}. We define $\rho_0$ such that the total charge is equal to $Z$. Due to the expected smallness of the correction, we continue in an analytical fashion and consider the charge distribution as a continuous superposition of uniform spheres of radius $r$ and density d$\rho/$d$r$. We write then a modified finite size correction
\begin{align}
L_0(Z, W, \beta_2)^* &= \frac{4}{3}\pi R^{2(1-\gamma)} \nonumber \\
&\times \int_0^{\infty} r^3 \frac{d\rho}{dr}L_0(Z, W, r)r^{2(\gamma-1)}dr
\label{eq:L0_deformed}
\end{align}
where $L_0(Z, W, r)$ denotes the use of Eq. (\ref{L0}) with $R$ replaced by the continuous variable. In the case of a uniformly charged sphere we have $d\rho/dr = 3/(4\pi R^3)\,\delta(R)$ with $\delta(x)$ the Dirac delta function, trivially satisfying Eq. (\ref{eq:L0_deformed}). The ratio
\begin{equation}
D_{\text{FS}}(Z, W, \beta) = \frac{L_0(Z, W, \beta_2)^*}{L_0(Z, W)}
\label{eq:D_FS}
\end{equation}
then constitutes the deformed nuclear shape correction to the Fermi function. Figure \ref{fig:D_FS} shows the magnitude of the effect for different $Z$ at several momenta for a reasonably large deformation $\beta_2 = 0.2$. Important to note is that this has to be combined with the other deformation-dependent effects, discussed in the nuclear structure Sec. \ref{sec:isospin_breakdown}. This decreases the overall effect, and cannot be neglected.

\subsection{$R_N(W, W_0, M)$ and $Q(Z, W, M)$: Finite nuclear mass corrections}
Up to now we have approximated $\beta$ decay as a two-body phase space, whereas it should in fact be a three-body phase space opened up by the recoil of the nucleus. It was thus previously tacitly assumed that the nucleus is infinitely massive. The deviation from an infinitely massive nucleus introduces a kinematic recoil correction, $R_N$, and a subsequent small electromagnetic correction, $Q$. We briefly discuss both in part.

\subsubsection{$R_N(W, W_0, M)$}
\label{sec:R_N}

The effect of the recoil after $\beta$ decay of a nucleus of finite mass $M$ ($M = AM_{\text{nucleon}} - B$, with $M_{\text{nucleon}} = 1837.4$ the mass of the nucleon in units of $m_ec^2$, and $B$ the binding energy) is to multiply the phase space by a factor\footnote{Note that we have added the subscript $'N'$ to this factor originally defined as $R(W, W_0, M)$ in \textcite{Wilkinson1989a} to differentiate it from the radiative corrections.} $R_N(W, W_0, M)$ \cite{Horowitz1948, Kofoed-Hansen1948, Shekhter1959, Wilkinson1982, Wilkinson1990}
\begin{equation}
R_N(W, W_0, M) = 1 + r_0 + r_1 / W + r_2  W + r_3 / W^2   ~ ,
\label{R_N}
\end{equation}
\noindent
where for vector decay (Fermi transitions)
\begin{subequations}
\begin{align}
r_0^V &= W_0^2 /(2 M^2) - 11 /(6 M^2), \\
r_1^V &= W_0 /(3 M^2),  \\
r_2^V &= 2/M - 4 W_0 /(3 M^2), \\
r_3^V &= 16 /(3 M^2) ~ ,
\end{align}
\end{subequations}
\noindent
and, for axial decay (Gamow-Teller transitions)
\begin{subequations}
\begin{align}
r_0^A &= -2 W_0 /(3 M) - W_0^2 /(6 M^2) - 77 /(18 M^2), \\
r_1^A &= -2 /(3 M) + 7 W_0 /(9 M^2), \\
r_2^A &= 10/(3 M) - 28 W_0 /(9 M^2), \\
r_3^A &= 88 /(9 M^2) ~ .
\label{ri-A}
\end{align}
\end{subequations}
\noindent
This is a small effect of order 10$^{-5}$ to 10$^{-3}$ at most (see Fig.~1  by \textcite{Wilkinson1990}). For mixed transitions these corrections can simply be added together with the appropriate weighting factors
\begin{equation}
\frac{1}{1+\rho^2} \quad \text{and} \quad \frac{1}{1+\rho^{-2}}
\end{equation}
for Fermi and Gamow-Teller transitions, respectively and with $\rho = \frac{C_AM_{GT}}{C_VM_F}$ as defined before.

\subsubsection{$Q(Z, W, M)$}
\label{sec:Q}
A final consequence of the finite nuclear mass and consequent recoil is a change in the Coulomb field in which the departing electron or positron moves. It is not fixed in space but is itself recoiling against the combined lepton momenta so that the field experienced by the $\beta$ particle differs with time from what it would have been if the nucleus would not be recoiling, as is assumed for the Fermi function in Eq.~(\ref{eq:F_0}). \textcite{Wilkinson1982, Wilkinson1993b} calculated the effect of this retreat of the source of the Coulomb field from the combined lepton momenta for a pure vector (Fermi) transition. This can be generalized for mixed Fermi/Gamow-Teller transitions to \cite{Wilkinson1982}:

\begin{equation}
Q(Z, W, M) \simeq 1 \mp \frac{\pi \alpha Z}{M} \frac{1}{p} \left(1 + a\frac{W_0-W}{3W} \right) ~ ,
\label{eq:recoil_coulomb}
\end{equation}

with $a$ the $\beta$-$\nu$ correlation coefficient, $a = (1-\rho^2/3)/(1+\rho^2)$ \cite{Jackson1957}. For pure vector transitions one has $a=1$, while for pure Gamow-Teller transitions we find $a=-1/3$ \cite{Jackson1957}. The size of this correction amounts at most to a few percent of the typical error in the phase space factor $f$ due to the uncertainty in the $Q_{EC}$-value, and so is typically negligible, although Wilkinson retained it in the calculation of $f$-values for the superallowed Fermi transitions \cite{Wilkinson1993b}.

\section{Radiative corrections}
\label{sec:radiative_corr}
We started this work with a discussion concerning the Fermi function, a consequence of the electrostatic interaction of the daughter nucleus with the departing $\beta$ particle. Within the context of quantum electrodynamics (QED), the continuous exchange of photons is not the only radiative process occurring after the decay. These radiative corrections to the $\beta$ spectrum shape have a long history \cite{Berman1958, Kaellen1967}, and took a leap forwards with the work by \textcite{Sirlin1967}. A distinction was made between `inner' and `outer' radiative corrections, where the former is sensitive to the actual underlying weak interaction whereas the latter carries some nuclear dependence. The inner corrections can be calculated to high precision with standard electroweak methods, and are typically incorporated into effective coupling constants. The outer corrections, on the other hand, include energy-dependent terms and is our main concern here. Writing the uncorrected $\beta$ spectrum as $d\Gamma_0/dW$ we have then
\begin{equation}
\frac{d\Gamma}{dW} = \frac{d\Gamma_0}{dW}(1+\Delta_R^{V/A})(1+\delta_R(W, W_0))
\end{equation}
where $\Delta_R^{V/A}$ and $\delta_R(W, W_0)$ stand for inner and outer radiative corrections, respectively. The former has played an important role in establishing the universality of the electroweak interaction through comparison of the decay strength of muon decay \cite{Sirlin2013}. Its calculation was addressed twice by Marciano and Sirlin \cite{Marciano1986, Marciano2006}, the second time with improved precision, leading to the value $\Delta_R^V = (2.361 \pm 0.038)$\%. For the axial vector case the inner radiative correction is simply incorporated into the experimental value of $g_A$ from neutron decay.

As mentioned above, we will only concern us here with the outer radiative corrections, seeing as to how they are both energy- and nucleus-dependent. These corrections, contributing on top of those already included in the Fermi function, typically amount to a few percent and clearly cannot be neglected. Contributions are treated in a Feynman-driagram fashion, and concern both virtual photon exchange as well as one or more real photons in the final state. Seeing as to how these are experimentally relevant due to their possible detection, we briefly elaborate on the so-called inner bremsstrahlung. As this section contains no new results, we limit ourselves to listing the results obtained through decades of intensive study by a series of different authors. Excellent reviews have been provided by \textcite{Wilkinson1995b, Wilkinson1997}, \textcite{Towner2008} and \textcite{Sirlin2013}. For completeness, we briefly comment on the radiative corrections for the neutrino due to its relevance in the reactor neutrino oscillation studies.

\subsection{Total spectral influence}
We discuss here the energy-dependent part of the outer radiative correction to various orders in $Z^n\alpha^m$. Outside of the terms for which $m=n$ which are already contained in the Fermi function, it was found that $m$ must always be larger than $n$ \cite{Beg1972}. This is a crucial result, as, for example, $Z^2\alpha$ terms can easily exceed unity and overthrow the conserved vector current (CVC) hypothesis. Great attention has been given to the three lowest $m=n+1$ terms, each written as $\delta_m$. The outer radiative correction is typically expressed as
\begin{equation}
R(W, W_0) = 1 + \delta_R(W, W_0).
\label{eq:radiative}
\end{equation}
Previously \cite{Wilkinson1982, Wilkinson1997, Towner2002, Hardy2005a, Hardy2005c}, $\delta_R^{\prime}$ was simply the sum of $\delta_1, \delta_2$ and $\delta_3$. Recently, however, significant improvements have been made by \textcite{Marciano1986, Czarnecki2004} where the leading order contributions $\mathcal{O}(\alpha^n\ln^n(M_N/2W_0))$ have been summed via renormalisation group analysis. This allows one to write \cite{Towner2008, Sirlin2013}
\begin{align}
1+\delta_R = &\left\{1+\frac{\alpha}{2\pi}\left[g(W_0,W)-3\ln\frac{m_p}{2W_0}\right]\right\} \nonumber \\
& \times \left\{L(2W_0, m_p)+\delta_2+\delta_3 \right\},
\label{eq:delta_R_new}
\end{align}
where $m_p$ is the proton mass and 
\begin{equation}
L(2W_0, m_p) = 1.026725\left[1-\frac{2\alpha}{3\pi}\ln2W_0 \right]^{\frac{9}{4}},
\end{equation}
and $g(W_0, W)$ is described below. We discuss each of the $\delta_m$ separately below.

\subsubsection{Order $\alpha$ correction}
The lowest order radiative correction corresponds to one (virtual)-photon Feynman diagrams. This includes a renormalisation of the the weak vertex, electron and proton propagators but also internal brehmsstrahlung, discussed below. 
\begin{figure}[h]
\centering
\includegraphics[width=0.15\textwidth]{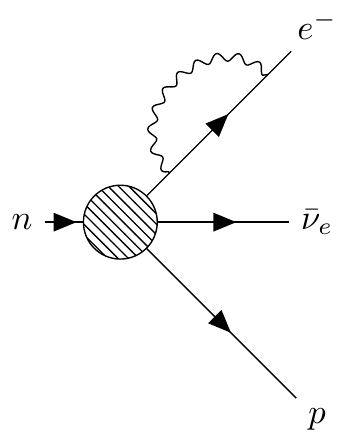}
\includegraphics[width=0.15\textwidth]{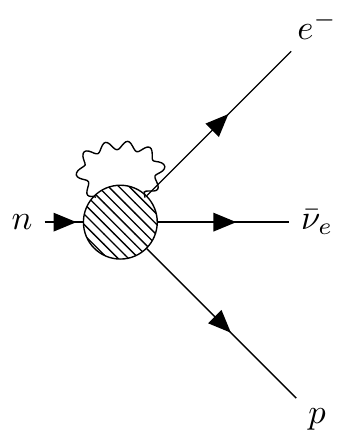}
\includegraphics[width=0.15\textwidth]{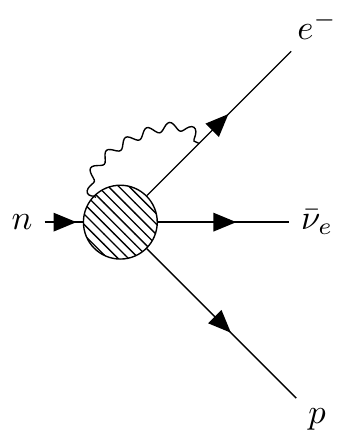}
\includegraphics[width=0.15\textwidth]{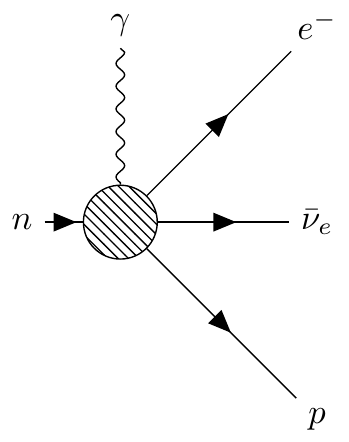}
\includegraphics[width=0.15\textwidth]{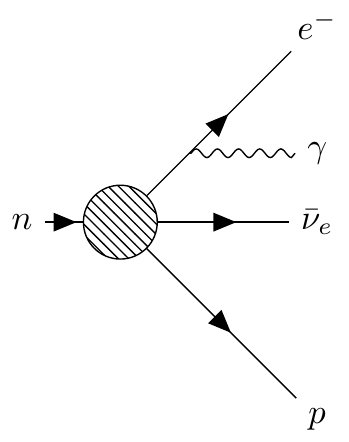}
\caption{Feynman diagrams for the order $\alpha$ corrections. The bottom row displays the inner bremsstrahlung corrections, discussed separately below (Sec. \ref{sec:IB}). This entails that when the final state photon goes undetected or can be distinguished from the $\beta$ particle, the spectrum of the latter is appropriately corrected for.}
\label{fig:feynman_order_alpha}
\end{figure}
The relevant Feynman diagrams are shown in Fig. \ref{fig:feynman_order_alpha}. Due to the zero photon mass, both the renormalization and inner bremsstrahlung processes individually create an infrared divergence which is cancelled when both contributions are added. Usually $\delta_1$ is noted with the well-known $g(W_0, W)$ function \cite{Sirlin1967, Sirlin2013}
\begin{widetext}
\begin{align}
g(W_0, W) =~ &3 \ln (m_p) - \frac{3}{4} + \frac{4}{\beta} \text{L}_s \left(\frac{2 \beta}{1 + \beta} \right) + 4 \left(\frac{\tanh^{-1} \beta}{\beta} - 1 \right) \left[\frac{W_0 - W}{3 W} - \frac{3}{2} + \ln [ 2 (W_0 - W)] \right]   \nonumber \\
 & + \frac{\tanh^{-1} \beta}{\beta} \left[ 2 (1+\beta^2) + \frac{(W_0-W)^2}{6W^2} - 4 \tanh^{-1} \beta \right] .
 \label{eq:g_sirlin}
\end{align}
\end{widetext}
\noindent with tanh$^{-1}$ the inverse hyperbolic tangent function, $M_N$ the nucleon mass, $\beta = p/W$ and
\begin{equation}
\text{L}_s = \int_0^x \frac{\ln(1-t)}{t} dt = - \sum_{k=1}^{k=\infty} \frac{x^k}{k^2}  \equiv -\text{Li}_2(x) ~ ,
\end{equation}
the Spence function, also known as the dilogarithm. Its large $W_0$-limit is \cite{Wilkinson1995b}
\begin{equation}
g(W_0 \rightarrow \infty, W) = \left[ 3 \ln \left(\frac{M_N}{2 W_0}\right) + \frac{81}{10} - \frac{4 \pi^2}{3} \right],
\end{equation}
\noindent which is dominated by the first term. Equation (\ref{eq:g_sirlin}) is universal in the sense that is the same for both electrons and positrons, and Fermi and Gamow-Teller decays independent of the nucleus. It is exact except for small terms of order $\mathcal{O}(\alpha (W/M)\ln(M/W))$ and $\mathcal{O}(\alpha q/M)$, where $q$ is the momentum transferred to the (anti)neutrino. 

Finally, we comment on a peculiar logarithmic divergence for $W=W_0$ in Eq. (\ref{eq:g_sirlin}). In the integration over the phase space this is clearly not a problem, but it points to a possible shortcoming in the analysis. It was shown that it is related to the emission of soft real photons \cite{Repko1983}. Two possibilities have been proposed to remove this divergence due to its relevance in the endpoint-sensitive tritium $\beta$-decay. The first \cite{Repko1983} proposes to sum these soft real photon contributions to all orders of perturbation theory, leading to the replacement
\begin{equation}
t(\beta)\ln (W_0-W) \to (W_0-W)^{t(\beta)}-1
\label{eq:radiative_exponentiation}
\end{equation}
where
\begin{equation}
t(\beta) = \frac{2\alpha}{\pi}\left[\frac{\tanh^{-1} \beta}{\beta} -1\right].
\end{equation}
The change in the tritium $\beta$ spectrum shape is negligible due to its low endpoint, however for higher energies the change in the $ft$ value can become several parts in $10^4$ and so cannot always be neglected. The second possibility of resolving the divergence takes into account the finite detector resolution \cite{Gardner2004}. In the treatment by Sirlin, the outgoing $\beta$ particle and $\gamma$ are always distinguishable, which is equivalent to an energy resolution of zero. If we instead consider a finite energy resolution, the distinguishability of the real photon depends on its energy, implying the same conclusion to the $\beta$ spectrum correction. This assumes an absorption of the $\gamma$ particle, however, an effect which is completely negligible in most experimental set-ups. We thus use Eq. (\ref{eq:radiative_exponentiation}) in our current description.

\subsubsection{Order $Z\alpha^2$ correction}
Spurred on by the high precision measurements of superallowed Fermi decays, two pioneering papers were published by \textcite{Jaus1970} and \textcite{Jaus1972} discussing the higher order radiative corrections. In subsequent years some tension arose between numerical and analytical methods \cite{Sirlin1986}, until finally both approaches agreed after a correction in the former \cite{Jaus1987}.
\begin{figure}[h]
\centering
\includegraphics[width=0.15\textwidth]{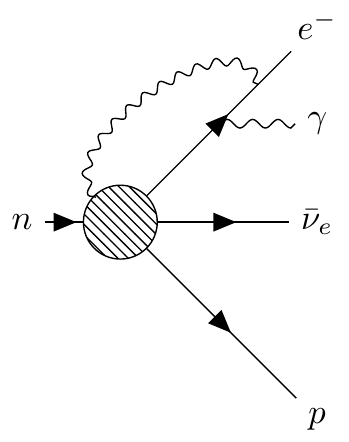}
\includegraphics[width=0.15\textwidth]{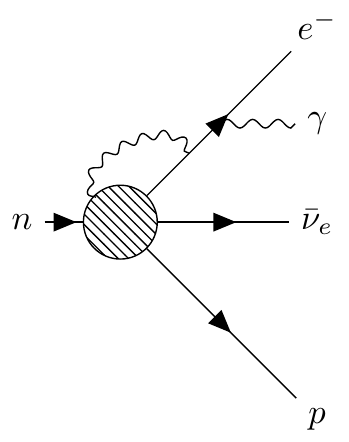}
\caption{Dominant Feynman diagrams for the order $Z\alpha^2$ corrections. The first of these already contains a correction present in the product $F_0\delta_1$, and has to be explicitly subtracted. Three more diagrams contribute to the order $\mathcal{O}(Z\alpha^2)$ correction, but are non-dominant. These can be found, for example, in \textcite{Sirlin1986}, and discuss the vacuum polarization of the brehmsstrahlung photon, and the two possibilities for the renormalization of the electron propagator.}
\label{fig:feynman_order_Zalpha2}
\end{figure}
Using the notation of the latter one writes
\begin{equation}
\delta_2(Z, W) = Z \alpha^2 \sum_{i=1}^{4} \Delta_i(W) ~ .
\label{eq:Zalpha2}
\end{equation}
The first three terms in the sum concern the graphs shown in Fig. \ref{fig:feynman_order_Zalpha2}, with $\Delta_3$ coming specifically from the axial vector component. All of these depend on the nucleus considered, as was to be expected from any correction involving interaction with the nucleus. The explicit weak interaction dynamics is, however, contained in the inner radiative correction, and the terms in Eq. (\ref{eq:Zalpha2}) depend only on the shape and radius of the charge distribution. The Feynman diagrams for the fourth term, on the other hand, are not explicitly shown, as they are non-leading and present the interaction of the photon with only the electron through vacuum polarization and renormalization of the $\beta$ particle propagator.

The terms of leading order in the nucleon mass, $M_N$, called $Z \alpha^2 \Delta_1$ in Eq. (\ref{eq:Zalpha2}) are difficult to evaluate and depend on the nuclear form factor $F(q^2)$. Using the trivial identity $F = 1 + (F-1)$, $\Delta_1$ was split up into $\Delta_1 = \Delta_1^0 + \Delta_1^F$, where the first term is now energy-dependent but nucleus-independent and vice versa for the second. The first is then typically combined with that other `purely QED' term, $\Delta_4$. Expressions have been derived for $\Delta_1^0$ both in the non-relativistic (assuming zero $\beta$ particle momentum) and extreme-relativistic approximation (neglecting terms of $\mathcal{O}(W/M)$) so that \cite{Sirlin1986}
\begin{align}
\Delta_1^0 + \Delta_4 &\eqtext{NRA} \ln M_N - \frac{2}{3}\ln (2W) + \frac{35}{9} + \frac{\pi^2}{6} - 6\ln2 \nonumber \\
&\eqtext{ERA} \ln M_N - \frac{5}{3} \ln(2 W) + \frac{43}{18}.
\label{eq:delta_1_4}
\end{align}
Results based on the extreme-relativistic approximation were compared to the numerical results of \textcite{Jaus1987} by \textcite{Sirlin1987b} and a general good agreement was found. Differences in the decay rate for $Z=26$ were found to be on the few $10^{-4}$ level. For higher $Z$, then, we expect a better agreement using a proper interpolation between the non-relativistic and extreme-relativistic results should this be needed.
The remaining terms have been evaluated using different models for the charge distribution $\rho(r)$. We list here the integral forms together with their evaluation for a uniformly charged sphere. Normalizing $\rho$ as $\int \rho(r) r^2 dr = 1$, we find \cite{Sirlin1987b}
\begin{align}
\Delta_1^F &= 1 - \gamma_E  - 4\pi\int_0^\infty \rho(r) r^2 \ln(M_Nr)dr - (8/M_N) \nonumber \\
&\times \int_0^\infty \rho(r) r [1+\gamma_E+\ln(M_Nr)]dr \\
\Delta_2 &= (4/M_N)\int_0^\infty \rho(r)r \left[1-\frac{\pi}{4M_Nr}\right] \\
\Delta_3 &= \frac{8g_Ag_M}{M_N}\int_0^\infty \rho(r)r \left[\gamma_E + \ln(M_Nr)-\frac{1}{2}+\frac{\pi}{8M_Nr} \right]
\end{align}
The other $\Delta_i$ have been evaluated using different models for the charge distribution $\rho(r)$. For the uniformly charged sphere of radius $R = \left( \frac{5}{3}\langle r^2 \rangle \right)^{1/2} = \sqrt{10}/ \Lambda$, one has \cite{Sirlin1987b}
\begin{align}
\Delta_1^F = &\ln\left[ \frac{\Lambda}{M_N} \right] - \kappa_2 - \frac{3}{\sqrt{10} \pi} \frac{\Lambda}{M_N} \nonumber \\
& \times\left[ \frac{1}{2} + \gamma_E + \ln \sqrt{10} + \ln \left[ \frac{M_N}{\Lambda} \right] \right] \noindent \\
\Delta_2 = &\frac{3}{2 \sqrt{10} \pi} \frac{\Lambda}{M_N} \left[ 1 - \frac{\pi}{2 \sqrt{10}} \frac{\Lambda}{M_N} \right]  \noindent \\
\Delta_3 = &\frac{3}{\sqrt{10}\pi} g_A ~ g_M \frac{\Lambda}{M_N} \left[ \gamma_E - 1 + \ln \sqrt{10} \right. \nonumber \\
& \left. + \ln \frac{M_N}{\Lambda} + \frac{\pi}{4 \sqrt{10}} \frac{\Lambda}{M_N} \right] .
\end{align}
with $\kappa_2 \equiv \gamma_E - \frac{4}{3} + \ln \sqrt{10} = 0.395$, $g_A$~=~1.270, $g_M$~=~4.706, $\gamma_E \simeq 0.5772$. Further, $\Lambda \equiv \sqrt{6} / \sqrt{\langle r^2 \rangle}$ with the rms nuclear charge radius $\langle r^2 \rangle$ in natural units, with $\Lambda$ ranging from about 400 at $A = 10$ to about 160 at $A = 250$.

For the modified Gaussian model for the charge distribution $\rho(r)$ \cite{Sirlin1987b} more extended expressions are obtained. Both models give, however, very similar results for the superallowed Fermi transitions ranging from $^{14}$O to $^{54}$Co corresponding to differences in the averages $\langle \delta_2(E) \rangle$ over the energy spectrum of less than 10$^{-4}$ \cite {Sirlin1987b}. Using a Fermi distribution for heavier nuclei, the difference can then be expected to be even smaller.

\subsubsection{Order $Z^2\alpha^3$ correction}
\begin{figure}[h]
\centering
\includegraphics[width=0.15\textwidth]{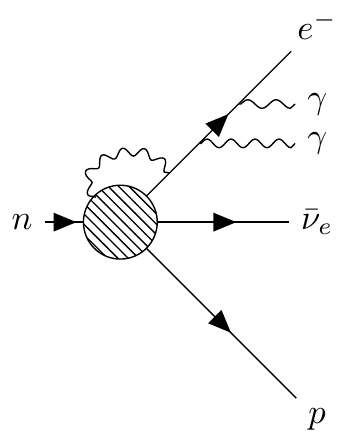}
\includegraphics[width=0.15\textwidth]{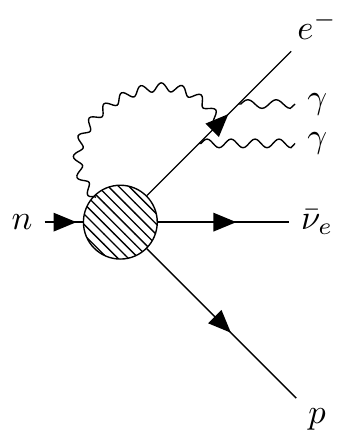}
\includegraphics[width=0.15\textwidth]{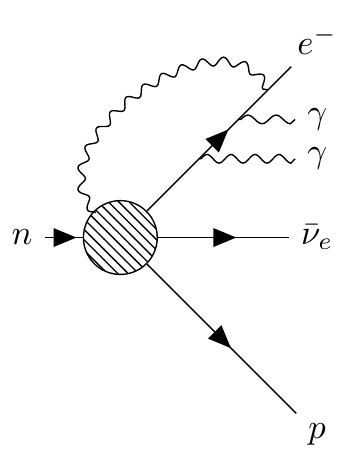}
\caption{Dominant Feynman diagrams for the order $Z^2\alpha^3$ corrections.}
\label{fig:feynman_order_Z2alpha3}
\end{figure}

The reasoning for the $Z^2\alpha^3$ corrections is analogous to that discussed above, with the relevant Feynman diagrams shown in Fig. \ref{fig:feynman_order_Z2alpha3}. Technically its evaluation is extremely challenging, however, and \textcite{Sirlin1987b} has only proposed a so-called `heuristic' correction.
This is written as
\begin{align}
\delta_3^{he} = &Z^2 \alpha^3 \left[ a ~ \ln \left(\frac{\Lambda}{W} \right) + b ~ f(W) \right. \nonumber \\
 &\left.  + \frac{4\pi}{3} ~ g(W) -0.649 ~ \ln(2W_0) \right] ,
\end{align}
where
\begin{align}
a &= \frac{\pi}{3} - \frac{3}{2\pi} \quad \text{(Ref.~10 of Ref.~\textcite{Sirlin1987b})}, \\
b &= \frac{4}{3 \pi}\left(\frac{11}{4} - \gamma_E -\frac{\pi^2}{6}\right), \\
f(W) &= \ln(2W)-5/6, \\
g(W) &= \frac{1}{2} \left[ \ln^2(R) - \ln^2(2W) \right] + \frac{5}{3} \ln(2RW).
\end{align}

\subsubsection{Higher order corrections}
Due to the computational complexity, little is known about the higher order corrections of the type $Z^m\alpha^{m+1}$. Based on the leading order behavior of the previously discussed terms, \textcite{Wilkinson1997} has put forward a very approximate estimate
\begin{equation}
\overline{\delta}_{Z^n \alpha^m} \approx Z^n \alpha^m ~ K_{nm} \overline{\ln^{m-n} \frac{\Lambda}{W}} ~ ,
\label{delta-nm}
\end{equation}
and suggested for $K_{nm}$ an average value of 0.50 from the above results. If $K_{nm}$ is given a fixed value the summed effect of all the higher-order terms is \cite{Wilkinson1997}
\begin{equation}
\overline{\delta}_{\text{higher}} \approx \sum_{n=3}^{n=\infty} \overline{\delta}_{Z^n \alpha^{n+1}} =  \overline{\delta}_{Z^3 \alpha^4}/(1-Z\alpha),
\end{equation}
with $\delta_{Z^3\alpha^4}$ evaluated using Eq. (\ref{delta-nm}). It is clear these corrections become relevant for higher $Z$.

\subsection{Neutrino radiative corrections}
\label{sec:neutrino_radiative}
Even though the outgoing (anti)neutrino has no direct interaction with the surrounding electric field, it is indirectly influenced through virtual photon exchange and energy conservation from inner brehmsstrahlung. Due to the recent interest in the proper conversion of the cumulative $\beta$ spectrum emerging from a nuclear reactor, these have been treated explicitly by \textcite{Sirlin2011}. The former case is essentially unchanged relative to the $e^{\mp}$ spectrum, after substitution of $W_e \to W_0 - W_{\nu}$ in all relevant quantities. The treatment of the internal bremsstrahlung differs, however, and the subsequent total $\mathcal{O}(\alpha)$ radiative correction is much smaller. One finds
\begin{align}
R_{\nu}&(Z, W, W_0) = 1 + \frac{\alpha}{2\pi}\left\{3\ln \left(\frac{m_p}{m_e} \right) +\frac{23}{4} \right. \nonumber \\
&-\frac{8}{\hat{\beta}}\text{Li}_2\left(\frac{2\hat{\beta}}{1+\hat{\beta}} \right) + 8\left(\frac{\tanh^{-1} \hat{\beta}}{\hat{\beta}}-1 \right)\ln\left(\frac{2\hat{W}\hat{\beta}}{m_e} \right) \nonumber \\
&\left. +4\frac{\tanh^{-1}\hat{\beta}}{\hat{\beta}}\left[\frac{7+3\hat{\beta}^2}{8} - 2\tanh^{-1}\hat{\beta}\right] \right\}
\end{align}
where analogous to the electron we have
\begin{equation}
\hat{W} = W_0 - W_{\nu}; ~~~ \hat{p} = \sqrt{\hat{W}^2-m_e^2}; ~~~ \hat{\beta} = \frac{\hat{p}}{\hat{W}}
\end{equation}
with $W_{\nu}$ the energy of the outgoing (anti)neutrino.

\subsection{Radiative $\beta$ decay - Internal bremsstrahlung}
\label{sec:IB}

The radiative corrections discussed above include the effects of inner bremsstrahlung, otherwise known as radiative beta decay where an additional photon is created in the final state
\begin{equation}
n \rightarrow p + e^- + \bar{\nu}_e + \gamma,
\end{equation}
where the energy of the $\gamma$ ray follows a continuous spectrum thereby reducing the $\beta$ particle energy. This is typically understood as a two-step process, where an electron is first ejected with energy $W'$, and subsequently emits a photon with energy $\omega$. Classically, this can simply be understood from the sudden acceleration of the beta particle during which it emits radiation. This effect has been extensively described by \textcite{Bloch1936} and \textcite{Knipp1936}, and is commonly referred to as the KUB theory. Using results from classical electrodynamics one can write the photon ejection probability as \cite{Schopper1966}
\begin{equation}
\Phi(W, \omega) = \frac{\alpha p}{\pi \omega p^{\prime}}\left\{\frac{W^2+W'^2}{W'p}\ln (W+p)-2\right\},
\label{eq:phi_omega}
\end{equation}
with $W = W^{\prime}-\omega$ and $p^{\prime} = \sqrt{W'^2-1}$ its correspondingly redefined momentum. In order to find the photon spectrum we must average over all intermediate virtual states to find
\begin{equation}
S(\omega) = \int_{1+\omega}^{W_0}N(W')\Phi(W', \omega)dW',
\label{eq:S_IB}
\end{equation}
where $\omega$ is the final photon energy and $N(W')$ is the $\beta$ spectrum. Equations (\ref{eq:S_IB}) and (\ref{eq:phi_omega}) immediately reveal the dependence on the endpoint energy of the beta transition and its rapid fall-off towards higher photon energies.

In this initial approach Coulomb effects with the daughter nucleus were neglected, however, and all results were evaluated using plane waves. \textcite{Nilsson1956}, \textcite{Lewis1957} and \textcite{Spruch1959} all introduced this correction and obtained expressions for its energy spectrum and angular correlation. This roughly translates to introducing the Fermi function into Eq. (\ref{eq:S_IB}). \textcite{Felsner1963} has produced more rigorous work on this, and shows a much improved agreement with experimental data. Very recent theoretical work has been performed by \textcite{Ivanov2014} describing the QED tree-level contributions to the internal bremsstrahlung spectrum of $^{35}$S. This amounts to changing Eq. (\ref{eq:phi_omega}) by Eq. (16) in the aforementioned work, where additional terms of order $\mathcal{O}(\omega / W)$, $\mathcal{O}(\alpha^2 Z^2/W^2)$ and $\mathcal{O}(W/M)$ are introduced. Further research was performed specifically for the neutron to investigate the influence of recoil order terms and weak magnetism \cite{Ivanov2013, Ivanov2017}. Even though its inclusion shows up stronger in the internal brehmsstrahlung compared to the regular branching ratio, its magnitude is still far below the currently available experimental precision.

Detection of this radiation reveals underlying weak interaction physics, and is both correlated with the outgoing lepton momenta and carries a degree of circular polarization \cite{Cutkosky1954a, Berman1958, Kinoshita1959, Mao2011, Batkin1992}. Consequently, its experimental detection has been an interesting observable for several decades \cite{Boehm1954, Goldhaber1957, Basavaraju1983, Budick1992, Khalil2011a, Khalil2011b, Singh2014, Singh2015}. Recent experimental branching ratios for the neutron find BR$_{\beta \gamma} \sim 3 \cdot 10^{-3}$ \cite{Bales2016}, and also for higher $Z$ significant branching ratios were found such as for $^{32}$P at $2 \cdot 10^{-3}$ \cite{Berenyi1969a}. It has been studied intensively following electron capture, reviewed extensively by \textcite{Bambynek1977}. Its relevance in the analysis of the tritium beta spectrum measurements has been theoretically discussed by \textcite{Gardner2004}, and continues to garner interest through the study of correlation parameters \cite{Gardner2012, Gardner2013} looking for $T$ and $CP$ violation.

\section{Nuclear structure effects - the shape factor}
\label{sec:nuclear_structure}
In the foregoing, nuclear structure effects have been ignored and we focused instead on electromagnetic corrections and kinematics. When considering the nucleus as a non-trivial system with a finite size, nuclear structure and spatial variations of leptonic wave functions become deeply intertwined. The exact treatment of this fact has undergone careful study by several authors and different formalisms over several decades \cite{Konopinski1941, Rose1954, Weidenmuller1961, Huffaker1963, Buhring1963, Schulke1964, Buhring1965a, Buhring1965b, Schopper1966, Konopinski1966, Huffaker1967, Behrens1969, Behrens1970, BlinStoyle1973, Morita1973, Behrens1978, Behrens1982, Holstein1974a, Holstein1974b, Holstein1974c, Calaprice1977a, Kleppinger1977}. In essence, we have finally arrived at the heart of \emph{nuclear} $\beta$ decay. Our starting point lies in the elementary particle approach, and from this we branch out into the Behrens-B\"uhring and Holstein formalisms. We utilize the accurate calculational machinery of the former and transform our results to the clean notation of the latter.

\subsection{Introduction}
\label{sec:shape_factor_intro}
The $\beta$ decay Hamiltonian is traditionally constructed as a current-current interaction
\begin{equation}
H_\beta(x) = \frac{G \cos \theta_C}{\sqrt{2}}[J_\mu^\dagger(x) L^\mu(x) + \text{H.c.}]
\end{equation}
where $\theta_C$ is the Cabibbo angle and $J_\mu(x)$ and $L^\mu(x)$ are the nuclear and lepton currents, respectively. This is plugged into the S matrix of Eq. (\ref{eq:transition_general}) as a first order perturbation. As we discussed in Sec. \ref{Fermi function}, the introduction of electromagnetism requires a change in the lepton current $L^\mu$ exchanging the $\beta$ particle wave function. So the strong interaction requires a change in the nuclear current from its pure $V$-$A$ shape. We have then the generalizations
\begin{align}
J_\mu^\dagger(x) &= \langle f | V_\mu(x) + A_\mu(x) | i \rangle \\
L^\mu(x) &= i\bar{\phi}_e(x)\gamma^\mu(1+\gamma_5)v_\nu(x),
\end{align}
where $\phi_e$ is the solution of the $\beta$ particle wave function in a Coulomb potential. The transition matrix element (TME) constructed from this Hamiltonian is constrained by angular momentum coupling rules for the spin transition $J_i^{\pi_i}\rightarrow J_f^{\pi_f}$, allowing decays of different multipole orders $K$ within the vector triangle $(J_i, J_f, K)$. It is natural then to write the TME as a sum over all possible $K$ with complementary projection operators acting on both lepton and nuclear spaces. This approach allows for the definition of form factors $F_K$ that absorb all the nuclear structure information. These form factors are a function of $q^2 \equiv (p_f-p_i)^2$, the only Lorentz invariant scalar available. With form factors the treatment can continue without any model dependence, which is also known as the elementary particle approach practiced by many authors \cite{Kim1965, Holstein1974b, Armstrong1972, Behrens1982}. Specifically this translates into an expression of the following type \cite{Schulke1964, Schopper1966}
\begin{align}
\langle f | V_{\mu}-A_{\mu} | i \rangle &\propto \sum_{KM}\sum_{s,L=K-1}(-1)^{J_f-M_f+M}(-i)^L \nonumber\\
&\times \sqrt{4\pi}\sqrt{2J_i+1}\left(\begin{array}{ccc}
J_f & K & J_i\\
-M_f & M & M_i
\end{array}\right) \nonumber \\
&\times \frac{(qR)^L}{(2L+1)!!}F_{KLs}(q^2),
\end{align}
where $K$ corresponds to the multipole order of the transition and $K, L$ and $s$ have to form a vector triangle. The form factors $F_{KLs}(q^2)$ contain the actual nuclear information and can be expanded as a function of $(qR)^2$. Since for $\beta$ decay this quantity is very small, the expansion can typically be stopped after the first order. A very similar expression is obtained, for example, in the Holstein formalism (see \textcite{Holstein1974b} for a discussion and comparison). In case of a non-trivial lepton current, a similar expansion\footnote{Non-trivial here meaning Coulomb corrected wave functions rather than simple free Dirac spinors. In this case an expansion can be made in terms of $r$, $(\alpha Z)$, $(m_eR)$ and $(W_eR)$. This is discussed in the next section.} is made such that the TME can be written down to different orders of precision.
In the aforementioned formalisms, all information other than the phase space factor and Fermi function are typically grouped in a so-called shape factor $C(Z, W)$
\begin{equation}
N(W)dW \propto pW(W_0-W)^2F(Z, W)C(Z, W).
\label{eq:shape_factor}
\end{equation}
In some formalisms the Fermi function and/or the finite size effect is also included in $C(Z, W)$. Here we will however adhere to the notation in Eq. (\ref{eq:shape_factor}).
Now $C(Z, W)$ includes effectively a combination of two effects: The first is the spatial variation of the leptonic and nuclear wave functions inside the finite nuclear volume, the second is the effect of raw nuclear structure. It should be understood that all additional corrections discussed in the previous sections are not included in this original formulation, and can be considered higher order corrections. Before we continue with an analytical description, we discuss the nature and evaluation of the form factors $F_{KLs}(q^2)$.

\subsection{Form factors}
\label{sec:form_factors}

\subsubsection{Introduction}
\label{sec:form_factors_introduction}
The form factors describe rather generally the nuclear structure and allow for a model independent analysis of $\beta$ decay observables. These can then be deduced from experimental measurement, and in theory we could stop the discussion. Using the impulse approximation we can however continue and express these form factors in terms of nuclear matrix elements. This entails that we treat the nucleus as an assembly of non-interacting nucleons that all couple to the $\beta$ decay Hamiltonian as if they were free particles. Generally speaking, we have then a collection of one body operators $O_{KLs}$ that can be written as
\begin{align}
O_{KLs} = \sum_{\alpha,\beta}\langle \alpha | O_{KLs} | \beta \rangle a_{\alpha}^{\dagger}a_{\beta},
\end{align}
where $a_{\alpha}^{\dagger}$ creates a nucleon in state $\alpha$, and $a_{\beta}$ annihilates a nucleon in state $\beta$. We have then
\begin{equation}
\langle f | O_{KLs} | i \rangle = \sum_{\alpha,\beta}\langle \alpha | O_{KLs} | \beta \rangle \langle f | a_{\alpha}^{\dagger}a_{\beta} | i \rangle~.
\label{eq:operator_impulse}
\end{equation}
The last factors in this equation are called the one body density matrix elements which can be calculated by e.g. the shell model. In special cases this formulation can be further simplified and allow for immediate analytical evaluation. An example will be given below. After the brief discussion of the impulse approximation, we are able to explicitly write the form factors in terms of nuclear matrix elements (see Sec. 6.2 in \textcite{Behrens1982} for a more thorough discussion). As an example, consider Table \ref{table:matrix_elements_BB} where the correspondence between form factor coefficients\footnote{These are related to the more general form factors via the expansion $F_{KLs}(q^2) = \sum_nA(n, L)(qR)^{2n}F_{KLs}^{(n)}$, where $A(n, L)$ is a trivial pre-factor. As $F_{KLs}^{(0)}$ and $F_{KLs}^{(1)}$ are typically of similar magnitude, it is clear that $F_{KLs}(q^2)$ is dominated by the former.} and nuclear matrix elements is shown, as well as the possible spin changes and `forbidden' transitions for which the form factor coefficient is identically zero. For the sake of simplicity, induced currents (see Sec. \ref{sec:induced_currents}) have been neglected here. A more extensive overview of $F_{KLs}^{(n)}$ with selection rules can, for example, be found in \textcite{Behrens1969}.

\begin{table*}
\caption{Definitions of the nuclear matrix elements (from \textcite{Behrens1978}) for allowed transitions. Here the form factors are written in their symbolic integral notation. The possible spin changes are shown explicitly, together with `forbidden' transitions for which the form factor coefficient is automatically zero. Here $\bm{\alpha}$ is constructed with the Pauli matrices, $\bm{\sigma}$ on its off-diagonal elements. We have neglected second-class contributions and terms of order $\mathcal{O}(g_M/(M_NR)^2)$.}
\begin{ruledtabular}
{\renewcommand{\arraystretch}{1.8}
\begin{tabular}{cllcc}
Form Factor (BB) & Cartesian Form & $\Delta J$ & Forbidden & Type\\
\hline
$^VF_{000}^{(0)}$ & $+g_V\int \mathbf{1}$ & 0 & $-$ & \multirow{2}{*}{Allowed} \\
$^AF_{101}^{(0)}$ & $\mp g_A\int \bm{\sigma}$ & 0,1 & $0-0$ & \\
\hline
$^VF_{000}^{(1)}$ & $+g_V\int \left(\frac{r}{R}\right)^2$ & 0 & $-$ & \multirow{3}{*}{Main correction terms}\\
$^AF_{101}^{(1)}$ & $\mp g_A\int \bm{\sigma}\left(\frac{r}{R}\right)^2 \pm \frac{g_P}{2(M_NR)^2}\int \bm{\sigma}$ & 0,1 & $0-0$ & \\
$^AF_{121}^{(0)}$ & $\mp g_A \frac{3}{\sqrt{2}}\int \frac{(\bm{\sigma}\cdot \bm{r})\bm{r}-\frac{1}{3}\bm{\sigma}\cdot r^2}{R^2} \pm \frac{g_P5\sqrt{2}}{(2M_NR)^2}\int \bm{\sigma}$ & 0,1 & $0-0$ & \\
\hline
$^VF_{011}^{(0)}$ & $+g_V\int i\frac{\bm{\alpha} \cdot \bm{r}}{R}$ & 0 & $-$ & \multirow{3}{*}{Relativistic correction terms}\\
$^VF_{111}^{(0)}$ & $-g_V\sqrt{\frac{3}{2}}\int \frac{\bm{\alpha} \times \bm{r}}{R} - \frac{g_M-g_V}{2M_NR}\sqrt{3}\int \bm{\sigma}$ & 0,1 & $0-0$ &\\
$^AF_{110}^{(0)}$ & $\pm g_A \sqrt{3}\int \gamma_5 \frac{i\bm{r}}{R} \pm \frac{g_P\sqrt{3}}{(2M_NR)^2}\left[W_0R\pm \frac{6}{5}\alpha Z\right]\int \bm{\sigma}$ & 0,1 & $0-0$ & \\
\end{tabular}
}
\end{ruledtabular}
\label{table:matrix_elements_BB}
\end{table*}

\subsubsection{Induced currents}
\label{sec:induced_currents}

Before we continue with the explicit calculation of the shape factor $C(Z, W)$, we address the issue of induced currents. Because the decaying nucleon sits inside a nuclear potential, influences from QCD seep into the weak vertex. Assuming the weak interaction to be purely $V-A$, several Lorentz invariant terms can be added that transform in the same way. For the simple neutron decay this is written in the Behrens-B\"uhring (BB) \cite{Behrens1982} and Holstein (HS) \cite{Holstein1974} formalisms as\footnote{Here $\mathbf{q}=-i(\bm{\nabla}_f-\bm{\nabla}_i)$  when written in operator form. In the presence of an electromagnetic field gauge invariance requires $q_{\mu}\rightarrow\frac{\partial}{\partial\chi_{\mu}}\rightarrow \frac{\partial}{\partial\chi_{\mu}}-ieA_{\mu}$ where $A_{\mu}=(\mathbf{A}, i\phi)$ is the potential of the electromagnetic field. See Refs. \textcite{Behrens1971} and \cite{Holstein1974b}.}
\begin{widetext}
\begin{align}
J_{\mu}^{BB} &= i\langle \bar{u}_p | C_V\gamma_{\mu}-f_M\sigma_{\mu\nu}q^{\nu} + if_Sq_{\mu} - \frac{C_A}{C_V}\gamma_{\mu}\gamma_5-f_T\sigma_{\mu\nu}\gamma_5q^{\nu}+if_P\gamma_5q_{\mu} | u_n \rangle, \\
J_{\mu}^{HS} &= i\langle \bar{u}_p | g_V\gamma_{\mu}-\frac{g_M-g_V}{2M}\sigma_{\mu\nu}q^{\nu}+i\frac{g_S}{2M}q_{\mu}+g_A\gamma_5\gamma_{\mu}-\frac{g_{II}}{2M}\sigma_{\mu\nu}\gamma_5q^{\nu}+i\frac{g_P}{2M}\gamma_5q_{\mu} | u_n \rangle,
\label{eq:currents_BB_HS}
\end{align}
\end{widetext}
where\footnote{The original results by Holstein are written using the conventions by \textcite{Bjorken1964}, which differs both in the metric used, as well as the sign of $q$. We have written Eq. (\ref{eq:currents_BB_HS}) in the metric and notation by \textcite{Behrens1982} to show the correspondence in the definition of the constants.} $\sigma_{\mu\nu} = -\frac{i}{2}[\gamma_{\mu},\gamma_{\nu}]$ and all $C$ and $g$ are functions of $q^2$. It is here the elementary particle treatment shines, as the entire formulation in terms of form factors can be retained simply by redefining them and including these induced currents \cite{Buhring1965a}. The two approaches, while inherently different, are completely equivalent\footnote{This can be seen from Eqs. (7-8) in \textcite{Holstein1974a} and is thoroughly discussed in Sec. 9.2 in \textcite{Behrens1982}.}, and the redefinitions of the different $F_{KLs}^{(n)}$ can be found in Refs. \textcite{Buhring1965b, Behrens1971, Behrens1982}. As an example $^VF_{000}^{(0)}$ now becomes
\begin{equation}
^VF_{000}^{(0)} \rightarrow ~^V\mathcal{M}_{000}^{(0)}\pm \frac{f_S}{R}(W_0R\pm\frac{6}{5}\alpha Z)~^V\mathcal{M}_{000}^{(0)},
\label{eq:transform_form_factor_fs}
\end{equation}
where $\mathcal{M}_{KLs}$ are the actual nuclear matrix elements as written in the second column of Table \ref{table:matrix_elements_BB}. When neglecting induced currents it is clear that both notations coincide. Just as the work of Behrens and collaborators continues with the more general formulation using $^{V/A}F_{KLs}(\mathbf{q}^2)$, so intruduces Holstein general form factors. Here, the total TME is expanded into a combination of form factors labelled $a, b, c, d, e, f, h, g$ and $j_{2,3}$, all of which have some non-trivial $q^2$ dependence\footnote{A careful distinction must be made as the form factors are expanded as a function of 3-momentum $\mathbf{q}^2$ in the BB formalism, while it is performed with the 4-momentum $q^2 = W_0^2-\mathbf{q}^2$ in that of Holstein.} \cite{Holstein1971a, Holstein1971, Holstein1974a}. These are now the Fermi ($a$), weak magnetism ($b$), Gamow-Teller ($c$), induced tensor ($d$), induced scalar ($e$) and further higher order corrections, respectively. When expanding in terms of $q^2$ the influence of higher order terms can be neglected however, except for the leading Fermi and Gamow-Teller form factors written as
\begin{align}
a(q^2) &= a_1 + a_2 q^2 + \ldots \\
c(q^2) &= c_1 + c_2 q^2 + \ldots
\end{align}
Each of these form factors is a combination of the BB form factors discussed above, with the additional advantage that unlike the BB form factors, these are manifestly covariant and more clearly reveal underlying symmetries.

\begin{table*}
\centering
\caption{Summary of the $a$, $e$, $b$, $c_1$, $c_2$, $d$ and $h$ form factors and their relation to the nuclear matrix elements defined in Table~\ref{table:matrix elements}. Here CVC and SCC refers to conserved vector current hypothesis and second class currents (see Sec. \ref{sec:symmetries}), respectively. In this table the impulse approximation is only given to first order, and the relativistic matrix elements are neglected as is done in \cite{Holstein1974a}.}
\begin{ruledtabular}
{\renewcommand{\arraystretch}{1.4}
\begin{tabular}{cllr}
Form factor  & Formula in impulse approx. & Remark & Type \\
\hline
\textit{a}  & $a\cong g_V\mathcal{M}_F$ & $g_V=1$ (CVC) \cite{Ademollo1964} & \multirow{3}{*}{Vector} \\
$e$   &  $e\cong g_V ( \mathcal{M}_F \pm A g_S)$&  $e=0$ (CVC, SCC) & \\
$b$   & $b\cong A(g_M\mathcal{M}_{GT}+g_V\mathcal{M}_L)$  & $g_M\cong4.706$ & \\
\hline
$c_1$  & $c_1\cong g_A \mathcal{M}_{GT}$ & $g_A \rightarrow g_{A,\text{eff}}=1$ \cite{Towner1987} & \multirow{5}{*}{Axial vector}\\
$c_2$  & $c_2\cong \frac{1}{6} g_A \left[ \mathcal{M}_{\sigma r^2} + \frac{1}{\sqrt{10}} \mathcal{M}_{1 y} \right]$ &$c_2 \sim R^2$ & \\
$d$  & $d\cong A(g_A \mathcal{M}_{\sigma L} \pm g_{II} \mathcal{M}_{GT})$ & $g_{II} \sim g_T\cong0$ (SCC) & \\
 & ~~$\equiv d^I \pm d^{II}$  & $d^I = 0$ (analog states) & \\
$h$  & $h\cong\frac{-2}{\sqrt{10}}M^2g_A\mathcal{M}_{1y}-A^2g_P \mathcal{M}_{GT}$ & $g_{P,\text{free}} \approx -229 \to g_{P,\text{eff}} =$ ?  & \\
\end{tabular}
}
\end{ruledtabular}
\label{table:form factors}
\end{table*}

Table \ref{table:form factors} lists these form factors in terms of the nuclear matrix elements which are listed in Table \ref{table:matrix elements}, as predicted by the impulse approximation. These again have to be provided by the shell model unless we can treat them approximately as is explained in Sec. \ref{sec:analytical_me}.

\begin{table}
\centering
\caption{Definitions of the nuclear matrix elements by \textcite{Calaprice1977}. Here $\tau$ denotes the typical isospin ladder operator, and an explicit sum over all nucleons is included. Note that the BB matrix elements from Table \ref{table:matrix_elements_BB} are dimensionless due to appropriate powers of $R$, unlike those of Holstein presented here.}
\begin{ruledtabular}
{\renewcommand{\arraystretch}{1.4}
\begin{tabular}{c  r}
Matrix element & Operator form \\
\hline
$\mathcal{M}_F$ & $\langle\beta\|\Sigma\tau^{\pm}_i\|\alpha\rangle$\\
$\mathcal{M}_{GT}$ & $\langle\beta\|\Sigma\tau^{\pm}_i\overrightarrow{\sigma}_i\|\alpha\rangle$ \\
$\mathcal{M}_L$ & $\langle\beta\|\Sigma\tau^{\pm}_i\overrightarrow{l}_i\|\alpha\rangle$ \\
$\mathcal{M}_{\sigma r^2}$ & $\langle\beta\|\Sigma\tau^{\pm}_i\overrightarrow{\sigma}_ir_i^2\|\alpha\rangle$ \\
$\mathcal{M}_{\sigma L}$ & $\langle\beta\|\Sigma\tau^{\pm}_ii\overrightarrow{\sigma}_i\times\overrightarrow{l}_i \|\alpha\rangle$ \\
$\mathcal{M}_{Ky}$ & $\left(\frac{16\pi}{5}\right)^{\frac{1}{2}} \langle\beta\|\Sigma\tau^{\pm}_ir_i^2C^{nn'k}_{12k}\sigma_{in}Y^{n'}_2(\hat{r}_i)\|\alpha\rangle$ \\
\end{tabular}
}
\end{ruledtabular}
\label{table:matrix elements}
\end{table}

\subsubsection{Validity of the impulse approximation}
In the previous section we have discussed the possibility of transforming the model-independent form factors to one-body matrix elements using the impulse approximation. The latter assumes that all nucleons in the nucleus interact with the weak vertex as if they were free, thereby neglecting many-body effects such as meson exchange and core-polarization. It has been shown \cite{Armstrong1972a} that meson exchange need not even be invoked to show the breakdown using only PCAC. It comes then as no surprise that the impulse approximation breaks down once sufficient accuracy is required. For the vector matrix element $^VF_{000}$ the impulse approximation is a valid approach, as the current is conserved. Below, we will discuss the divergence of the axial current and its relation to the pion field, for which we have to calculate QCD matrix elements. Corrections to the simple shell model methods are historically split up into two categories: meson exchange effects and `nuclear' effects, the latter of which comprises core-polarization, relativistic effects and configuration mixing stemming from an insufficient knowledge of the nuclear wave functions. The distinction is artificial, however, as the nuclear potential relies on an incorporation of mesonic degrees of freedom. This fact was emphasized by \textcite{Wilkinson1974b}. The study of mesonic contributions has a long history, and includes pioneering works by \textcite{Blin-Stoyle1959, Chemtob1971, Delorme1971, Rho1974}, but falls outside of the scope of this work. Reviews can be found in the excellent papers by \textcite{Ejiri1978, Blin-Stoyle1978, Towner1987}. Effects of core-polarization on the other hand, result from a truncation of the set of basis states by considering the nucleus as an inert core with some valence nucleons \cite{Shimizu1974}. In many cases this is not an appropriate approximation, and significant corrections occur from allowing particle-hole excitations across shell gaps \cite{Barroso1975, Koshigiri1981}. This is a complicated matter, however, as severe cancellations occur between core-polarization and meson exchange effects, while remaining careful for double counting \cite{Towner1979}. Individual effects can easily range to 40\%, but due to cancellations in allowed decays these largely correspond to the impulse approximation results \cite{Morita1985}. For forbidden decays, however, this cancellation is much less pronounced and significant deviations are encountered when using the impulse approximation \cite{Baumann1998}.

For allowed decays then, one typically remedies the situation in an \emph{ad hoc} way by renormalising the coupling constants to produce effective couplings \cite{Wilkinson1973, Wilkinson1973a, Wilkinson1974b, Towner1992, Warburton1992}.
Based on the above reasoning, we expect the effective coupling constant to decrease for increasingly heavy nuclei. This is the case for the $sd$-shell, where $g_A$ is set to $1.1$ and to unity in $fp$-shell nuclei \cite{Martinez-Pinedo1996}. New experimental measurements of highly forbidden decays push this value even further downwards to $g_A \approx 0.9$ \cite{Haaranen2017}. Statistical methods have been applied by \textcite{Deppisch2016}, where a nice overview of the quenching of $g_A$ is provided for different methods in different mass regions. Because of the $g_A^4$ dependence of the $0\nu \beta\beta$ cross section, its study has received renewed appreciation \cite{Kostensalo2017}. Interesting to note here is the use of the $\beta$ spectrum as the prime means of deducing $g_A$ as its shape depends only on the relative values of the involved nuclear matrix elements.

Less known is the subsequent renormalisation of the induced pseudoscalar coupling, for which the original interaction is typically seen as an exchange of a virtual pion \cite{Ericson1973}. The subsequent interaction then depends on the pion propagator, which is however modified by its interaction with the surrounding nuclear matter. This is a delicate matter, and contains severe model dependence \cite{Delorme1976}. In an extreme case the pseudoscalar coupling constant can be quenched by as much as $80\%$, which is significantly larger than the connected quenching of $g_A$. Great care is then required when evaluating matrix elements containing a pseudoscalar component.

Finally, we stress the model-dependence of the quenching of the coupling constants, as it originates from a failure to directly take into account many-body effects and correlations. This can, for example, clearly be seen from a difference in effective $g_A$ when using the shell model or the interacting boson model \cite{Haaranen2017}. Ab initio methods, such as Green's function Monte Carlo provide a more systematic framework and circumvent the need for quenching by taking into account many-body correlations \cite{Brida2011}. This falls outside of the scope of this text, however, and the reader is referred to excellent reviews by \textcite{Pieper2001} and \textcite{Carlson2015}.

\subsection{Symmetries}
\label{sec:symmetries}
In dealing with allowed $\beta$ decay, one finds that results are significantly constrained and simplified through the application of symmetries. The first we discuss here is the conserved vector current (CVC) hypothesis \cite{Feynman1958}, after which we briefly comment on second class currents. Discussions on $T$-invariance and partially conserved axial current can be found, for example, in works by \textcite{Holstein1974} and \textcite{Behrens1982}.

\subsubsection{Conserved vector current}
As with the electromagnetic interaction, the conservation of the vector current allows us to derive simple relations for different form factors. In the most trivial example one finds
\begin{align}
\langle J_fM_f | V_0(0) | J_iM_i \rangle &= \,^VF_{000}(q^2)\delta_{J_iJ_f}\delta_{M_iM_f} \nonumber \\
&= \langle T_f T_{3f} | T_{\mp} | T_i T_{3i} \rangle \delta_{J_iJ_f}\delta_{M_iM_f} \nonumber \\
&= \sqrt{(T_i\pm T_{3i})(T_i \mp T_{3i}+1)} \nonumber \\
&\times \delta_{J_iJ_f}\delta_{M_iM_f}
\label{eq:MF_isospin}
\end{align}
with $V_0$ the temporal part of the vector current and $T_3$ the third component of the isospin vector, meaning
\begin{equation}
^VF_{000} = \sqrt{(T_i\pm T_{3i})(T_i \mp T_{3i}+1)} = \,^V\mathcal{M}_{000}^{(0)}.
\end{equation}
Comparing this to Eq. (\ref{eq:transform_form_factor_fs}), we find that CVC excludes any induced scalar currents.

Further, this can be used to find relationships between different vector form factors, as discussed, for example, by \textcite{Behrens1971}. A relationship relevant to our further discussion is the following\footnote{'This formula is a simplication of the general form factor coefficient $^VF_{011}^{N-1}(k_e, m, n, \rho)$ which can be found in \textcite{Behrens1971}.}
\begin{align}
-2N^VF_{011}^{N-1} &= (W_0 \mp (m_n-m_p))R^VF_{000}^N \nonumber \\
&+\pm \alpha Z \left\{\int \left(\frac{r}{R}\right)^{2N}U(r)Y_0^0\right\}.
\label{eq:F011_CVC}
\end{align}
Here $m_n$ and $m_p$ stand for the neutron and proton masses, respectively, and the integration is performed with the nuclear wave functions which were omitted for notational convenience. Further, $U(r)$ is defined via $V(r) = -\alpha Z/R U(r)$ with $V(r)$ the electronic potential, and $Y_0^0$ is the constant spherical harmonic function coming from the expansion of the lepton current.

Through its likeness with the electromagnetism, a deeper connection can be constructed between weak and electromagnetic observables. As both divergenceless currents behave quasi-identically, matrix elements can be interrelated. \textcite{Behrens1982} discuss this in our adopted form factor formalism, and show that $^VF_{KK0}(q^2)$ and $^VF_{KK1}(q^2)$ are related to elastic electron scattering reduced transition strengths for electric and magnetic types, respectively. This entails that we can evaluate these matrix elements using the \emph{charge} distribution rather than the weak charge distribution. 

The weak magnetism form factor, mentioned in the previous section, provides the largest contribution of the form factors induced by the strong interaction. Its main contribution comes from the $^VF_{111}(q^2)$ form factor which is related to a magnetic matrix element, hence the name. In the general nuclear current decomposition as written by \textcite{Holstein1974}, this is now written as $b(q^2)$ rather than $g_M$. For $\beta$ transitions between analog states the CVC hypothesis relates this form factor to electromagnetic properties of the $\beta$ transition, allowing one to calculate $b$ on the basis of experimental data. Thus, for the so-called mirror $\beta$ transitions between isospin $T = 1/2$ mirror nuclei we write \cite{Calaprice1976}
\begin{align}
b^{\mp} &= A \sqrt{\frac{J+1}{J}} \mu^{\mp},  \nonumber \\
\mu^{\mp} &= \mp(\mu_1 - \mu_2) ~ ,
\end{align}
\noindent where $A$ is the mass number, $J$ the angular momentum, and $\mu_1$ and $\mu_2$ are the magnetic moments of the mother and daughter nuclei, respectively.

Further, for the pure Gamow-Teller transitions within isospin triplet states the CVC hypothesis allows calculating the weak magnetism form factor from the width, $\Gamma_{M1}^{\text{iso}}$, of the analog \textit{isovector} part of the $\gamma$ transition with M1 multipolarity, via \cite{Calaprice1976}
\begin{equation}
b_{\gamma}^2 = \eta \,6 \frac{\Gamma_{M1}^{\text{iso}} M^2}{E^3_{\gamma} \alpha} ~ ,
\end{equation}
with $E_{\gamma}$ the energy of the $\gamma$ transition, $M$ the average nuclear mass of mother and daughter, $\alpha$ the fine-structure constant, and $\eta$ a constant. The latter is unity if the final states of $\beta^{\pm}$ and $\gamma$ processes are equal, while it becomes $(2J_i^\gamma+1)/(2J_f^\gamma+1)$ when reversed to correct for the proper degeneracy in the phase space of the $M1$ transition.

A new compilation of experimental data and subsequent discussion on weak magnetism behavior throughout the mass range was performed by \textcite{SeverijnsTBP}.

\subsubsection{Partially conserved axial current}
The application of partially conserved axial current typically results in the Goldberger-Treiman relation \cite{Goldberger1958}, relating the pion-nucleon coupling to $g_A$. This is generally not of much use unless one knows the pion matrix element. This relation can however be translated into a condition for $g_P$
\begin{equation}
g_P(q^2) = -g_A(0)\frac{(2M_n)^2}{m_{\pi}^2-q^2}.
\label{eq:gP}
\end{equation}
Simply using experimental results for nucleon and pion masses at $q^2=0$ one obtains $g_P(0) \approx -229$. This expression is however only valid assuming isospin invariance, and much work has been done by several authors (see the works by \textcite{Gorringe2004} and \textcite{Bhattacharya2012} and references therein) to provide higher order corrections to this result. These corrections are small, however, and change $g_P$ on the $5\%$ level. Experimentally measurements have been performed for muon capture, for which $q^2=0.88m_{\mu}^2$, and so can be roughly translated using Eq. (\ref{eq:gP}). Currently, however, the most accurate results from chiral perturbation theory are in sufficient agreement with experimental results to warrant using the PCAC results \cite{Gorringe2004}. We point again to the importance of the quenching of $g_P$ due the meson exchange effects in the nuclear medium. Thus, Eq. (\ref{eq:gP}) is valid only for the free nucleon, and significant care must be taken when evaluating pseudoscalar currents for nuclear decays.

\subsubsection{Second class currents}
We can classify terms in the expansion of the nuclear current following their transformation properties under the G-parity operation, $G = C e^{i \pi T_y}$, i.e. the product of a charge conjugation operation, $C$, and a rotation by $\pi$ around the $y$-axis in isospin space \cite{Weinberg1958}. Only two terms in Eq. (\ref{eq:currents_BB_HS}) transform differently compared to their main vector or axial vector analogues. These are called second class currents, as opposed to their first class counterparts. The first is the induced scalar interaction, which was already eliminated because of CVC. The second is the induced tensor interaction, for which no similar constraint is available. As can be seen from Table \ref{table:form factors} the axial tensor form factor, $d$, contains both a so-called first class term, i.e. $d^I$, and a second class term, $d^{II}$. If both the vector and axial-vector weak nucleon currents, $V_{\mu}$ and $A_{\mu}$, have a definite $G$-parity the induced terms (and thus also the one related to the $d$ form factor) are expected to hold the $G$-symmetry, that is, the decay of a proton and a neutron in a nucleus should be symmetric. Experimentally no indication for the existence of second-class currents has been found as yet (see e.g. \cite{Grenacs1985, Shiomi1996, Wilkinson2000, Minamisono2001, Minamisono2011a, Sumikama2011}), so that second-class currents can be ignored, i.e. $d^{II} \equiv 0$. The $d$ form factor then reduces to its first class part, i.e. $d = d^I$, which is only non-zero for transitions between non-analog states (i.e. belonging to a different isotopic multiplet), as the matrix element $M_{\sigma L}$ in $d^I$ can be shown to vanish for transitions between analog states \cite{Holstein1974b}.

\subsection{Analytical matrix elements}
\label{sec:analytical_me}
In the introduction (Sec. \ref{sec:form_factors_introduction}) we mentioned the decomposition of a general operator $O_{KLs}$ into its single-particle constituents. Using the form factor formalism, this now has a more concrete meaning, and using the impulse approximation we can build on this foundation. Following the work of \textcite{Shalit1963} and its discussion by \textcite{Behrens1982}, we can write many-particle matrix elements in terms of \emph{single}-particle matrix elements by introducing a factor $C(K)$ that absorbs all many-particle effects and which depends only on the tensor rank $K$. For a simple configuration with two nucleons in initial and final states in an orbital with total angular momentum $j$ one finds
\begin{align}
&\langle \phi(j^2, J_fT_fT_{3f})||\sum_{n=1,2}\left\{O_{KLs}t_-\right\}_n||\phi(j^2, J_iT_iT_{3i})\rangle \nonumber \\
&= C(K)\langle j || O_{KLs} || j \rangle
\label{eq:CK_double_state}
\end{align}
where $C(K)$ is a simple combination of spin and isospin variables. For a non-trivial system of an actual nucleus, we typically employ the shell model and consider the interacting nucleons to be in different orbitals outside a fixed core. Writing the total wave function for a nucleus with spin $J$, isospin $T$ and $N$ active particles as
\begin{equation}
\Psi(NJT) = \sum_{k}a_k\phi_k(NJT)
\label{eq:shell_model_state}
\end{equation}
with $\phi_k(NJT)$ anti-symmetric single particle wave functions, we find Eq. (\ref{eq:CK_double_state}) generalises directly to
\begin{align}
&\langle \Psi_f(NJ_fT_f)||O_{KLs} ||\Psi_i(NJ_iT_i)\rangle \nonumber \\
&=\sum_{k,l}a_ka_lC_{kl}(K)\langle j_l || O_{KLs} || j_k \rangle,
\end{align}
where $C_{kl}$ describes the coefficients of fractional parentage relevant for $kl$ configuration \cite{Shalit1963}. Comparing this to Eq. (\ref{eq:operator_impulse}) we identify the prefactors as the one body density matrix elements. In the following discussion we are however not concerned with the absolute calculation of matrix elements, but rather in the ratio between two different ones. When taking a ratio of two many-particle matrix elements of rank $K$, we see the complex many body dynamics described by $C(K)$ in Eq. (\ref{eq:CK_double_state}) drops out if both operators are of the same rank $K$. If the shell model state of Eq. (\ref{eq:shell_model_state}) is dominated by a single configuration this result holds approximately. We will extensively use this fact in the analytical calculation of the shape factor $C(Z, W)$ below.

It remains then to write down the shape factor in an analytical form. Here we stand before a crossing, as we can attempt to split up $C(Z, W)$ into a purely leptonic convolution part and a nuclear structure part, or make no attempt at decoupling and stick with the full formulation. The former is done by Holstein, whereas in the seminal work by \textcite{Behrens1982} the two parts remain coupled. We continue here with this approach as it provides the greatest precision results, but present the final results in the notation of Holstein. This allows for easy inspection and a clear connection to the aforementioned symmetries. In Appendix \ref{app:em_comparison} we compare both approaches.

\subsection{Isospin invariant shape factor}
\label{sec:coupled_BB}
In the beginning of this section we have introduced the concept of a shape factor, denoted by $C(Z, W)$. This is not to be confused with the many-particle coupling coefficients in, for example, Eq. (\ref{eq:CK_double_state}). It encompasses the information introduced by the expansion of the nuclear and lepton current in the transition matrix element, and as such for a large part determines the shape of the spectrum. In the formalism by Behrens and B\"uhring $C(Z, W)$ is written as \cite{Behrens1982}
\begin{align}
C(Z, W) = &\sum_{k_e,k_{\nu}, K}\lambda_{k_e}\left\{M_K^2(k_e, k_{\nu})+m_K^2(k_e,k_{\nu}\right. \nonumber \\
& \left. - \frac{2\mu_{k_e}\gamma_{k_e}}{k_eW}M_K(k_e,k_{\nu})m_K(k_e,k_{\nu})\right\},
\label{eq:C_BB}
\end{align}
where
\begin{align}
\lambda_{k_e} &= \frac{\alpha^2_{-k_e}+\alpha^2_{+k_e}}{\alpha^2_{-1}+\alpha^2_{+1}}, \\
\mu_{k_e} &= \frac{\alpha^2_{-k_e}-\alpha^2_{+k_e}}{\alpha^2_{-k_e}+\alpha^2_{+k_e}}\frac{k_eW}{\gamma_{k_e}},
\end{align}
are Coulomb functions depending on $\alpha_k$ (see Eq. \ref{eq:fermi_buhring}), while $M_K(k_e, k_{\nu})$ and $m_K(k_e, k_{\nu})$ contain the convolution of leptonic wave functions and nuclear structure information encoded as form factors, discussed in the previous section. The integers $k_e, k_{\nu}$ are defined as $|\kappa_{e,\nu}|$ where $\kappa_{e, \nu}$ is related to the angular momenta in the usual way\footnote{Here $\kappa$ is the eigenvalue of the operator $K = \beta (\bm{\sigma} \bm{L} + 1)$, such that $k=|\kappa|=j+\frac{1}{2}, \kappa=-l-1$ if $l=j+\frac{1}{2}$, and $\kappa=l$ if $l=j-\frac{1}{2}$.}. It is a well-known fact that that the solution to the Dirac equation does not contain a definite parity, such that we consider the outgoing leptonic wave functions as spherical waves. The integer $K$ corresponds to the multipolarity of the transition, and must form a vector triangle with $j_e$ and $j_{\nu}$ as well as with the nuclear spins $J_i$ and $J_f$. For allowed transitions we have then $|J_i-J_f| \leq K \leq J_i+J_f$ from the nuclear vector triangle.

In this coupled approach all leptonic information is still contained within $M_K$ and $m_K$. Even though the nuclear decompositions are completely equivalent in both the HS and BB formalisms, the treatment of the leptonic current is not. In the latter, a rigorous expansion of the radial wave functions is made in $r^2, (m_eR)^a, (WR)^b$ and $(\alpha Z)^c$ thereby introducing a function $I(k_e, m, n, \rho;r)$ that is sensitive to nuclear shape information\footnote{Here $m=a+b+c$ represents the total power of $(mR), (WR)$ and $(\alpha Z)$, $n=b+c$ is the total power of $(WR)$ and $(\alpha Z)$, and $\rho=c$ is the power of $(\alpha Z)$.}. This function is tabulated both in the general case as well as for a uniformly charged sphere in Tables 4.2 and 4.3 in \textcite{Behrens1982}, respectively. As it also includes nuclear structure information, this is typically combined with the original form factors $F_{KLs}^{(n)}$ and is written as\footnote{This is presented in Eqs. (6.159)-(6.166) in \textcite{Behrens1982}. As $I(k_e, m, n, 0) = 1$, we have $F^{(n)}_{KLs}(k_e, m, n, 0) = F^{(n)}_{KLs}$.} $F^{(n)}_{KLs}(k_e, m, n, \rho)$.

This now allows to better understand the structure of $C(Z, W)$ and continue with its analytical formulation. In the notation by Behrens and B\"uhring, capital letters are used for large components, while lower case terms represent small components. Developing $M_K$ and $m_K$ in terms of $(WR)$ we can after a tedious but straightforward calculation write a general shape factor \cite{Behrens1978, Behrens1982}. In the approach by BB, all constant factors are divided out and one arrives at
\begin{equation}
C(Z, W) = 1+aW+b\frac{\mu_1\gamma }{W}+cW^2,
\label{eq:C_explicit_BB}
\end{equation}
where $a$, $b$ and $c$ are given by Eqs. (14.117)-(14.119) in \textcite{Behrens1982}. We will however only divide out the main matrix elements, $^VF_{000}^{(0)} \equiv g_V\mathcal{M}_F$ and $^AF_{101}^{(0)}\equiv \mp g_A\mathcal{M}_{GT}$, and adjust our shape factor accordingly. 

This leaves us with a series of other, often more complicated, nuclear matrix elements which require evaluation somehow. Assuming isospin invariance and CVC, however, we can link these matrix elements to \emph{electromagnetic} matrix elements. This entails that instead of using initial and final state nuclear wave functions, one can use the full charge distribution as discussed in Sec. \ref{sec:symmetries}. This corresponds to using $F1111 = \frac{27}{35}$, $F1221 = \frac{57}{70}$, $F1222 = \frac{233}{210}$ and $F1211 = -\frac{3}{70}$. This approach can however be improved when using a more realistic charge distribution. Several of these possible replacements have been discussed in the electrostatic finite size corrections in Sec. \ref{size-and-mass}, specifically when discussing the $U$ correction factor. This is elaborated upon in Appendix \ref{app:general_shape_factor}.

\subsubsection{Superallowed $0^+ \to 0^+$ Fermi decay} 
In the case of superallowed Fermi decay, only terms with $K=0$ contribute. We deal then only with different form factor coefficients of the form $^VF_{000}^1(1, m, n, \rho)$. We use the expansion of $M_0(1,1)$ and $m_0(1,1)$ valid to orders $(\alpha Z)^2$, $R^2$, $\alpha Z R$ to calculate $C(Z, W)$ from Eq. (\ref{eq:C_BB}). Here the prefactor $\gamma \mu_1$ can be safely assumed to correspond to unity to our current order of precision. After extraction of $^VF_{000}^0$ we can write the shape factor $C(Z, W)$ in the following expansion
\begin{align}
^{V}C(Z, W)_0 &\simeq 1 + ^{V}C_0 + ^{V}C_1 W \nonumber \\
&~~~ + ^{V}C_{-1}/W + ^{V}C_2 W^2,
\label{V-CZW}
\end{align}
where
\begin{subequations}
\begin{align}
^{V}C_0 &= -\frac{233}{630}(\alpha Z)^2 - \frac{1}{5}(W_0R)^2\nonumber \\
& ~~~\mp \frac{6}{35} \alpha ZW_0 R, \label{eq:V_C_0} \\
^{V}C_1 &= \mp \frac{13}{35} \alpha ZR + \frac{4}{15} W_0 R^2,  \\
^{V}C_{-1} &= \frac{2}{15} W_0 R^2 \pm \frac{1}{70} \alpha ZR \\
^{V}C_2 &= -\frac{4}{15} R^2 ~ .
\end{align}
\end{subequations}

\subsubsection{Pure Gamow-Teller decay}
In pure Gamow-Teller decay the situation becomes more complicated, and we now have contributions from $K \geq 1$ terms. For notational convenience and clarity we first introduce the Holstein variables and the translation used between Holstein's and the Behrens-B\"uhring formalism in which the calculations were performed. We have used
\begin{subequations}
\begin{align}
^AF_{101}^{(0)} &= \mp c_1 \\
^VF_{111}^{(0)} &= -\sqrt{\frac{3}{2}}\frac{1}{MR}b \\
^AF_{110}^{(0)} &= - \frac{\sqrt{3}}{2MR}d
\end{align}
\end{subequations}
where the relevant Holstein form factors to the appropriate order in $\mathbf{q}^2$ are defined in Table \ref{table:form factors}. Extracting now the main Gamow-Teller form factor $^AF_{101}^{(0)}$ we can derive a similar result\footnote{Care must be taken when precisely comparing the evaluation of $h$ and $^AF_{121}$. The latter can be written as
\begin{equation}
^AF_{121}^0 = \mp g_A \mathcal{M}_{121}^0 \mp \frac{g_P}{2M_nR}5\sqrt{\frac{2}{3}}\mathcal{N}_{110}^0,
\label{eq:AF121}
\end{equation}
where upon careful evaluation we find $\mathcal{M}_{121}^0 \propto -\mathcal{M}_{1y}$, and $\mathcal{N}_{110}^0$ is the relativistic matrix element
\begin{equation}
\mathcal{N}_{110}^0 = \sqrt{3}\int \beta \gamma_5 \frac{i\mathbf{r}}{R}.
\end{equation}
We use either a Foldy-Wouthuysen transformation \cite{Foldy1950, Roman1965, Bjorken1964}, or reduce $f(r)$ using the non-relativistic limit of the Dirac equation, to reduce $\beta \gamma_5$. Converting the angular momenta coupling, we finally obtain $\mathcal{N}_{110}^0 = -\sqrt{3}/(2M_NR)\mathcal{M}_{101}^0$. In order to precisely show the influence of the oft-ignored induced pseudoscalar current, we separate them in Eqs. (\ref{eq:AC0})-(\ref{A-C0-C1-C2}).}
\begin{align}
^{A}C(Z, W)_0 &\simeq 1 + ^{A}C_0 + ^{A}C_1 W + ^AC_{-1}/W\nonumber \\
&~~~ + ^{A}C_2 W^2  + \Phi \mathcal{P}(Z,W),
\label{A-CZW}
\end{align}
where
\begin{subequations}
\begin{align}
^AC_0 &= -\frac{1}{5}(W_0R)^2+\frac{4}{9}R^2\left(1-\frac{1}{20}\Lambda\right) \nonumber \\
&~~~+\frac{1}{3}\frac{W_0}{Mc_1}(\mp 2b+d) \pm \frac{2}{5}\alpha Z\frac{1}{MRc_1}(\pm 2b +d) \nonumber \\
&~~~\pm\frac{2}{35}\alpha Z W_0R(1-\Lambda)-\frac{233}{630}(\alpha Z)^2 \label{eq:AC0} \\
^AC_1 &= \pm\frac{4}{3}\frac{b}{Mc_1} + \frac{4}{9}W_0R^2(1-\frac{1}{10}\Lambda) \nonumber \\
&~~~\mp \frac{4}{7}\alpha Z R\left(1-\frac{1}{10}\Lambda\right) \label{eq:AC1}\\
^AC_{-1} &= -\frac{1}{3Mc_1}(\pm 2b +d)-\frac{2}{45}W_0R^2(1-\Lambda) \nonumber \\
&~~~\mp \frac{\alpha Z R}{70} \\
^AC_2 &= -\frac{4}{9}R^2(1-\frac{1}{10}\Lambda)
\label{A-C0-C1-C2}
\end{align}
\end{subequations}
and
\begin{equation}
\Lambda = \frac{\sqrt{2}}{3}10\frac{\mathcal{M}_{121}^{(0)}}{\mathcal{M}_{101}^{(0)}}.
\label{eq:Lambda_CA}
\end{equation}
In order to stress its importance, we have separated out one part of Eq. (\ref{A-CZW}), written as $\Phi \mathcal{P}(Z, W)$. Here $\Phi$ is
\begin{equation}
\Phi = \frac{g_P}{g_A}\frac{1}{(2M_NR)^2}
\end{equation}
and represents the typically ignored induced pseudoscalar contribution. Using the free nucleon values obtained from PCAC we find $\Phi \approx -0.13$ for a typical medium-$Z$ with $R \sim 0.01$. This raises serious questions about the validity of neglecting induced psuedoscalar effects when high precision is required, even when assuming a strongly quenched $g_P$ value (see also the work by \textcite{Gonzalez-Alonso2014}). For this reason we wish to stress the importance Now $\mathcal{P}(Z, W)$ is a polynomial arising from the contributions of $^AF^1_{101}(k_e, m, n, \rho)$ and $^AF_{121}^0(k_e, m, n, \rho)$ as can be seen from Table \ref{table:form factors}. The generalization from the bare form factors $^AF_{1L1}$ to the form factor coefficients is straightfoward and is for example tabulated in Table 7 by \textcite{Behrens1971}. The non-relativistic approximation is derived using standard methods, and $\mathcal{P}$ is then given by
\begin{equation}
\mathcal{P}(Z,W) = P_0 + P_1W + P_{-1}/W
\end{equation}
where
\begin{subequations}
\begin{align}
P_0 &= \frac{2}{3}(W_0R)^2 - \frac{20}{27}R^2 \pm \frac{868}{675}\alpha Z W_0R \nonumber \\
&~~~~ + \frac{63}{50}(\alpha Z)^2 \\
P_1 &= \frac{80}{27}W_0R^2 \pm \frac{195}{1215}\alpha ZR \\
P_{-1} &= -\frac{4}{3}W_0R^2 \pm \frac{1}{25}\alpha Z R
\end{align}
\end{subequations}
Many of the factors in these equations can be large enough to by significant on the few parts in $10^4$ or even more. Clearly, care has to be taken when performing high precision spectrum shape measurements, and in particular when extracting $ft$ values from mirror transitions. The quenching of $g_P$ in nuclear matter is expected to be nucleus-dependent, just like $g_A$, such that its inclusion can introduce shifts in the overall $\mathcal{F}t$ values.

Contributions coming from $K\geq 2$ contribute only at the few $10^{-6}$ level, such that for our current precision we do not take it into account.

\subsection{Isospin breakdown and nuclear structure}
\label{sec:isospin_breakdown}
The results presented in the previous section depended on isospin invariance to replace the nuclear wave functions in the matrix elements by the complete charge distribution. Isospin is, however, not an exact symmetry and its breakdown is significant on the level of precision we aim for. This breakdown manifests itself in the form of an isovector correction, and further requires the explicit calculation of several single-particle matrix elements in the Gamow-Teller shape factor.

\subsubsection{Isovector correction}
The use of CVC and isospin invariance links weak matrix elements to electromagnetic variables, specifically to the so-called isovector component. Due to the breakdown of isospin, we introduce a further modification called the isovector correction based on the approach by \textcite{Wilkinson1993b}. In order to put the correction on a clearer footing, we take a look at the problem before using a perturbative expansion of the wave functions. This can more easily be seen in the notation of \textcite{Hardy2005c} where $M_K$ is written as
\begin{equation}
M_K(k_e, k_{\nu}) = \frac{\sqrt{4\pi}}{\hat{K}\hat{J}_i}\sum_{Ls}(-)^{K-L}\langle j_{\alpha} ||F(r) \hat{T}_{KLs} || j_{\beta} \rangle,
\end{equation}
with $F(r)$ a combination of lepton wave functions. The dominant contributions for the Fermi and Gamow-Teller matrix elements come from its temporal and spatial components, respectively. In this notation, both are proportional to
\begin{equation}
L_L = C \int_0^{\infty} R_\alpha (r)F(r) R_\beta(r)r^2 dr
\label{eq:L_L_Towner}
\end{equation}
where all constants related to angular momentum coupling have been absorbed by $C$ and $R_{\alpha (\beta)}$ stands for the initial (final) nuclear wavefunction. These states tend to have a higher $\langle r^2\rangle$ value compared to the entire charge distribution. Combined with the decrease of the leptonic wave functions inside the nuclear volume, this results in an overestimation of the ratio of form factors in Eqs. (\ref{A-CZW}) and (\ref{V-CZW}). We generalize the correction proposed by \textcite{Wilkinson1993b} and write
\begin{equation}
C(Z, W)_I = \frac{\left[\int_0^{\infty}dr\Psi_e\Psi_{\nu}\rho_{w}\right]^2}{\left[\int_0^{\infty}dr\Psi_e\Psi_{\nu}\rho_{\text{ch}}\right]^2},
\label{eq:C_I_wilkinson}
\end{equation}
where $\rho_{w}$ is the nuclear density distribution participating in the decay process, the so-called \emph{weak} charge distribution, and $\rho_{\text{ch}}$ the charge distribution. An expansion of the lepton wave functions gives rise to terms of order $r^2$, such that estimates of the \emph{weak} rms radii enter in the equation, denoted by $\langle r^2 \rangle_{\text{w}}$. We write the radial part of the weak charge distribution as a product of initial and final wave functions, each consisting out of a sum of weighted harmonic oscillator functions\footnote{The harmonic oscillator wave functions can be trivially extended to include a spin-$\frac{1}{2}$ contribution, and leaves the radial part of the Hamiltonian unchanged. The possible $l(l')$ will thus be restricted to $j\pm \frac{1}{2}$ of initial and final particle states. In case $j$ is not any more a good quantum number, such as an axially deformed potential, more $l$ values enter. This is discussed in Sec. \ref{sec:deformation}.}
\begin{equation}
\rho_{w}(r) = C\sum_{nln'l'} \alpha_{nl}\beta_{n'l'}R_{nl}(r)R_{n'l'}(r)
\end{equation}
where $C$ is a normalization constant, $R_{nl}(r)$ is the harmonic oscillator radial wave function for quantum numbers $n$ and $l$, such that $\int R_{nl}^2dr = 1$. In closed form these are written as
\begin{subequations}
\begin{align}
R_{nl}(r) &= Nr^{l+1} ~ \exp{(-\nu r^2)}~ L^{l+1/2}_k(2\nu r^2) \label{eq:R_nl_HO} \\
N &= \left[\left(\frac{2\nu^3}{\pi}\right)^{\frac{1}{2}}\frac{2^{k+2l+3}~ k! ~ \nu^l}{(2k+2l+1)!!} \right]^{\frac{1}{2}} \\
k &= n-1
\end{align}
\end{subequations}
where $L_{n}^{(\alpha)}(x)$ is the generalized Laguerre polynomial, and $\nu$ is a free parameter. An additional advantage of these functions is the availability of a closed formula for the radial integrals which will be necessary in the treatment below \cite{Nilsson1955}
\begin{align}
\langle n_f l_f | &r^L | n_i l_i \rangle = (-1)^{n_i+n_f}(2\nu)^{L/2} \nonumber \\ &\times\sqrt{\frac{\Gamma(n_i)\Gamma(n_f)}{\Gamma(n_i+t-\tau_i)\Gamma(n_f+t-\tau_f)}} \tau_i!\tau_f! \nonumber \\
&\times \sum_{\sigma} \left[\frac{\Gamma(t+\sigma+1)}{\sigma ! (n_i-\sigma-1)!(n_f-\sigma-1)!} \right. \nonumber \\
&\left. \times \frac{1}{(\sigma+\tau_i-n_i+1)!(\sigma+\tau_f-n_f+1)!} \right]
\end{align}
where
\begin{subequations}
\begin{align}
\tau_i &= \frac{1}{2}(l_f-l_i+L) \\
\tau_f &= \frac{1}{2}(l_i-l_f+L) \\
t &= \frac{1}{2}(l_i+l_f+L+1)
\end{align}
\end{subequations}
with the upper and lower limits of $\sigma$
\begin{align}
&\text{max}(n_i-l_i-1,n_f-\tau_f-1) \leq \sigma \nonumber \\
&\leq \text{min}(n_i-1, n_f -1).
\end{align}
For the specific case when $n=n'$ and $l=l'$, we find the well-known relation
\begin{equation}
\langle r^2 \rangle_{nl} = \frac{1}{4\nu}(4n+2l-1).
\label{eq:r2_HO}
\end{equation}
The free parameter $\nu$ is related to the traditional oscillator parameter through $b = \sqrt{2\nu}$. It can now be constrained through its relation with the nuclear radius. Summing the contribution of Eq. (\ref{eq:r2_HO}) for all nucleons and equating it to the nuclear rms radius, one finds $b = 2^{7/6}3^{-1/6}5^{-1/2}RA^{-1/6} \approx 0.836 R A^{-1/3}$. This corresponds to the traditional approximation $\hbar \omega = 41 A^{-1/3}$\,MeV when using $R = r_0 A^{1/3}$.


The $C_I$ correction as defined in Eq. (\ref{eq:C_I_wilkinson}) can be calculated using a more explicit formulation of the lepton wave functions. In the non-relativistic approximation we write the latter as the large part of the radial behavior. For the $j=1/2$ $\beta$ particle and (anti)neutrino this is $g_{-1}(r)$ and $j_0(qr)$, respectively, where $j_0$ is a spherical Bessel function and $q$ the (anti)neutrino momentum. In the standard way of expanding in powers of $r$ and equating terms, one finds to order $r^2$
\begin{subequations}
\begin{align}
\Psi_e (r) &\approx 1 - \frac{1}{6}\left[(W\pm V_0)^2-1 \right]r^2 \label{eq:psi_e_exp}\\
\Psi_{\nu} (r) &\approx 1 - \frac{1}{6}(W_0-W)^2r^2, \label{eq:psi_v_exp}
\end{align}
\end{subequations}
for a uniformly charged sphere, with $V_0 = \pm 3\alpha Z/(2R)$ the electrostatic potential at the origin. Equation (\ref{eq:C_I_wilkinson}) then reduces to
\begin{align}
C(Z, W)_I &= \sum_{nln'l'}\alpha_{nl}^2\beta_{n'l'}^2 I_{nl}^{n'l'}(I_{nl}^{n'l'}-2\xi \langle r^2 \rangle^{n'l'}_{nl}) \nonumber \\
&\times (1+\frac{6}{5}\xi R^2)
\label{eq:C_I_approx_HO}
\end{align}
to first order in $\xi$, where $I_{nl}^{n'l'} = \int_0^{\infty} R_{nl}R_{n'l'}dr$ and 
\begin{equation}
\xi = \frac{1}{6}\left[(W_0-W)^2+(W+V_0)^2-1\right].
\end{equation}
Here we used our definition that $\langle r^2 \rangle = \frac{3}{5}R^2$ for any charge distribution. In the extreme-single particle approximation in $j$-$j$ coupling (1s, 1p, 1d$_{\frac{5}{2}}$, 2s, ...) this expression simplifies drastically and the sum in Eq. (\ref{eq:C_I_approx_HO}) disappears. Further assuming the initial and final particle to be in the same $nl$ state, we find
\begin{equation}
C_I = 1-\frac{1}{2\nu}\xi (4n+2l-1)+\frac{6}{5}\xi R^2
\label{eq:C_I_Intermediate}
\end{equation}
When considering the modified Gaussian function of Eq. (\ref{eq:mod_gauss}) as a charge distribution, we can relate $\nu$ to the $A$ fit parameter using the same idea as that of \textcite{Wilkinson1993b}. We can interpret the constant term in Eq. (\ref{eq:mod_gauss}) as the $l=0$ contribution while the $A(r/a)^2$ then necessarily stands for $l\neq 0$. Assuming $n=1$ and $l=0$ as done there, we must have $3/(4\nu)=3a^2/2$ after proper normalisation of the $l=0$ part of $\rho_{MG}$. Reordering terms we find then from Eq. (\ref{eq:C_I_Intermediate}) his result
\begin{equation}
C(Z, W)_I = 1-\frac{8}{5}w\xi R^2\frac{1}{5A'+2}
\label{eq:wilkinson_CI}
\end{equation}
where
\begin{equation}
w = (4n+2l-1)/5.
\end{equation}
Our more general expression (Eq. (\ref{eq:C_I_approx_HO})) does not rely on a specific charge distribution, and should employ the same charge distribution as was used to evaluate the $^{V/A}F1mn\rho$ terms in the previous section. This is discussed further in Appendix \ref{app:general_shape_factor}.


In case of vector transitions we can simply apply $C_I$ to the entire shape factor $^VC$ of the previous section. It is slightly more complicated for Gamow-Teller transitions as we have assumed the nuclear wave functions to correspond to the charge distribution only for form factors of the type $^AF_{101}^{(0)}(1, m, n, \rho)$, analogous to the vector result. In order to avoid double counting then, we can split up $^AC(Z, W)$ into a nuclear structure dependent part $^A_{\text{ns}}C(Z, W)$ containing all terms proportional to $b$, $d$ and $\Lambda$, and all other shape dependent terms $^A_{\text{sh}}C(Z, W)$. We properly apply the isovector correction using
\begin{equation}
^AC(Z, W) \longrightarrow C_I(Z, W)\,^A_{\text{sh}}C(Z, W) + \,^A_{\text{ns}}C(Z, W).
\end{equation}
The terms corresponding to $^A_{\text{sh}}C$ are, for example, written by \textcite{Wilkinson1990}, and discussed in Appendix \ref{app:general_shape_factor}.

\subsubsection{Single particle matrix elements}
\label{sec:single_particle}
In the nuclear structure shape function of pure Gamow-Teller decay, $^A_{\text{ns}}C$ we are sensitive to several ratios of non-trivial form factors. The weak magnetism contribution, denoted by $b$, can be constrained by CVC for specific transitions, while $d$ can be shown to vanish for analog transitions. We have no such constrictions for $\Lambda$, defined in Eq. (\ref{eq:Lambda_CA}), and so we must rely on the explicit calculation of the matrix elements. Specifically, we need to calculate the ratio between e.g. $\mathcal{M}_{121}^{(0)}$ and $\mathcal{M}_{101}^{(0)}$. Explicit expressions have been provided in Appendix \ref{app:single_particle_tables} for the convenience of the reader. As an example, we write the ratio $\mathcal{M}_{121}^{(0)}/\mathcal{M}_{101}^{(0)}$ here for several allowed scenarios
\begin{equation}
\frac{\mathcal{M}_{121}^{(0)}}{\mathcal{M}_{101}^{(0)}} = \left\{
\begin{array}{lr}
-l\frac{\sqrt{2}}{2l+3} \frac{\langle r^2 \rangle}{R^2}& j_f=j_i=l+\frac{1}{2} \\
-(l+1)\frac{\sqrt{2}}{2l-1}\frac{\langle r^2 \rangle}{R^2} & j_f=j_i=l-\frac{1}{2} \\
\frac{1}{2^{3/2}}\frac{\langle r^2 \rangle}{R^2} & j_f = l \pm \frac{1}{2} ~~ j_i = l \mp \frac{1}{2} 
\end{array} \right.
\label{eq:M_ratio_SP}
\end{equation}

We can test the validity of Eq. (\ref{eq:M_ratio_SP}) in the simpler cases where we expect our single-particle approximation through $j-j$ coupling to work. This can be done by comparing results for the weak magnetism contributions for which experimental data are available. As an example, we consider the mirror decays of 6 isotopes throughout the lower mass range (see \textcite{SeverijnsTBP} and references therein). In the extreme single-particle approximation for mirror nuclei the weak magnetism current is easily calculated, and one finds $b/Ac_1 = 1/g_A(l+1+g_M)$ and $-1/g_A(l-g_M)$ for $j_f=j_i=l\pm 1/2$, respectively.
Results are shown in Table \ref{tab:bAc_mirror_sp}.

\begin{table}[h!]
\centering
\caption{Examples of $T=1/2$ mirror transitions for which we compare experimental $b/Ac_1$ values to their pure single-particle analogues throughout the mass range $3 < A < 59$. We have some freedom in choosing $g_A$ for the higher lying states. For $^3$H we used $g_A=1.2723$ while for all $sd$ nuclei we used $g_A = 1.1$ based on standard works \cite{Brown1988, Wildenthal1984} and $g_A = 1$ for the $fp$ shell \cite{Siiskonen2001, Martinez-Pinedo1996}.}
\begin{ruledtabular}
{\renewcommand{\arraystretch}{1.4}
\begin{tabular}{l|ccc}
Nucleus & $J^{\pi}$ & $(b/Ac_1)_{\text{exp}}$ & $(b/Ac_1)_{\text{sp}}$ \\
\hline
$^3$H & $1/2^+$ & 4.2212(24) & 4.48 \\
$^{13}$N & $1/2^-$ & 3.1816(75) & 3.37 \\
$^{31}$S & $1/2^+$ & 5.351(14) & 5.18 \\
$^{39}$Ca $^{\text{a}}$ & $3/2^+$ & 1.2349(28) & 2.46 \\
$^{45}$V & $7/2^-$ & 7.51(23) & 8.70 \\
$^{59}$Zn & $3/2^-$ & 6.68(33) & 6.71
\end{tabular}}
\end{ruledtabular}
\label{tab:bAc_mirror_sp}
\begin{flushleft}
\footnotesize{$^{\text{a}}$ The electromagnetic moments of $^{39}$Ca do not agree very well with the Schmidt values \cite{Minamisono1976, Matsuta1999}, implying larger deviations were expected to occur.}
\end{flushleft}
\end{table}

We conclude that for this investigated mass range where the single-particle behavior can be reasonably well assumed, the non-relativistic impulse approximation evaluation yields results consistent within 15\% except for the $d_{3/2}$ orbital in the case of $^{39}$Ca. From its magnetic moment and that of its related neighbor $^{37}$Ar \cite{Pitt1988}, we could however already conclude that a single-particle approach is not justified. Indeed, despite having only one hole outside a double magic nucleus, core polarization and meson exchange effects are significantly enlarged \cite{Barroso1975}.

We can conclude this to be a very good agreement for a very simple approach. This assumes a careful study of the magnetic moments and ground state properties, however, taking into account possible orbital reversals through strong deformations. We assume the matrix element evaluation in Eq. (\ref{eq:M_ratio_SP}) to perform similarly, justifying our single-particle approach in this mass range given the previously discussed requirements.

\subsection{Relativistic terms in superallowed decays}
\label{sec:relativistic_terms}
In the previous section, when dealing with superallowed $0^+ \to 0^+$ decay, we have ignored contributions from relativistic matrix elements, as is typically done. For the Gamow-Teller decay we have made no such approximation, and immediately implemented their properties. To recapitulate, with relativistic we mean matrix operators directly or indirectly containing $\gamma_5$. This is because these operators connect the small, $f$, and large, $g$, radial wave functions when treating the transforming nucleon relativistically as in Eq. (\ref{eq:sol_dirac}). Specifically, we have to calculate matrix elements of the form
\begin{equation}
\int_0^{\infty}g_f(r)\phi(r)f_i(r)r^2dr,
\end{equation}
where $\phi(r)$ is equal to $j_L(qr)$ or $(r/R)^{L+2N}I(k_e, m, n, \rho; r)$, and $f$ and $g$ can be interchanged. One then typically introduces a further non-relativistic approximation \cite{Brysk1952, Talmi1953, Rose1954, Rose1954a, Eichler1963, Lipnik1966, Hardy2005a}. In this non-relativistic approach the small $f$ component can be transformed such that $g_{\kappa}(r)$ is a solution of the Schr\"odinger equation, as we have done in the footnote when discussing Eq. (\ref{A-CZW}). Typically, a factor $1/(M_NR) \approx 1/20$ appears, suppressing the relativistic influence significantly. In the simple case of superallowed Fermi decay we can write
\begin{align}
C(Z, W) &= C(Z, W)_{NR} + \frac{^VF_{011}^{(0)}}{^VF_{000}^{(0)}}f_2(W) \nonumber \\
&~~~ + \frac{^VF_{011}^{(1)}}{^VF_{000}^{(0)}}f_3(W) + \ldots,
\end{align}
in the notation of \textcite{Behrens1968}, where $f_i(W)$ are simple functions of the lepton momenta and slowly varying Coulomb functions, proportional to $R$, and where the subscript NR stands for non-relativistic. These Coulomb functions are described for Fermi transitions in Appendix \ref{app:relativistic_me}. Using the results in our discussion of CVC, we can approximate Eq. (\ref{eq:F011_CVC}) to write\footnote{This relation is not exact and can in fact vary significantly, as discussed by \textcite{Damgaard1966}. An exact treatment can, for example, be found in \textcite{Behrens1968}. We are here, however, only interested in an order of magnitude estimation.} \cite{Fujita1962}
\begin{equation}
2N~^VF_{011}^{(N-1)} = -\{(W_0\mp (m_n-m_p))R\pm \frac{6}{5}\alpha Z\}~^VF_{000}^{(N)}.
\end{equation}
It is then immediately clear why we have initially chosen to neglect these matrix elements, as the prefactor is exactly zero in the case of isospin invariance. Here $6\alpha Z /5R$ is the difference in Coulomb energy assuming a uniformly charged sphere. Isospin is however not an exact symmetry nor is the nucleus a perfectly homogeneous sphere, and \textcite{Wilkinson1993b} has treated this with much care, providing upper limits on the magnitude of the effect averaged over the full spectrum. This was again done using the assumption of an extreme single-particle interaction in the nucleus. In this case this results in a conservative estimate, as the effect increases for higher lying orbitals. For the heaviest nucleus considered, $^{54}$Co, the upper limit was set at 0.01\%, while for the lightest, $^{14}$O, it was put at 0.001\%. The endpoint energy for the former is 8243.12\,keV, so that an average slope would be in the $10^{-5}\,$MeV$^{-1}$ range as the dominant terms in $f_2(W)$ are linear in $W$. This gives a general idea of the involved magnitudes. As is done for Gamow-Teller decays discussed above, we can attempt to evaluate these matrix elements explicitly using single-particle values. This is elaborated upon in Appendix \ref{app:relativistic_me}.

\subsection{Nuclear deformation}
\label{sec:deformation}

As was previously discussed in Sec. \ref{sec:D_FS}, a deviation in the spherical shape of the nucleus introduces profound effects on the $\beta$ decay rate on the few $10^{-4}$ level. This influence extends towards the terms sensitive to nuclear structure discussed before. 

\subsubsection{Leptonic convolution}
The most obvious change occurs in the convolution of the lepton and nuclear wave functions, where we initially assumed a uniform charge density. Here we used the rough expansion of the lepton wave functions near the origin of Eqs. (\ref{eq:psi_e_exp})-(\ref{eq:psi_v_exp}).This uniform density is then replaced by that of Eq. (\ref{eq:charge_dist_deformed}), from which we can calculate the electric potential at the center of the nucleus, yielding \cite{Wilkinson1994}
\begin{equation}
V_{0d} = \frac{3\alpha Z}{2a}(\Phi^2-1)^{-1/2}\ln \left[(\Phi^2-1)^{1/2}+\Phi \right]
\end{equation}
for the prolate case and
\begin{equation}
V_{0d} = \frac{3\alpha Z}{2a}(1-\Phi^2)^{-1/2}\arcsin (1-\Phi^2)^{1/2}
\end{equation}
for the oblate case where
\begin{equation}
\Phi = \frac{b}{a}
\end{equation}
as discussed before. We follow the approach of \textcite{Wilkinson1994} and write 
\begin{equation}
\zeta_i = \frac{1}{10}\left[(W_0-W)^2+(W\pm V_{0i})^2-1 \right],
\end{equation}
such that we define
\begin{equation}
D_{C}(Z, W, \beta_2) = \frac{1-\zeta_d R^2}{1-\zeta_s R^2},
\label{eq:D_C0}
\end{equation}
where $V_{0s} = \pm 3\alpha Z / (2R)$. The effect of $D_{C}$ largely compensates that of $D_{FS}$ defined in Sec. \ref{sec:D_FS}, so that the effect is typically on the order of a few $10^{-4}$ for extreme deformations.

Finally, we discuss the influence of deformation on our so-called isovector correction $C_I$. Here, we account for the fact that the decaying nucleon sits in an orbital with limited overlap with the full charge distribution. We considered this nucleon to sit in the highest occupied orbital according to the typical $j$-$j$ coupling. In the deformed case, however, $j$ is not any more a good quantum number and the sum in Eq. (\ref{eq:C_I_approx_HO}) extends over several $l$ rather than just $j\pm \frac{1}{2}$. This can create significant deviations in the expectation values for $\langle r^2\rangle_w$. We briefly discuss the deformed harmonic oscillator in the following section and comment on the modification of $C_I$ in the final overview and crosscheck in Sec. \ref{sec:bss_overview}.

\subsubsection{Deformed single-particle matrix elements}
\label{sec:deformed_single_particle}
In the case of Gamow-Teller transitions we are sensitive to matrix elements which cannot be reduced to their electromagnetic analogs through the use of CVC, and are calculated using single-particle matrix elements. In the case of nuclear deformation, our initial harmonic oscillator wave functions are not any more good approximations of the nuclear wave function, as $l$ is not a good quantum number. The single-particle wave function is then a normalized linear combination \cite{Davidson1968}
\begin{equation}
\chi_{\Omega}(\bm{r}') = \sum_j C_{j\Omega}\phi(j \Omega)
\end{equation}
with $\Omega$ the projection of the single-particle angular momentum on the symmetry-axis of the deformed nucleus, and $\bm{r}'$ the radial coordinate in the body-fixed frame. The coefficients $C_{j\Omega}$ are tabulated by e.g. \textcite{Nilsson1955} and \textcite{Davidson1968} for different values of $\beta_2$. The $\phi(j\Omega)$ are solutions of the spherical single-particle orbital with the proper spin-angular functions
\begin{equation}
\phi(j\Omega) = g(r', \kappa)\chi^{\Omega}_{\kappa}
\end{equation}
such that we can keep using the harmonic oscillator radial wave functions in the calculation of matrix elements. We use the result of \textcite{Behrens1982} to find
\begin{align}
&\langle \phi(J_fK_f; \Omega_f) || O_{KLs} || \phi(J_iK_i; \Omega_i) \rangle \nonumber \\
&= \sqrt{\frac{(2J_f+1)(2J_i+1)}{(1+\delta_{K_f0})(1+\delta_{K_i0})}} \sum_{j_fj_i} C_{j_f\Omega_f}C_{j_i\Omega_i} \nonumber \\
&\times \left\{(-1)^{J_f-K_f+j_f-\Omega_f} \left( \begin{array}{ccc}
J_f & K & J_i \\
-K_f & \Omega_f-\Omega_i & K_i
\end{array} \right)  \right. \nonumber \\
&\times \left( \begin{array}{ccc}
j_f & K & j_i \\
-\Omega_f & \Omega_f-\Omega_i & \Omega_i
\end{array}\right) + \left( \begin{array}{ccc}
J_f & K & J_i \\
K_f & -\Omega_f-\Omega_i & K_i
\end{array}\right) \nonumber \\
&\left. \times \left(\begin{array}{ccc}
j_f & K & j_i \\
\Omega_f & -\Omega_f-\Omega_i & \Omega_i
\end{array} \right) \right\}\langle j_f || O_{KLs} || j_i \rangle
\label{eq:deformed_matrix_element}
\end{align}
where the last factor is again the single-particle matrix element, and $K$ is the projection of the combined angular momentum $\bm{I} = \bm{L} + \bm{j}$ on the symmetry-axis, with $\bm{L}$ the angular momentum of the deformed core, and $\bf{j}$ that of the single particle. In the rotational ground state we have then for an odd-$A$ nucleus $K = \Omega$. In both even-even and odd-odd nuclei the active particles couple to $K=0$ in the ground state. Equation (\ref{eq:deformed_matrix_element}) imposes an additional selection rule as $K \geq |K_f-K_i|$. As an example, consider the weak magnetism form factor $b/Ac_1$ for the heavily deformed $^{19}$Ne mirror isotope. Its ground state is $1/2^+$, consistent with a deformation of $\beta_2=0.269$ of \textcite{Moeller2015}, meaning the unpaired nucleon sits in the $1/2+[220]$ orbital. From \textcite{Davidson1968}, we find the wave function is dominated by the $1d_{5/2}$ harmonic oscillator function, combined with parts from $2s_{1/2}$ and $1d_{3/2}$. From our deliberations of Eq. (\ref{eq:M_ratio_SP}) we know that only cross-terms with $l_i = l_f$ will survive. As $^{19}$F has a nearly identical deformation, we assume $C_{j_f\Omega_f} = C_{j_i\Omega_i}$. We find then $b/Ac_1 = 5.55$ rather than $7.01$ when assuming a pure $d_{5/2}$ occupation. This is much closer to the value we get from CVC, $(b/Ac_1)_{CVC} = 4.9129(25)$, and entails that even for very strong deformations, we are able to predict these quantities in a single-particle estimate within 15\%. This is more extensively discussed elsewhere \cite{SeverijnsTBP}.

\subsection{Summary}
\label{sec:C_summary}
We have presented in a fully analytical form the so-called shape factor correction to the $\beta$ spectrum, both for Fermi and for the first time also for Gamow-Teller decays. Calculations were performed in the Behrens-B\"uhring formalism and coupled to the work of Holstein. This allows one to obtain the highest precision results due to an improved treatment of the lepton wave functions while maintaining notational clarity. In doing so, we expounded on the symmetry properties that underlie the relevant form factors, and discussed their calculation, focusing both on the reduction to single-particle matrix elements and their careful evaluation. As the move towards sub-per mille precision requires, we have further discussed the evaluation of deformation influences, the breakdown of isospin invariance and pointed to the relevance of induced interactions. We attempted to do this while preserving transparency in our calculations in the complex and at times convoluted methods required for the shape factor evaluation. The oft-neglected induced pseudoscalar interaction was kept alive in our formulation, and we have shown the need for its careful evaluation. Despite non-trivial quenching in the nuclear medium, its significance cannot be underestimated when high precision results are required.

The move towards the formulation of Holstein allows for an easier interpretation of and use with experimental results. In the appendices we show the correspondence between our improved results and the earlier conclusions of Holstein and co-authors to put aside any possibility for double counting, for the convenience of the reader. We believe then we have presented a more complete, transparent and clear-cut integration of the final state Coulombic effects without the artificial separation of nuclear structure and leptonic convolution effects.

For many experimentally interesting cases we have no use for CVC to derive precise values for the induced currents, raising the question of the required precision on these observables. As many spectral shape measurements are intent on measuring a possible Fierz contribution, the main induced spectral distortions come from the term linear in $W$. Given values of $b/Ac_1 \sim \mathcal{O}(5)$, the slope is influenced mainly by the weak magnetism contribution, averaging around $0.4\%$ MeV$^{-1}$. The next largest contribution comes from the $2\alpha Z R x/35$ term, where we both have to carefully evaluate the $\mathcal{M}_{121}^{(0)}$ matrix element as well as the induced pseudoscalar contribution (Eq. (\ref{eq:Lambda_CA})). For both weak magnetism and $\Lambda$ of Eq. (\ref{eq:Lambda_CA}), an extreme single-particle calculation can be sufficient to bring the required precision into the $10^{-4}$ domain, given that it is a valid approximation. We have seen in Sec. \ref{sec:single_particle} that this is not always the case. Its legitimacy can be deduced from a study of the magnetic moments. When we are not able to confidently trust the extreme single-particle limit, however, one requires shell model input. Care then has to be taken to avoid double counting with deformation effects. Spectral measurements in particular have an advantage, however, in that all terms linear in $W$ in the shape factor have a strongly suppressed influence from $\Lambda$, putting the focus on $b/Ac_1$ for which explicit experimental data is available.

\section{Atomic and chemical effects}
\label{atomic_effects}
The nucleus cannot be completely separated from its orbiting electrons as the decay is governed by the total Hamiltonian. Even though the interaction point lies within the nucleus, the emitted $\beta$ particle undergoes continuous interaction with the atomic electrons that surround it. The screened Coulomb field both changes the interaction density within the nuclear volume and opens possibilities for more discrete interactions such as shake-off and exchange effects. These discrete processes all originate from a non-orthogonality between initial and final atomic states. We discuss in turn the screening correction, the exchange effect, shake-up and shake-off processes, the atomic mismatch correction, possible bound state $\beta$ decay, and explore the molecular effects on the $\beta$ spectrum and decay rate. The bulk of the effects are typically located in the low energy region, as expected from the Compton wavelength of the outgoing electron. However, experiments sensitive to explicit energy dependencies in the $\beta$ spectrum shape can not ignore atomic effects even at higher energies. 

The description of screening has been known for several decades and is now combined with numerical results from atomic physics. In addition, a new analytical fit of the exchange effect is proposed here, based on precise numerical calculations.

Finally, we discuss and review molecular effects on the outgoing beta spectrum. As the decaying atom can reside in many different chemical environments, the consequent molecular effects have to be evaluated in a case by case manner. We discuss general features and point to possible problematic areas in an analytic fashion.

\subsection{Screening corrections}
\label{screening}

When the $\beta$ particle is created inside the nucleus, it sees not only the nuclear charge but also the atomic electrons surrounding it. The electromagnetic potential is altered at the site, enhanced by the non-zero probability of finding an $s$ electron inside the nucleus. The effective charge as seen by the $\beta$ particle is screened because of atomic electrons. Whereas the inclusion of the Fermi function is a first order Coulomb correction, effects coming from a screened nuclear charge are higher order effects. This is because the largest deviation in the screened electronic potential lies at the atomic rather than nuclear radius. Considering the Fermi function equivalent to switching the electron radial wave function in the $S$ matrix from plane wave to Coulomb function, an analytical treatment of screening requires an analytical solution of the Dirac equation in a screened Coulomb potential. Unfortunately, this is inherently impossible \cite{Rose1961}. The Schr\"odinger and Klein-Gordon equation allow for an analytical solution in a simple screened Coulomb potential \cite{Durand1964}, but the validity of this can be questioned when considering a high-precision description.

To overcome this hurdle, initial results were obtained by Rose \cite{Rose1936} using a WKB argument, arguing a simple rescaling of the total $\beta$ particle energy
\begin{equation}
\tilde{W} = W- V_0, \qquad \tilde{p} = \sqrt{\tilde{W}^2-1},
\end{equation}
where $V_0$ is the potential shift due to screening at the origin. The screening correction is then $\tilde{p}\tilde{W}\tilde{F}/pWF$. Even though this approach is strictly only valid for energies going toward infinity, a very good behavior was nevertheless found when comparing to numerical results \cite{Matese1966}. However, due to the re-scaling of the energy the correction is not defined for energies lower than $V_0$, resulting in discontinuities for energies typically lower than 10\,keV. Further study of electronic screening has been a subject of intensive research over several decades \cite{Longmire1949, Brown1964, Good1954, Durand1964, Buhring1965a, Wilkinson1970, Buhring1984, Lopez1988}, including using the Feynman diagram technique \cite{Durand1987}. There appears to be some confusion, however, in the evaluation of $V_0$, specifically whether it corresponds to the potential shift relative to the daughter or mother atom as it was not clearly stated in the initial article by Rose \cite{Rose1936}. As described by different authors, $V_0$ is defined as \cite{Lopez1988, Saenz1997a}
\begin{equation}
V_0 = \pm \alpha  \langle \chi | \sum_{i=1}^{Z_p} \frac{1}{r_i} | \chi \rangle,
\label{eq:def_V0_simple}
\end{equation}
where $Z_p$ and $|\chi \rangle$ correspond to the charge and electronic wave function of the \emph{parent} atom, rather than the daughter atom. This has been noted as well by \textcite{Buhring1984a}. The result of Eq. (\ref{eq:def_V0_simple}) is obtained after applying the sudden approximation, i.e. the charge change of the nucleus is instantaneous compared to the static electronic distribution, and closure for the final atomic states. To first order in $\alpha Z$ these authors then find
\begin{equation}
S(Z, W) = 1+\frac{WV_0}{p^2} + \frac{V_0}{W},
\label{eq:screening_simple}
\end{equation}
which corresponds to the result initially obtained by Rose \cite{Rose1936}. Instead of performing a Lippmann-Schwinger expansion of the Coulomb interaction, one can, however, move beyond perturbation theory, as discussed by Halpern \cite{Halpern1970}. The latter shows that Coulomb interaction before and after the $\beta$ decay has the effect of changing the beta wave functions from a plane wave to an exact solution of the Dirac equation in a screened Coulomb field. This fact will be used when deriving a more elaborate screening correction below.

Of paramount importance for a consistent description of screening is an accurate atomic potential in the region of the nucleus. Several potentials were proposed over the past few decades, many of which are based on the Thomas-Fermi or Thomas-Fermi-Dirac potential describing the atom with a non-relativistic statistical model \cite{Fermi1927, Thomas1927a, Dirac1930, Bethe1997, Gross1979, Bonham1963}. These potentials typically fail, however, for regions of large potential gradients and low electron densities. This makes them unreliable for distances close to but also far away from the nucleus \cite{Salvat1987}, precisely the region where screening will be most important.

Of particular theoretical importance due to its simplicity is the Hulth\'en screening potential
\begin{equation}
V_H(r) = \mp \alpha Z \lambda [\exp(\lambda r)-1]^{-1},
\label{eq:hulthen}
\end{equation}
with a single parameter $\lambda$ that depends on $Z_p$, the charge of the parent nucleus. This dependence is typically written as \cite{Buhring1984a, Buhring1983a, Buhring1983}
\begin{equation}
\lambda(Z_p) = 2C(|Z_p|)\alpha |Z_p|^{1/3}m_e,
\label{eq:buhring_lambda}
\end{equation}
with $C(|Z_p|)$ a slowly increasing function of $Z_p$ of order unity\footnote{Here we have implicitly used the sudden approximation in using $Z_p$ rather than $Z_p\pm 1$ of the daughter atom since the atomic electrons are considered to be in parental orbitals.}. This potential has been used in several theoretical treatments of screening, with the analysis by B\"uhring \cite{Buhring1984} being of significant interest to us. In contrast to the work by Durand \cite{Durand1964}, here the Dirac equation was used to investigate the effect.

This is made possible because of a distinct advantage in the analysis of electronic screening. The wave functions have to be well known only in the nuclear region rather than the entire space. The birth of the beta particle happens inside the nucleus, so that the decay rate is affected by the depth of the potential well at this site. As electromagnetism is conservative, this means any potential energy lost while travelling through the atom will be regained at infinity. It is exactly this reason why the electronic screening effect is a second order effect compared to the Fermi function. This locality of the interaction allows for a power expansion of the wave function near the origin. In this way an analytical description for the electron radial wave function over the full space can be avoided and one does not have to rely on approximate WKB methods. Such a power expansion was performed by B\"uhring \cite{Buhring1984} using the Hulth\'en potential defined in Eq. (\ref{eq:hulthen}). The screening correction was found to be
\begin{align}
S(Z, W) =~ &X(\tilde{W}/W)|\Gamma(\gamma+i\tilde{y})/\Gamma(\gamma+iy)|^2 \nonumber \\
&\times |\Gamma(\gamma+2i\tilde{p}/\lambda)/\Gamma(1+2ip/\lambda)|^2 \nonumber \\
&\times \exp (-\pi y)(2p/\lambda)^{2(1-\gamma)},
\label{eq:buhring_screening}
\end{align}
where \footnote{Note that $X = 1 + \mathcal{O}(\lambda^2)$ as $\lambda \rightarrow 0$.}
\begin{align}
X &= \left[1+\frac{1}{4}(\lambda/p)^2\right]^{-1}\left\{1+\frac{1}{8}\left[(\tilde{W}+\gamma)/\tilde{W}\right](\lambda/p)^2 \right. \nonumber\\
&+\frac{1}{2}\gamma^2\left[1+(1\mp\alpha Z\lambda/(W+1))^{1/2}\right]^2 \nonumber\\
&\times \left[(W-1)/\tilde{W}\right](\lambda/p)^2 \nonumber \\
&\left.\times\left[1-\frac{1}{8}(1-\gamma)(1/\gamma)(\lambda/p)^2\right]\right\},
\label{eq:buhring_X}
\end{align}
and
\begin{equation}
\begin{array}{ccc}
\tilde{W} = W \mp \frac{1}{2}\alpha Z \lambda, & & y = \pm\frac{\alpha Z W}{p}, \\ \\
\tilde{p} = \frac{1}{2}p + \frac{1}{2}[p^2\mp 2\alpha Z \tilde{W}\lambda]^{1/2}, & & \tilde{y} = \pm\frac{\alpha Z \tilde{W}}{\tilde{p}}
\end{array}.
\label{eq:buhring_substitions}
\end{equation}
Here $\frac{1}{2}\alpha Z \lambda$ is the potential shift due to screening at the origin. This choice of potential has several inherent weaknesses however. Firstly, using only a single parameter severely limits its utility over the entire space, but the main obstacle lies in the evaluation of $\lambda$ from Eq. (\ref{eq:buhring_lambda}). The correct calculation of $C(|Z_p|)$ is problematic, since estimates vary wildly, as discussed for instance by \textcite{Buhring1984} and references therein.

The potential used in this work is that by \textcite{Salvat1987}. The screening potential is numerically calculated using Dirac-Hartree-Fock-Slater techniques for $Z=1-92$, and is then fitted with a sum of three Yukawa potentials
\begin{equation}
V_{S}(r) = \mp\frac{\alpha Z}{r}\sum_{i=1}^3\alpha_i \exp(-\beta_i r).
\label{eq:salvat_potential}
\end{equation}
It includes relativistic effects, and obtains high precision results. Even though it provides a better agreement with the spatial electronic distribution than aforementioned potentials, it cannot fully account for the oscillatory features of individual orbitals. The influence hereof can, however, safely be assumed to be negligible as its dominant contribution occurs within the nucleus.

\begin{figure}[h!]
\centering
\includegraphics[width=0.51\textwidth]{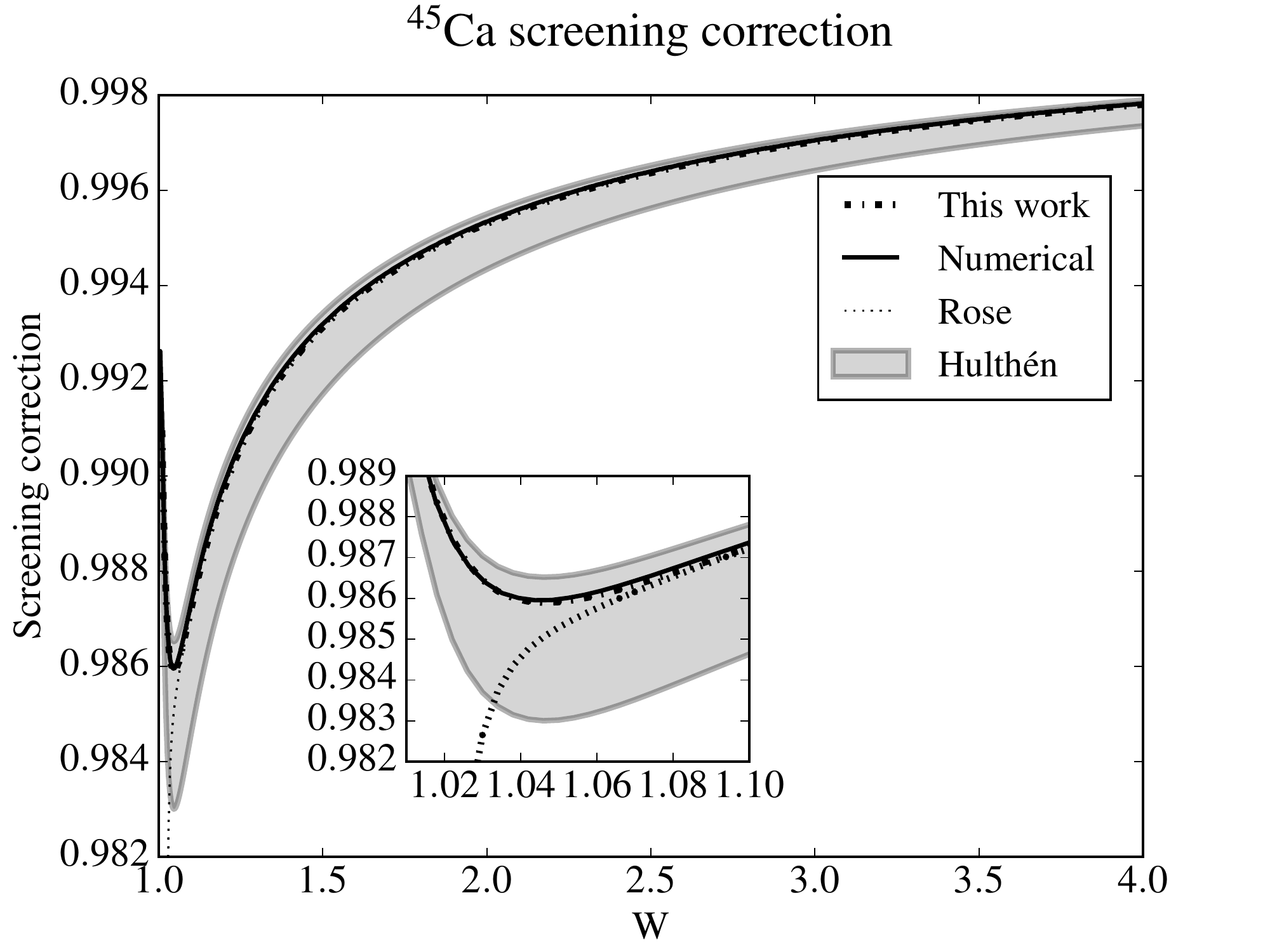}
\caption{Comparison of the screening correction for four different methods: Rose correction, screening evaluated with Eq. (\ref{eq:buhring_lambda}) (Hulth\'en) showing the area of uncertainty due to the uncertainty in $C(|Z|)$, screening evaluated with Eq. (\ref{eq:salvat_screening_shift}) (Salvat), and numerical results using the potential from Eq. (\ref{eq:salvat_potential}).}
\label{fig:compare_screening}
\end{figure}

The potential shift  due to screening for $r \rightarrow 0$ is then
\begin{equation}
\pm\alpha Z \sum_{i=1}^3 \alpha_i\beta_i,
\label{eq:salvat_screening_shift}
\end{equation}
where $\alpha_i, \beta_i$ belong to the parent atom, such that via the substitution $\lambda = 2\sum_i\alpha_i\beta_i$ the entire machinery developed in \textcite{Buhring1984} and Eqs. (\ref{eq:buhring_screening}) and  (\ref{eq:buhring_substitions}) transfers identically. Figure \ref{fig:compare_screening} shows the comparison of precise numerical calculations using Eq. (\ref{eq:salvat_potential}), and analytical descriptions for both screening strengths.
It was originally noted by \textcite{Buhring1984} that setting $X$ (Eq. (\ref{eq:buhring_X})) equal to unity in Eq. (\ref{eq:buhring_screening}) gives better results at low energies for $Z=80$. We have checked this for a series of isotopes after comparison to numerical calculations, and this appears to indeed give a better description. We therefore assume $X=1$ to be the better approximation. This could be understood by considering the limitations of the analytical description relative to the numerical procedure. In the former, the influence of the atomic potential is limited to the value at the origin through Eq. (\ref{eq:salvat_screening_shift}). While this is certainly the dominant effect, a further radial dependence of the screening potential induces additional change of the wave function at the origin. This effect can easily be seen to be small, as $\alpha_i\beta_i^2 \sim \mathcal{O}(\alpha^2)$. The combination of Eqs. (\ref{eq:salvat_potential}) and (\ref{eq:U_analytical}) can be utilized to further show the smallness of the effect. Analogous to this, the effect of screening on the spatial integration over the lepton wave functions contained in the shape factor, $C(Z, W)$, discussed above can be seen to be of similar magnitude. Further comments can be found in Appendix \ref{app:general_shape_factor}.

Recently a different screening correction by \textcite{Mougeot2014} has been put forward, proposing instead
\begin{equation}
S(Z, W) = \frac{f_{sc}}{f_{unsc}}
\label{eq:screening_Xavier}
\end{equation}
where $f$ is defined as\footnote{A typing error is present in the original work \cite{Mougeot2014} in the sign of $\kappa$ for $f_\kappa$, as it does otherwise not agree with the formalism by \textcite{Harston1992} on which it is based, and loses its interpretation as the probability of emerging in a $s$ state when evaluated at the nuclear radius for $\kappa = -1$.}
\begin{equation}
f = \frac{\overline{g_\kappa^2}}{\overline{g_\kappa^2}+\overline{f_{-\kappa}^2}}
\end{equation}
where the bar stands for the average over the entire space. Starting rigorously from the beta decay Hamiltonian and the corresponding $S$ matrix, it is seemingly impossible to arrive at an expression resembling Eq. (\ref{eq:screening_Xavier}). Further, the proposed energy dependence is completely opposite in sign to that of Eqs. (\ref{eq:screening_simple}) and (\ref{eq:buhring_screening}) and of larger magnitude. As applying this correction for superallowed Fermi decays would be in direct violation of the CVC hypothesis and CKM unitarity, a critical comment is necessary. This screening correction was originally proposed as a possible explanation in order to reproduce the measured spectra of 63Ni and 241Pu, the first measurements ever of such high precision at low energy, referring only to the interpretation of $f$ as the probability of an electron emerging in an $s$ state after decay when $\kappa = -1$ . While this is approximately true for $f$ evaluated at the nuclear radius, the same cannot however be said for an average over the entire space as the electron's creation lies within the nucleus. As it is ejected from the nucleus as a  $j=1/2$  particle, both  $\kappa = \pm 1$  states should contribute since the solution of the Dirac equation, Eq. (\ref{eq:sol_dirac}), does not have a defined orbital angular momentum. A screening correction must then also have contributions from both, analogous to the Fermi function, Eq. (\ref{eq:F_general}). It is noteworthy that the exchange correction in \textcite{Mougeot2014} only considered the contribution from the atomic $\kappa = -1$ orbitals. The good agreement with the measured spectra of 63Ni and 241Pu appears then a compensation of the neglect of atomic exchange effects with  $\kappa = 1$ orbitals, as well as further contributions from nuclear structure and atomic corrections discussed below. Therefore, the use of this screening correction is not recommended for high precision studies of beta decays.

\subsection{Atomic exchange effects}
\label{exchange}

The non-orthogonality of initial and final state atomic wave functions in $\beta$ decay allows for additional indirect processes through which electrons can be emitted into a continuum state. In case of the exchange effect, this non-orthogonality leaves a possibility for a $\beta$ particle to be emitted directly into a bound state of the daughter atom, thereby expelling an initially bound electron into the continuum. Experimentally it is impossible to distinguish this indirect process from regular $\beta^-$ decay, so that an additional correction to the experimentally measured spectrum is required \cite{Bahcall1962a, Bahcall1963b, Bahcall1963a, Bahcall1963, Haxton1985}.

\subsubsection{Formalism and procedure}

We present here an analytical parametrisation of this exchange correction fitted to precise numerical data. The formalism laid out here follows that of \textcite{Harston1992}. Whereas in principle all occupied orbitals in the mother atom can contribute, we consider here only allowed $\beta$ decay, so that only contributions from $s$ and $\bar{p} \equiv p_{1/2}$ orbitals play a part. In this case, the exchange effect is written as
\begin{equation}
X(E) = 1+\eta^T_{ex}(E),
\end{equation}
with
\begin{equation}
\eta_{ex}^T(E) = f_s(2T_s+T_s^2)+(1-f_s)(2T_{\bar{p}}+T_{\bar{p}}^2).
\label{eq:exchange_eta_T}
\end{equation}
Here $f_s$ is defined as
\begin{equation}
f_s = \frac{g_{-1}^c(R)^2}{g_{-1}^c(R)^2+f_{1}^c(R)^2}
\label{eq:exchange_fs}
\end{equation}
and
\begin{align}
T_s &= -\sum_{ns' \in \gamma}\langle Es'|ns\rangle \frac{g_{n,-1}^b(R)} {g_{-1}^c(R)} \label{eq:exchange_T_s}\\
T_{\bar{p}} &= - \sum_{n\bar{p}' \in \gamma}\langle E\bar{p}'|n\bar{p}\rangle \frac{f_{n,1}^b(R)} {f_{1}^c(R)}
\label{eq:exchange_T_p}
\end{align}
sum over the different occupied $s$ and $\bar{p}$ shells in the parent state, denoted by $\gamma$. We implicitly used that $\kappa$ is -1 and 1 for $s$ and $\bar{p}$ orbitals, respectively\footnote{Here $\kappa$, as before, is the eigenvalue of the $K=\beta(\bm{\sigma}\bm{L}+1)$ operator. Further $\bm{\sigma}$ is the typical combination of Pauli matrices, $\bm{L}$ is the orbital angular momentum operator, and $1$ is a $4\times 4$ unit matrix.}. The quantities in Eqs. (\ref{eq:exchange_fs})-(\ref{eq:exchange_T_p}) are evaluated at the nuclear radius $R$ using bound (superscript $b$) and continuous (superscript $c$) solutions of the Dirac equation, as written in Eq. (\ref{eq:sol_dirac}). Continuum wave functions are normalized on the energy scale, while bound states are normalized to unity. The factor $\langle Es'|ns\rangle$ stands for the integration of the overlap of bound and continuum state wave functions over the full space. In the non-relativistic hydrogenic approximation the general overlap integral can be written down analytically as \cite{Harston1992}
\begin{align}
\langle Z_f, El | Z_i nl \rangle &= \frac{(Z_i-Z_f)e^{2i(l+1-n)\arctan (p/\bar{Z}_i)}}{(-1)^{n-l-1}(2l+1)!p(p^2+\bar{Z}_i^2)^{l+2}} \nonumber \\
&\times 2^{2l+3}(p\bar{Z}_i)^{l+3/2}|\Gamma(l+1-i(Z_f/p))| \nonumber \\
&\times e^{\pi Z_f/2p}e^{-(2Z_f/p)\arctan (p/\bar{Z}_i)} \nonumber \\
&\times \,_2F_1(l+1+i(Z_f/p), l+1-n, \nonumber \\
& ~~~~~~~~~~ 2l+2, -4ip\bar{Z}_i/(\bar{Z}_i-ip)^2)
\label{eq:hydrogenic_exchange_analytical}
\end{align}
where $Z_i$ ($Z_f$) is the initial (final) effective charge seen by the participating orbitals, $\bar{Z}_i = Z_i/n$ and $_2F_1$ the confluent hypergeometric function. Some examples for $l=0$ are given by \textcite{Harston1992}, and will not be repeated here. A quick glance at Eq. (\ref{eq:hydrogenic_exchange_analytical}) reveals the downward trend for increasing $n$, meaning exchange process is mainly dominated by the $1s$ orbital, as is intuitively expected. While a useful first estimate, both approximations made to achieve Eq. (\ref{eq:hydrogenic_exchange_analytical}) are much too coarse to provide the precision we are after. We will move from a hydrogenic approximation to a proper atomic potential, and use the solution of the Dirac equation in this potential to further include relativistic effects.

As will be shown explicitly, the choice of potential is of crucial importance for a consistent description of the exchange effect. Here it was chosen to use the potential described by Behrens and B\"uhring, in the explicit formulation by \textcite{Mougeot2014}. It is a combination of a spherical intra-nuclear part, a numerical screening potential by \textcite{Salvat1987} and an exchange potential as was introduced historically by \textcite{Slater1951}. The latter is evaluated using the same electron density functions described in \textcite{Salvat1987} and combined with a more elaborate approach to the approximation by \textcite{Latter1955} to ensure good behavior for $r\rightarrow \infty$. The exchange potential, introduced to remove electrostatic self-interaction and ensure antisymmetry of the fermion wave function, is not to be confused with the exchange effect described here. One free parameter in this potential was varied to give agreement with numerically calculated binding energies\footnote{Here we prefer the precise relativistic local density approximation calculations by \textcite{Kotochigova1997, Kotochigova1997a} tabulated up to $Z=92$. From this point onwards we switch to those calculated by \textcite{Desclaux1973}, which appear to be in agreement with the former within 1\% for the innermost orbitals.} within 0.1\% or 0.1 eV. These binding energies can easily be interchanged with experimental data when available. Results for screened and unscreened Coulomb functions tabulated by \textcite{Behrens1969} were cross-checked and showed an overall excellent agreement.

The electron radial wave functions $rg_{\kappa}(r), rf_{\kappa}(r)$ are calculated numerically using a power expansion for both singular points of the Dirac equation at $r=0, \infty$ and connected via rescaling of the inner solution with $\alpha_{\kappa}$ (see Eqs. (\ref{eq:erwf_extensive}) and (\ref{eq:fermi_buhring})). The general approach is outlined in Ref.~\textcite{Behrens1982}, and more specifically in Ref.~\textcite{Salvat1995}. The Fortran 77 package \texttt{RADIAL} \cite{Salvat1995} was slightly modified and interfaced with a custom code to calculate the exchange effect.
As the free and bound state wave functions are typically solved on different grids, a common grid is constructed and evaluated using exact results or via Lagrangian three point interpolation. The integral $\langle Es^\prime | ns \rangle$ in Eqs. (\ref{eq:exchange_T_s}) and (\ref{eq:exchange_T_p}) can then be evaluated directly. Care has to be taken to make the grid sufficiently dense such that interpolation on the common grid does not introduce systematic errors.

\begin{figure}[h!]
\centering
\includegraphics[width=0.51\textwidth]{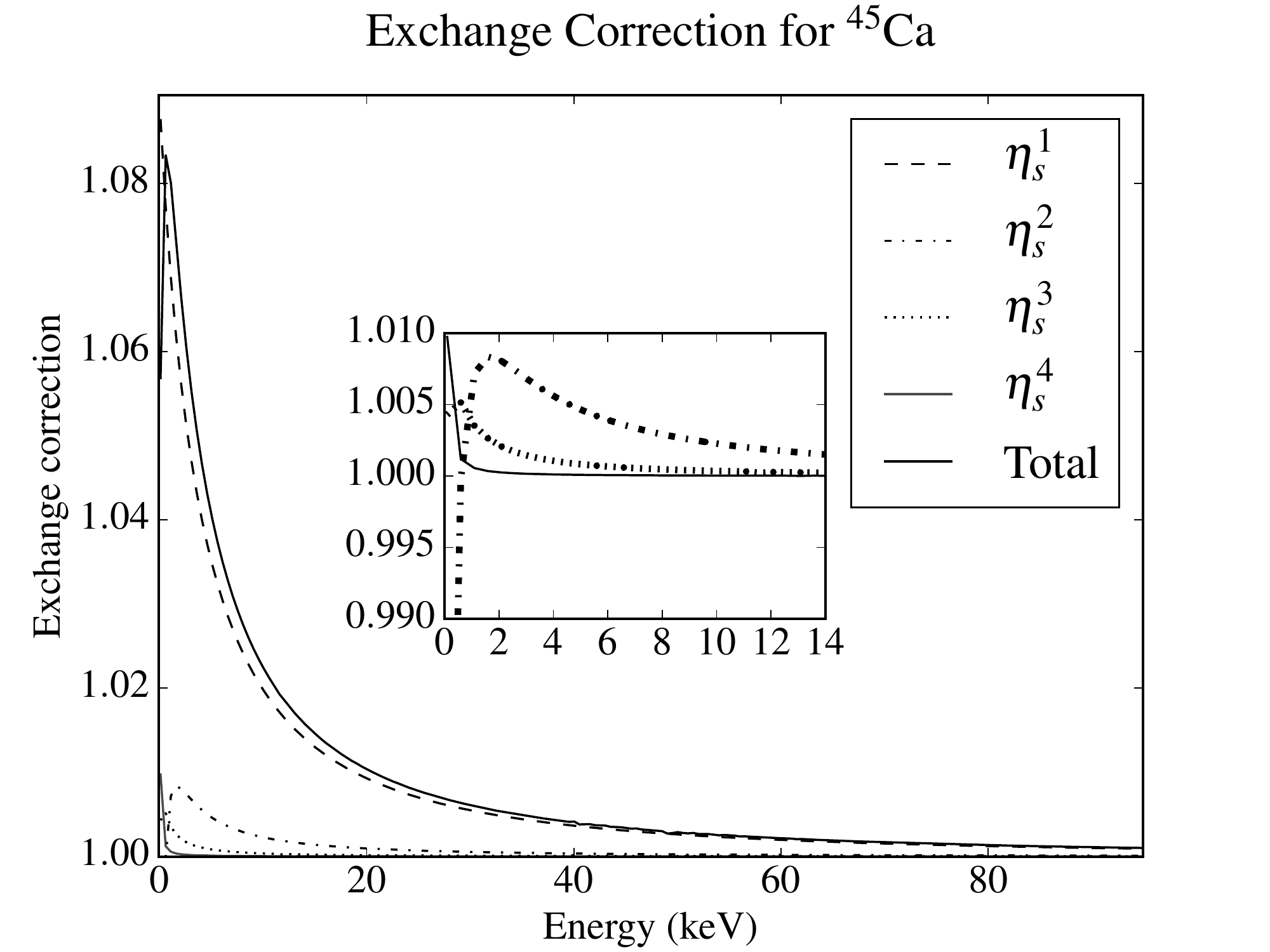}
\caption{Exchange correction to $\beta^-$ decay of $^{45}$Ca showing the explicit contributions from different $s$ orbitals. The total effect rises to nearly 10\% in the few-keV region, and drops significantly in the 100\,eV region due to negative contributions from higher lying orbitals, in this case $2s$.}
\label{fig:exchange_45Ca}
\end{figure}

As an example, the full exchange correction to the $\beta^-$ decay of $^{45}$Ca is presented in Fig. \ref{fig:exchange_45Ca}, explicitly showing the contributions from different orbitals. At very low energies of typically about 1 keV or lower, contributions from higher-lying orbitals can become negative, lowering the total correction by several percentage points, as is seen at the extreme left in Fig. (\ref{fig:exchange_45Ca}). It is clear from the magnitude of the effect at the lowest energies (few keV region) that an accurate description is a necessity for any work performed in the low energy region. For low endpoint energy transitions the phase space integrals can be significantly altered, e.g. by up to 30\% for transitions like $^{241}$Pu. The effect drops off quickly, however, and is typically negligible at energies of several hundreds of keV.

\subsubsection{Contribution from exchange with $p_{1/2}$ orbitals}
\begin{figure}[h!]
\centering
\includegraphics[width=0.48\textwidth]{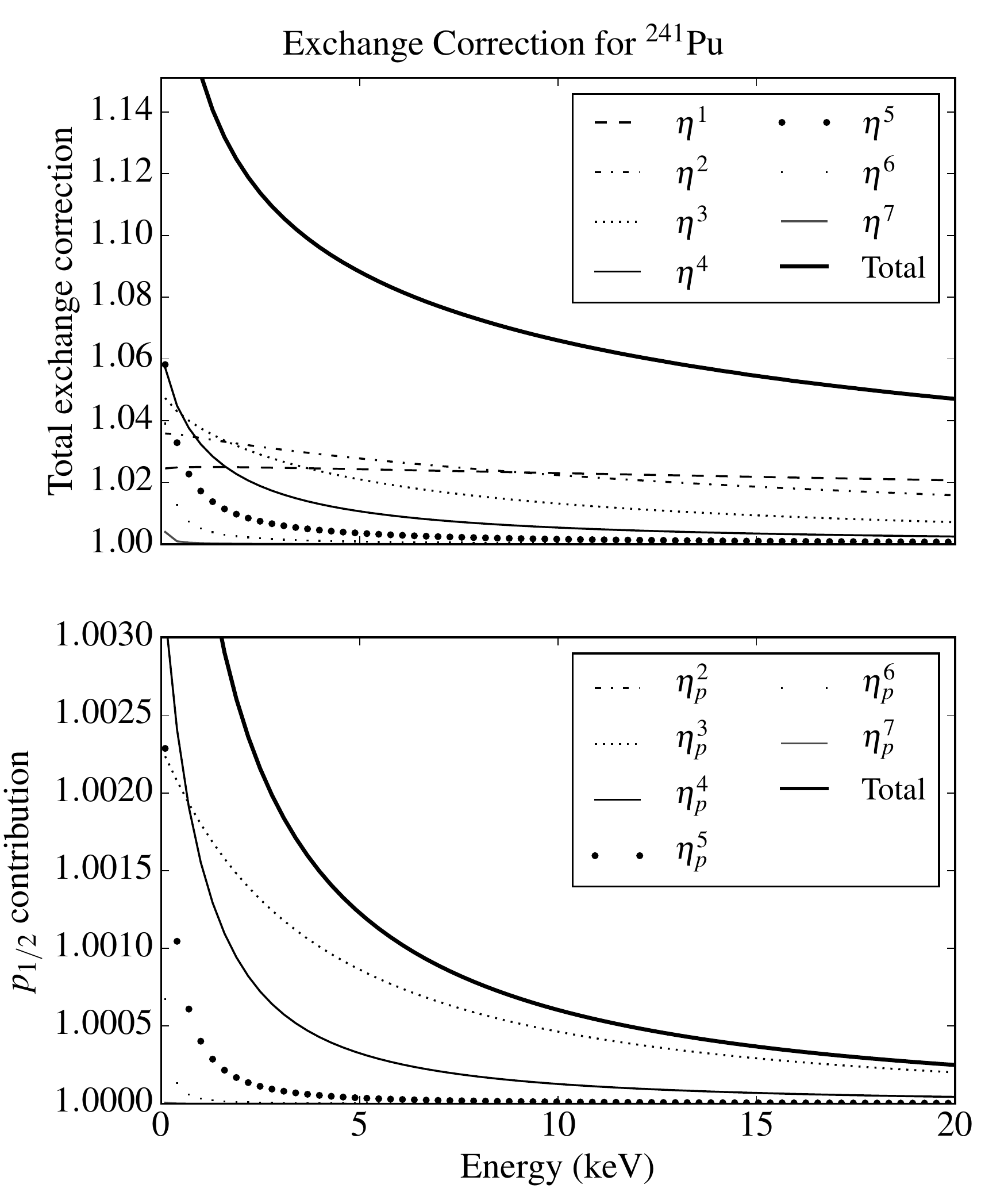}
\caption{Total exchange correction to the $\beta^-$ decay of $^{241}$Pu showing the explicit contributions from the different orbitals (top), together with the specific contribution of the $p_{1/2}$ orbitals (bottom). The former reaches $\sim$\,29\% at 100\,eV, while the latter rises to nearly 2\% at 100\,eV, indicating that it is definitely not to be neglected. The $p_{1/2}$ influence continues to be felt over a range of tens of keV at the precision we are aiming for.}
\label{fig:exchange_241Pu_p}
\end{figure}

As the electron is ejected with $j=1/2$ after allowed $\beta^-$ decay, the exchange effect occurs with all bound $j=1/2$ states. Typically the $p_{1/2}$ exchange contribution is ignored for small to medium masses \cite{Harston1992, Mougeot2014}, however this approximation is not any more viable for the precision we aim for over the full nuclear chart. Examining the behaviour of Eq. (\ref{eq:hydrogenic_exchange_analytical}) for different $l$ reveals the overlap integrals to be of similar magnitude, using $_2F_1(\alpha, \beta, \gamma, z) \approx 1 + (\alpha \beta / \gamma) z$ for $|z| \ll 1$. The $\bar{p}$
contribution is fundamentally suppressed, however, by the factor $1-f_s$ with $f_s$ close to unity, since $T_s$ and $T_{\bar{p}}$ are of similar magnitude. To our knowledge, this contribution has not been explicitly discussed in the literature. As an example we consider the exchange correction to the $\beta^-$ decay of $^{241}$Pu, by all standard a high mass decay. The result is shown Fig. \ref{fig:exchange_241Pu_p}, together with the specific contributions from the occupied $p_{1/2}$ orbitals.

The contribution of the latter is highly relevant in the lowest energy regions, where its contribution can grow up to 2\%, or nearly 10\% of the full exchange effect. It continues to be non-negligible for our purposes throughout the entire decay spectrum. As an example of a light mass nucleus, the maximum contribution of exchange with $p_{1/2}$ for the decay of $^{45}$Ca never exceeds a few parts in $10^5$, leaving it completely negligible. We include the contribution, small though it may be, in all following results.

\subsubsection{Influence of the atomic potential}

\begin{figure}[h!]
\centering
\includegraphics[width=0.48\textwidth]{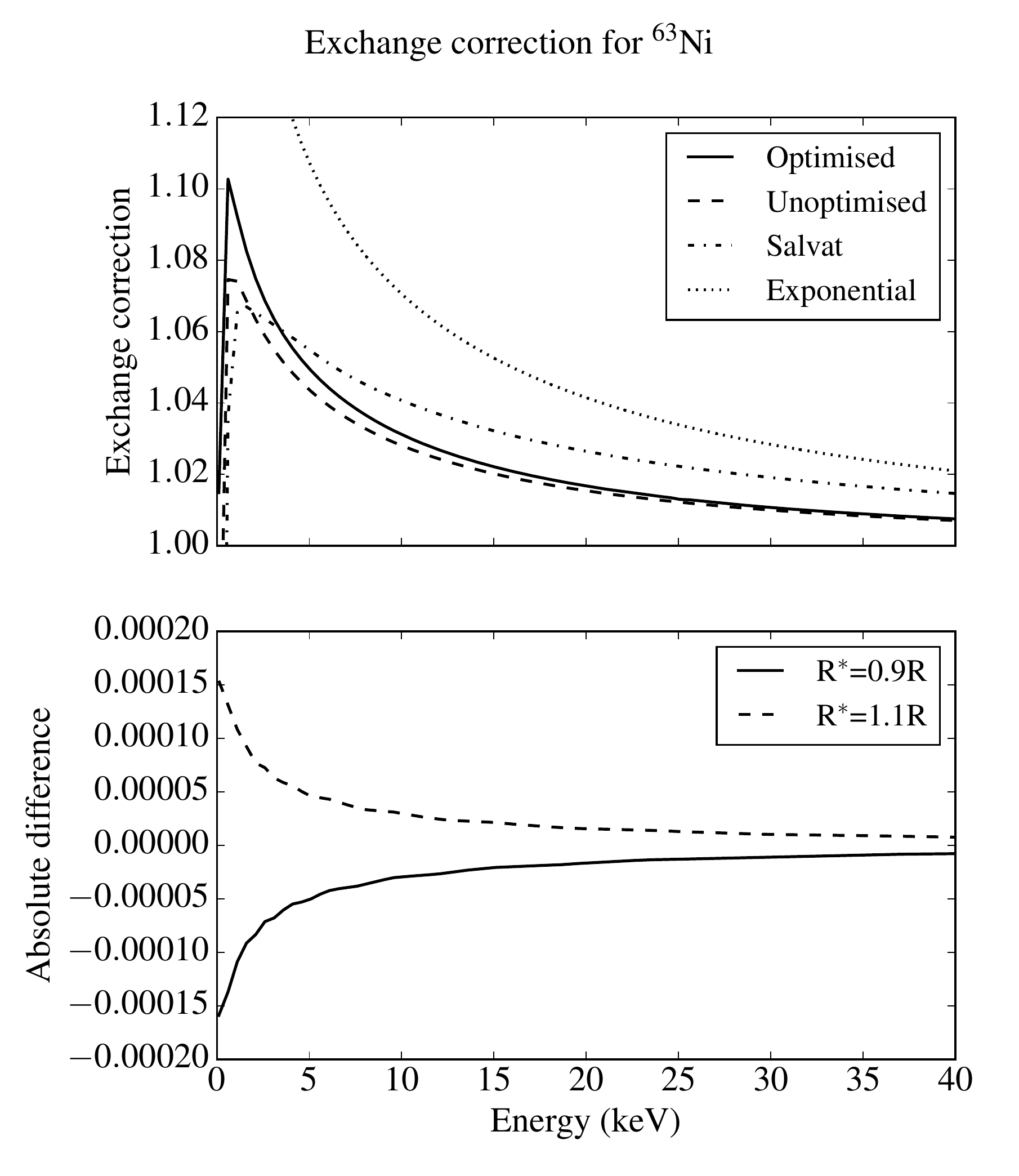}
\caption{Comparison of the exchange effect for $^{63}$Ni shown for different atomic potentials (top), and absolute differences in the total exchange correction when varying the nuclear radius by 10\% with the optimized potential (bottom). Results vary wildly, stressing the need for an accurate atomic potential. Particularly interesting is the large discrepancy stemming from the optimization of the binding energy. The different potentials used are defined in the text.}
\label{fig:exchange_compare_potentials}
\end{figure}

As mentioned above, the choice of potential is a critical component in the description of the exchange correction. Three different potentials were compared, with increasing steps of complexity. Initial theoretical results used simple Coulomb solutions of Schr\"odinger \cite{Harston1992} and Dirac \cite{Mougeot2012} equations and introduced a rough effect of screening by assuming an effective charge for each orbital. The simplest potential utilized here considers a simple exponentially screened field. The screening strength can be adjusted to give best agreement with bound state energies, with decreasing importance for increasing main quantum numbers\footnote{It can frequently occur in this approach that tuning the screening parameter to match the $1s$ binding energies results in unbound higher $ns$ states. As the lowest $s$ states give the highest contribution throughout most of the spectrum, we choose to neglect this detrimental effect in this very rough method.}. A second potential was constructed as a slight extension of the first, consisting of three Yukawa potentials, the coefficients of which were fitted to numerical data from Dirac-Hartree-Fock-Slater calculations \cite{Salvat1987} as in the previous section. Finally, the complete potential described above was used. For the latter the optimization of the exchange term was turned both on (Optimized) and off (Unoptimized). Results are shown in Fig. \ref{fig:exchange_compare_potentials} for the exchange corrections thus obtained for the $\beta$ decay of $^{63}$Ni.
The general trend is replicated by all potentials, apart from the very lowest energies where some orbitals can give large negative contributions. The total net magnitude is, however, a delicate quantity in the lowest energy regions, as even for the most complex potential an optimization of the exchange potential can have significant effects. Without this optimization the binding energies for higher lying orbitals can be seriously in error and consequently give incorrect contributions to the exchange correction.

The choice of the atomic potential reveals another key ingredient when looking at Eqs. (\ref{eq:exchange_T_s}) and (\ref{eq:exchange_T_p}) in the nuclear radius. It is imperative to know how strongly the exchange corrections depend on its precise value, as many charge radii are not experimentally known \cite{Angeli2013}. For this purpose, we take up the case of $^{63}$Ni as in the previous section, and vary the radius by 10\% in both directions. This is far larger than what can be predicted using current methods \cite{Bao2016}, where the uncertainty is estimated to be around 0.03\,fm, or about 0.8\% in the case of $^{63}$Ni. The result is shown in the bottom half of Fig. \ref{fig:exchange_compare_potentials}. The discrepancy grows towards lower energies, and only crosses $10^{-4}$ at roughly 1\,keV. In the extreme case of $^{241}$Pu, spectral differences only cross the $10^{-4}$ level at roughly 5\,keV, but introduce a constant offset that can reach several parts in $10^{-4}$ for the same procedure. The difference shows a reasonably linear behavior on $\Delta R$, such that extrapolation to reasonable uncertainties brings it to below the few $10^{-5}$ effect. 

\subsubsection{Analytical parametrisation}

\begin{figure}[h!]
\centering
\includegraphics[width=0.51\textwidth]{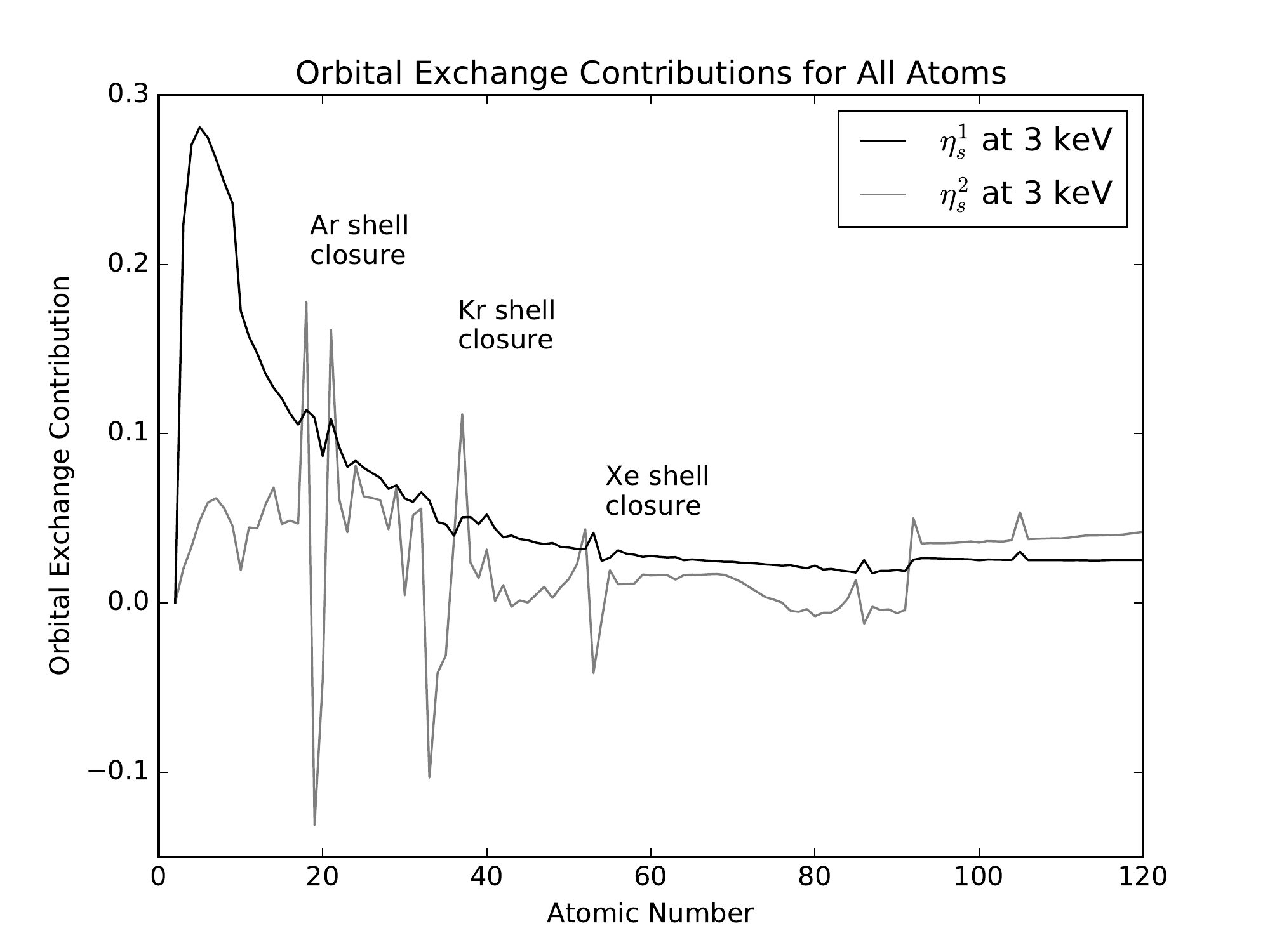}
\caption{Demonstration of atomic shell effects on the exchange contributions from the $1s$ and $2s$ orbitals evaluated at 3\,keV. The increased binding energy at higher $Z$ reduces the spatial extension of the bound states and thus decreases the spatial overlap between bound and continuum electron radial wave functions at low energies.}
\label{fig:shell_effects_exchange}
\end{figure}

The calculations required to arrive at the results shown in Fig. \ref{fig:exchange_45Ca} are involved and in the spirit of this work we would like to have a completely analytical description of the $\beta$ spectrum shape. The immediate issue with the case at hand resides in the integration in Eqs. (\ref{eq:exchange_T_s}) and (\ref{eq:exchange_T_p}). Whereas for most other effects, the radial wave functions have to be known only near the origin, thereby allowing power expansions even for non-trivial potentials, the exchange correction requires knowledge of the radial wave functions of both continuum and bound states for the entire space and for arbitrary potentials. The results in Fig. \ref{fig:exchange_compare_potentials} show that a hydrogenic approach with simple screening potentials is not sufficient when high precision is required, and the analytical results of Eq. (\ref{eq:hydrogenic_exchange_analytical}) for pure Coulomb fields do not allow for easy insertion of fit parameters. This is further hindered by the evaluation of the confluent hypergeometric function, the evaluation of which is in itself a significant hurdle. More importantly, the magnitude of the exchange effect is a complex interplay between atomic shell effects (through, e.g., the binding energy sensitivity; see Fig. \ref{fig:exchange_compare_potentials}), spatial increase of the potential and spatial extension of the bound states for different $n$. An example of this can be seen in Fig. \ref{fig:shell_effects_exchange}. Here the Argon, Krypton and Xenon shell closures have been specified.
This complex $Z$ dependence poses a significant challenge to a sufficiently precise analytical fit valid over the full $Z$ and $W$ range. Other external parameters based on tabulated values would have to be provided as input to the fit function. We therefore choose to tabulate the required fit parameters for each $Z$ individually. The analytical fit for the exchange correction as a function of $W$ contains 9 fit parameters, and is written as
\begin{align}
X(W) &\approx ~ 1 + a/W'+b/W'^2+c\cdot \exp(-dW') \nonumber \\
 &~~~+e\cdot \sin[(W-f)^{g}+h]/W^{i},
\label{eq:exchange_fit}
\end{align}
with $W' \equiv W-1$.
The fit parameters have been tabulated in Appendix \ref{app:exchange_fit_coeff} and show excellent agreement over the full tested energy range. The latter was chosen at 1\,MeV, where all contributions dip below the $10^{-4}$ level. The maximal differences are on the order of a few $10^{-4}$, with average residuals located at the $<10^{-5}$ level. The influence on the phase space integrals agree typically on the order of a few $10^{-5}$ in absolute terms. Special care was taken to avoid steep slopes over small energy ranges in the relative differences to allow for precise measurements of the energy dependence in the beta spectrum shape.

\subsection{Shake-off and shake-up processes: The endpoint shift}
\label{sec:shake-Off}
The change of atomic orbitals due to the aforementioned effects also has a direct consequence on the internal housekeeping of the atom. Initial and final states belonging to different quantum numbers are not any more orthogonal, allowing discrete excitations into higher allowed states (shake-up) and even the continuum (shake-off). This then reduces the available energy for the lepton pair, and changes the final state interaction with the $\beta$ particle. In deriving the results in the previous sections, these effects have typically been ignored on the basis of their relative importance compared to the magnitude of the higher order effect. For the precision aimed at in this work, this is not always valid. The probabilities for shake-up and shake-off depend strongly on $Z$, with the former (latter) becoming subsequently less (more) important for increasing $Z$.

\subsubsection{Shake-up}
\label{sec:shake_up}
Instead of double ionization of the shake-off process, atomic electrons in the final states can simply be excited into higher states. The probability of this process can typically be reduced to integrals of the radial wave functions of initial and final bound states, where the shake-off probability can simply be calculated as the deviation from unity when summing all aforementioned probabilities. Decays of few-electron systems are particularly prone to this effect, as for instance in the tritium system. Here, the single electron in the final state has a significant probability to end up in the $2s$ orbital or even higher ones \cite{Williams1983, Arafune1986, Hargrove1999}. Additionally, the recoil momentum of the daughter nucleus can induce excitations due to the sudden acceleration. One can show \cite{Feagin1979} that the average excitation energy due to the aforementioned effect is equal to
\begin{equation}
\Delta E_R = \frac{1}{2}Zm_ev_R^2 = ZE_R\frac{m_e}{M},
\label{eq:delta_E_recoil}
\end{equation}
with $M$ the nuclear mass and $v_R$ and $E_R$ the recoil velocity and energy, respectively. For $Z=50$ and a maximum recoil energy of 1\,keV, there is an average excitation energy of roughly 0.25\,eV. Considering atomic excitations are on the order of tens of eV, it is clear then that incomplete overlap between initial and final atomic states is the dominant effect in the full shake-up picture.

The $Q$ value of the decay is subsequently reduced by the mean excitation energy
\begin{equation}
\overline{\Delta E}_{\text{ex}} = \sum_{f} P_{f}E_{f},
\end{equation}
where $E_f$ is the excitation energy of the final state $f$ and $P_{f}$ the probability to populate that level. 

\paragraph{Effect on screening and exchange corrections}
Possible excitations in the final state also change the screening correction, where instead of Eq. (\ref{eq:screening_simple}), we should write now in the notation of Saenz and Froelich \cite{Saenz1997a}
\begin{align}
S(Z, W) &= 1 -\left(\frac{W}{p^2} + \frac{1}{W}\right) \alpha \sum_{n} \langle \varphi_0 | \phi_n \rangle^2 \langle \phi_n |\sum_{i=1}^{Z_p} \frac{1}{r_i} | \varphi_0 \rangle \nonumber \\
&= 1 -\left(\frac{W}{p^2} + \frac{1}{W}\right)V_0',
\label{eq:screening_shakeup}
\end{align}
where $|\phi_{n'} \rangle$ represents the $n'$-th final electronic state, and $|\varphi_0 \rangle$ is the initial electronic state. The value for the screening potential can change by as much as 20\% in the special case of tritium $\beta^-$ decay, simply because the atomic Coulomb interaction is limited to a single electron \cite{Hargrove1999}. For systems with a higher number of electrons, the relative change in the electronic distribution decreases. The magnitude of the effect is, however, also proportional to $Z$, such that errors introduced from neglecting shake-up are of the order of $1\cdot 10^{-4}$.

In the case of the exchange correction the situation is slightly more complicated. For medium to high $Z$ nuclei, the probability for shake-up is negligible at the $10^{-3}$ level, as explicitly calculated by, e.g., \textcite{Harston1992}. For lower $Z$ nuclei shake-up probabilities can be significant (e.g., around 25\% in tritium $\beta^-$ decay), and exchange with higher lying $ns$ orbitals should be taken into account even if they were not occupied in the initial state \cite{Harston1993}. In the extreme case of tritium, this inclusion changes the exchange correction at the few $10^{-4}$ level down to 0.5\,keV. Except for this special case then, we can safely neglect shake-up influences on the exchange correction.

\subsubsection{Shake-off}
\label{sec:shake_off}
Just as in the previous section, non-orthogonality of initial and final atomic states allows for excitations in the final state, including the continuum. Whereas in the tritium system the probability for additional ionization is an order of magnitude smaller than that of shake-up, this balance quickly changes when going to higher $Z$. One of the underlying reasons for this is the following: The possibility for shake-up depends on atomic overlap integrals between shells with differing main quantum numbers. Due to the Pauli principle, excitations are possible only to unfilled shells, for which the overlap integrals quickly disappear for inner electrons. Outer shells are close to the continuum, and electronic excitations are typically unbound for higher $Z$.

The description of shake-off has seen significant theoretical \cite{Levinger1953, Schwartz1953, Green1957, Carlson1968, Stephas1967, Stephas1971, Freedman1974, Suzuki1982, Law1982, Frolov2010, Ruiz2013a, Ruiz2013} and experimental \cite{Snell1957, Carlson1963, Carlson1963a, Carlson1963b, Schupp1980, Scielzo2003, Couratin2012, Couratin2013} effort. For practical reasons, this has mainly focused on the calculation of electron ejection from the $K$ shell, which is a small effect in all nuclei \footnote{The probability for $K$-ejection for $^{3}$H decay is on the order of $2 \cdot 10^{-3}$, while for higher $Z$ it is of the order of a few $10^{-4}$.} and not in our current interests. The seminal work by \textcite{Carlson1968}, even though theoretically significant improvements have been made, illustrates the general principles we are concerned with: (i) For a given main quantum number the shake-off probability decreases with increasing $Z$, due to the reasoning mentioned earlier. (ii) For a given atom, the shake-off probability increases with increasing main quantum number. (iii) The total shake-off probability is reasonably independent of $Z$, occurring for approximately 20-30\% on all decays\footnote{With exception to the lightest systems such as $^3$H and $^6$He, where shake-up and non-dissociative resonances are much more important, see the previous section.}.

Neglecting shake-up and collisional ionization \cite{Freedman1974, Carlson1963}, this can be described by the one-electron ionization probability \cite{Couratin2013}
\begin{widetext}
\begin{equation}
p_i \simeq 1 - \sum_{n^{\prime} \leq n_{max}} \left\{ |\langle \phi_{n^{\prime}l_i}^{D}|\phi_{n_il_i}^{M}\rangle |^2 + K^2|\langle \phi_{n^{\prime}l_i\pm 1}^{D}|\mathbf{r}|\phi_{n_il_i}^{M}\rangle |^2 - K^2 \text{Re}\langle \phi_{n^{\prime}l_i}^{D}|\phi_{n_il_i}^{M}\rangle^*\langle \phi_{n^{\prime}l_i}^{D}|r^2|\phi_{n_il_i}^{M}\rangle\right\},
\end{equation}
\end{widetext}
with $\phi^{D(M)}_{nl}$ the daughter (mother) wave function with primary quantum numbers $(n,l)$ and $K=\sqrt{2E_R/M}$ for a nucleus with recoil energy $E_R$ and mass $M$. Rather than calculate the probabilities of excitation into the continuum, it is easier to simply consider the atomic overlap and approximate all deviations from unity as coming from shake-off. Holes created by shake-off are filled through emission of X-rays or Auger electrons. The rate of these two competing processes is regulated by so-called fluorescence yields and is reviewed by, e.g., Hubbell \emph{et al.} \cite{Schonfeld1996, Krause1979, Hubbell1994}.

For simplicity, we first briefly discuss the single ionization process with a $\beta$ particle and an electron in the final state. In the case of $\beta^-$ decay, the two outgoing electrons are indistinguishable, which has to be reflected in the decay amplitude through a coherence term \cite{Law1972, Law1972a}. The decay is properly treated as a single step such that the full energy available has to be shared between all final states. The probability for shake-off is thus dependent on the $\beta$ energy and decreases for increasing energy. This was first emphasized by \textcite{Stephas1967}, even though one typically forgets about the phase-space dependent effects, and instead determines the integrated ejection probability. The mean energy of the ejected electron is similar to its binding energy \cite{Feinberg1941}, such that the largest spectral change occurs in the low energy region (see, e.g., the Kurie plots in \textcite{Law1972}). In case the ejection occurs from one of the outer shells, the typical binding energy is on the order of a few tens of eV, such that to first order the change in spectral shape can be approximated as an additional decrease in $W_0$ by the mean excitation energy weighted with the probability for ejection. From the results of \textcite{Couratin2013} combined with the binding energies of \textcite{Desclaux1973}, we find the mean energy loss due to shake-off per decay around 5 eV for chlorine\footnote{Charge states created due to emission of Auger electrons do not contribute to our correction as they emerge with a fixed energy independent of the $\beta$ decay. The mean energy is then obtained by summing over the holes created for different charge states, using $\overline{E}_{SO} \approx 1.8 B_{nl}$ with $B_{nl}$ the binding energy \cite{Feinberg1941}.}. Due to the arguments listed above, we expect this effect to be of similar magnitude for all $Z$.
 

\paragraph{Effect on screening and exchange corrections}

The screening and exchange corrections presented above (Eqs. (\ref{eq:buhring_screening}) and (\ref{eq:exchange_fit})) depend strongly on the final atomic potential. In the case of shake-off electron ejection, at least one additional hole is created through a single-step process, thereby altering the electronic density at the nucleus as well as the wave functions of all remaining occupied orbitals. For both corrections, the magnitude of the contribution decreases with increasing $n$, while the probability for shake-off is proportional with $n$. We can then reasonably approximate the shake-off influence as coming from events originating from the outer shell as we have done before. Using the screening potential as defined in Eq. (\ref{eq:screening_shakeup}) (see also, e.g., \textcite{Pyper1988}), the change in screening due to a hole in the outer shells is
\begin{align}
\frac{V_{0}-V_{\text{SO}}}{V_0} &= \frac{\langle f | r^{-1} | i \rangle - \sum_{f'} \langle f' | i \rangle^2 \langle f' | r^{-1} |  i\rangle}{\langle f | r^{-1} | i \rangle} \nonumber \\
&\approx \frac{\sum_{f'} \langle f' | i \rangle^2 (1-\Delta Z_{\text{eff}}) \langle nl | r^{-1} | nl \rangle}{\langle f | r^{-1} | i \rangle},
\label{eq:change_screening_shake-Off}
\end{align}
where $| f' \rangle$ is the final state with a hole in an orbital denoted by $n, l$. We have assumed the inner orbitals to remain unchanged in the event of shake-off in the outer shell, and allowed for a difference in effective charge seen by the remaining electron in the outer shell through $\Delta Z_{\text{eff}}$. Using results by Law \emph{et al.}, this change is approximately 0.3-0.4. We use the general result for hydrogenic orbitals $\langle r^{-1} \rangle = Z_{\text{eff}}/n^2$, such that Eq. (\ref{eq:change_screening_shake-Off}) transforms to
\begin{equation}
\frac{V_0-V_{\text{SO}}}{V_0} \approx \frac{1}{\langle f | r^{-1} | i \rangle}\sum_{n}\frac{\langle f' | i \rangle^2Z_{\text{eff}}(1-\Delta Z_{\text{eff}})}{n^2}.
\label{eq:change_screening_shake-Off_final}
\end{equation}
Effective charges seen by orbitals can be calculated from the mean radii as noted in, e.g., \textcite{Harston1992}. For a specific example of Ru$^{1+}$, Eq. (\ref{eq:change_screening_shake-Off_final}) yields approximately 0.1\%, which subsequently pushes the relative change in spectral shape to the few $10^{-5}$ level in the lowest energy range. The approximations used above are crude, but only have to yield results precise within a factor of 2 for us to neglect it.

For exchange the situation is again more complicated, both because of the magnitude of the effect and the sensitivity to the wave function over the entire space. Rigorously, we have to combine the approach taken by \textcite{Law1972, Law1972a} with that of \textcite{Harston1992}, by introducing the indistinguishability of the outgoing electrons from the former approach in the exchange terms described by the latter. Harston and Pyper have described this situation for the tritium system \cite{Harston1993}, but this has been ignored in their former work \cite{Harston1992}. Exchange can only occur because of the combination of the indistinguishability of electrons and the Pauli principle, such that in allowed $\beta^-$ decay only exchange with bound states having angular momentum $j=1/2$ can occur. In the medium to high $Z$ nuclei, the probability for shake-off to occur in an $s_{1/2}$ or $p_{1/2}$ state is smaller than 0.1\% \cite{Harston1992}. The effect of shake-off on exchange then decreases for increasing $Z$. For lower $Z$ the effect of a single shake-off electron of energy $W'$ can be approximated by including the following term in Eq. (\ref{eq:exchange_eta_T}), using the notation of \textcite{Harston1992}
\begin{equation}
\chi_{\text{ex}}^{\text{cont}}(E) = \int_{E' = 0}^{E_{\text{max}}} \phi(E, E') dE',
\label{eq:shakeoff_exchange_integral}
\end{equation}
where
\begin{align}
\phi(E, E') = &\sum_{A}\left(-\langle \gamma_u | \gamma_v' \rangle \langle \gamma_v'(A\rightarrow E's) | \gamma_u \rangle \frac{g^c_{E',s}}{g^c_{E, s}} \right. \nonumber \\
& \left. + \langle \gamma_v'(A \rightarrow E's) | \gamma_u \rangle^2 \left[ \frac{g^c_{E', s}}{g^c_{E, s}} \right]^2\right)\frac{W'}{p'},
\label{eq:shakeoff_exchange_phi}
\end{align}
where the summation over $A$ runs over all $s$ states, neglecting the $p_{1/2}$ state. When making the approximation that shake-off mainly occurs for the outer shell, we can limit the sum over $A$ to the final $ns$ shell. The evaluation of the integral in Eq. (\ref{eq:shakeoff_exchange_integral}) is explained in \textcite{Harston1993}. For the tritium decay, $\chi_{\text{ex}}^{\text{cont}}$ has a maximum value of -0.01\% at 1\,keV. Using again the decay of $^{35}$Ar as an example \cite{Couratin2013} for hole creation in $ns$ states in the final state, we can safely assume this correction can be neglected at the current order of precision.

\subsection{Atomic overlap: alternative atomic excitation correction}
\label{sec:atomic_mismatch}
The $\beta$ decay of a nucleus results in a sudden change of the nuclear potential, both due to a charge difference as well as a recoil effect. Both effects on the spectrum of the emitted $\beta$ particle are described elsewhere in this text. As the eigenstates for initial and final states belong to slightly different Hamiltonians, the initial and final atomic orbital wave functions only partially overlap. This allows for discrete effects such as shake-off and shake-up discussed in the previous section, which decrease the decay rate as the phase space becomes smaller.

The description of the effect can in a first approach be reduced to a difference in atomic binding energies. This analysis was first performed by \textcite{Bahcall1963, Bahcall1963a, Bahcall1965}, and is also recently included in detailed $\mathcal{F}t$ analyses \cite{Hardy2009}. A correction is constructed by looking at the relative change in the spectrum shape when changing $W_0$ with $W_0-\overline{\Delta E}_{\text{ex}}$. In this formalism, the correction can then be written as
\begin{equation}
r(Z, W) = 1-\frac{1}{W_0-W}\frac{\partial^2}{\partial Z^2}B(G),
\label{eq:atomic_mismatch_simple}
\end{equation}
where $B(G)$ is the total atomic binding energy for a neutral atom with $Z\pm 1$ protons in the case of $\beta^{\mp}$ decay. The second derivative of this quantity is related to the average excitation energy due to orbital mismatch via $\overline{\Delta E}_{\text{ex}} = -\frac{1}{2}\frac{\partial^2}{\partial Z^2}B(G)$. It can easily be parametrized as a function of $Z$ using numerical values by \textcite{Desclaux1973}, \textcite{Carlson1970} or \textcite{Kotochigova1997},
\begin{equation}
\frac{\partial^2}{\partial Z^2}B(G) = 44.200 \,Z^{0.41}+2.3196\cdot 10^{-7} Z^{4.45} \text{\,eV}.
\end{equation}
\textcite{Bambynek1977} have reviewed the approach by Bahcall and its effect on electron capture ratio's and found generally good agreement. Discussion and improvements have been presented by \textcite{Vatai1970} and \textcite{Faessler1970} who are mainly concerned with proper evaluation of the overlap integrals in the case of holes in the final state.

For completeness we also discuss a small correction to $r(Z, W)$. In the analysis performed by \textcite{Bahcall1963}, one implicitly uses the \emph{sudden approximation}, i.e. the change in the nuclear charge is instantaneous compared to atomic orbital velocities. For higher $Z$ this is no longer true for low-lying electrons, as $K$ electrons can achieve relativistic speeds. This has been treated by \textcite{Feagin1979}, who introduce an additional quantity to be combined with $\overline{\Delta E}_{\text{ex}}$ discussed above. This is written as
\begin{equation}
\overline{\Delta E}_K = 2(C_0+C_1),
\end{equation}
where $C_0$ and $C_1$ are described in detail by \textcite{Wilkinson1993a}. Its influence on the $Q$ value is on the order of a few eV \cite{Feagin1979}. In the notation of Wilkinson we extend the parametrisation of $K(Z)$ to include the full $Z$ range
\begin{equation}
K(Z) = -0.872+1.270Z^{0.097}+9.062\cdot 10^{-11}Z^{4.5}.
\end{equation}
The correction $r(Z,W)$ then becomes
\begin{equation}
r(Z, W) = 1-\frac{2}{W_0-W}\left(\frac{1}{2}\frac{\partial^2}{\partial Z^2}B(G)+2(C_0+C_1)\right).
\label{eq:atomic_mismatch_full}
\end{equation}
Its influence is felt mainly near the endpoint of the transition, where it cannot be neglected, as $\frac{\partial^2}{\partial Z^2}B(G)$ can become as large as a few hundreds of eV. It is thus particularly important for low energy transitions such as $^{63}$Ni (endpoint energy 67.2\,keV) and $^{241}$Pu (endpoint energy 20.8\,keV). For the former the correction reaches 1\% at 15\,keV before the endpoint and increases rapidly from that point onwards the endpoint

\subsection{Bound state $\beta$ decay}
\label{sec:bound_state}
Finally, we comment on the possibility of two-body bound state $\beta$ decay. As the electron is created inside an electronic potential well, there exists a possibility for the $\beta$ decay to be captured inside the potential well and produce an electron in a bound atomic state, effectively reducing the decay to a two-body problem. This was first studied by \textcite{Daudel1947}, and was later expanded upon by \textcite{Bahcall1961, Kabir1967} and finally \textcite{Budick1983} and \textcite{Pyper1988}. As this is a relevant issue in the analysis of $\mathcal{F}t$ values, we will briefly comment on it here. In the notation by Bahcall the ratio of probabilities can be written as
\begin{equation}
\frac{\Gamma_b}{\Gamma_c} = \frac{\pi(\alpha Z)^3}{f(Z, W_0)}(W_0-1)^2\Sigma,
\label{eq:bound_state_ratio}
\end{equation}
where $\Sigma$ can be found in \textcite{Bahcall1961} and depends on the binding energy of the bound orbital, atomic overlap integrals and the orbital wave functions evaluated at the nuclear surface. For free neutron decay, this ratio is approximately $4.2\cdot 10^{-6}$, and is completely negligible. For tritium decay, however, this ratio becomes roughly 1\% for $T^+$ and 0.5\% for T in the initial state. Higher $W_0$ values lead to smaller ratios, as the phase space integral $f$ is roughly proportional with $W_0$ to the fifth power. The kinematic dependence for $\Gamma_b/\Gamma_c$ then approximately follows a $W_0^{-3}$ behavior. Nuclei studied in $0^+\rightarrow 0^+$ superallowed decays have $Q$ values of several MeV, rendering the bound state decay probability completely negligible. For low energy decays and (partially) ionised initial states, this correction can grow significantly, however. This does not affect the $\beta$ spectrum shape, as it is a separate final state in the calculation of the $S$ matrix. It enters the equation when considering the $\mathcal{F}t$ analysis, however, and so cannot always be neglected.

\subsection{Chemical influences}
\label{sec:chemical_effects}

In many experiments, the decaying atom is bound within a molecule. Therefore, the electronic structure is modified as electrons rearrange themselves in \emph{molecular} orbitals. The presence of additional electrons and spectator nuclei influences the Coulombic final state interactions, while rotational and vibrational states open up more possibilities for energy transfer to the molecular final state \cite{Cantwell1956}. These effects can be considered even higher-order corrections, but prove essential in, for example, the determination of the antineutrino mass in the tritium system. After comments by \textcite{Bergkvist1971}, an extensive amount of literature was produced on the atomic and molecular effects on the endpoint energy of tritium \cite{Law1981, Kaplan1982, Williams1983, Budick1983, Strobel1984, Kolos1985, Jeziorski1985, Fackler1985, Lindhard1986, Arafune1986, Szalewicz1987, Durand1987, Kolos1988, Lopez1988, Claxton1992, Harston1993, Froelich1993, Froelich1996, Glushkov2009}. Driven by experimental discrepancies \cite{Backe1993, Belesev1995, Weinheimer1999, Lobashev1999}, this culminated in the seminal works by Saenz and Froelich \cite{Saenz1997a, Saenz1997b, Jonsell1999, Saenz2000, Doss2006, Doss2007}, describing in an ab initio and analytical way the influence of additional molecular electrons and spectator nuclei. There the effects of all electrons and nuclei within the molecule were treated on equal footing. The main results will be summarized here.

\subsubsection{Recoil corrections}
In case of $\beta$ decay inside a molecule, the recoiling daughter nucleus moves inside the molecular potential rather than a vacuum. This potential is typically described with a Born-Oppenheimer energy curve, which can be modelled using a Lennard-Jones type potential. When placed inside, the recoiling daughter nucleus kicks the molecule in a (predissociative) rovibrational state. Dissociation can occur for rovibrational states with a total energy larger than the dissociation energy, or through electronic excitation into a resonant continuum state. The typical energy scale of the former is of the order of a few eV \cite{Jeziorski1985, Jonsell1999} while for the latter is on the order of tens to a few hundred of eV \cite{Kaplan1982, Claxton1992, Saenz1997b, Jonsell1999, Saenz2000, Doss2006}. The integrated probability of the molecular continuum depends on the endpoint energy of the transition, as a higher endpoint energy implies a higher recoil energy, thereby typically ending up in the dissociative regime \cite{Cantwell1956}. In case of a final bound state with some angular momentum $J$, the energy of the $\beta$ particle is reduced due to the rovibrational excitation and the center-of-mass movement of the entire molecule rather than merely that of the recoiling atom. For isotropic emission of the $\beta$ particle relative to the molecular orientation, the majority of the energy deposited will be placed into rovibrational motion rather than center-of-mass of movement. We assume then that the mass used in the recoil corrections does not significantly change. At the precision aimed for in this work we neglect this effect, as it forms a small correction on already small corrections (see Sec. \ref{sec:R_N} for $R_N$ and Sec. \ref{sec:Q} for $Q$).

In case the recoiling daughter atom is to be detected (e.g., in the measurement of the beta-neutrino correlation $a_{\beta\nu}$), the situation is not so simple. As the recoil energy is inversely proportional to its mass, the probability for dissociation decreases with increasing $Z$. There is also a clear dependence on the $\beta$-$\nu$ angular correlation, as higher recoil momenta will tend to populate higher rotational and vibrational bands and vice-versa. This intuitive picture is also confirmed for diatomic molecules as shown, e.g., by \textcite{Cantwell1956}. Its evolution in the molecular potential has the effect of partially randomizing the outgoing angle, as well as an energy transfer to its molecular partners. This effect can at least partially be included by simulating the response of a recoiling nucleus in a Lennard-Jones potential using Monte Carlo techniques, as was done by, e.g., \textcite{Vorobel2003, Vetter2008}. In the most precise analysis, this effect introduces a systematic error of 0.05\%, pointing to the need for a more detailed understanding of molecular and dynamic effects in the future.

\subsubsection{Influence on the $Q$ value}
Whereas previously, the $Q$ value of the decay was decreased because of shake-up and shake-off within the atom, inclusion of molecular effects adds a further decrease because of rotational and vibrational excitations. Numerical results have mainly been published for the di-tritium molecule $T_2$, due to the high precision required in the determination of the antineutrino mass. The molecular influence then has the effect of changing the excitation possibilities to a broad continuum with resonances, and provide further broadening through population of rovibrational states. As mentioned before, the width of the populated rovibrational energy spectrum lies in the few eV region, and can typically be neglected. Excitation into the continuum now becomes non-trivial, however, as was explicitly demonstrated in the case of T$_2$, e.g., by \textcite{Doss2006}. Averaged over the entire continuum spectrum, differences between several tritium-substituted molecules is on the order of a few eV, including atomic tritium \cite{Kaplan1982}. The spectral shape depends on $W_0^2$, such that the relative error goes like $2\times \sigma_Q/Q$. For the study of $\mathcal{F}t$ values, this dependence is even heightened as the statistical rate function depends on $Q$ through a fifth power. The required precision on $Q$ should then be studied case by case to determine the required accuracy. In the lowest energy transitions, such as those of tritium, $^{63}$Ni or $^{241}$Pu, a substantial error on the $Q$ value on the few $10^{-4}$ to $10^{-3}$ level can be crudely expected for $\sigma_Q$ on the order of a few tens of eV. The change in the decay rate was treated approximately by \textcite{Pyper1988}, who find
\begin{equation}
\frac{\Delta \lambda}{\lambda} \approx \frac{3 \Delta W_0}{W_0}\left(1+\frac{1}{6}\gamma W_0\right)
\end{equation}
where $\Delta W_0$ is the difference in mean endpoint energy after averaging over all final states between two chemical states.

\subsubsection{Molecular screening}
In the approach by \textcite{Saenz1997a}, all Coulombic effects are treated equally to first order. This effect is typically split into an electronic and nuclear part, where the former can be written down as in Eq. (\ref{eq:def_V0_simple}) with the sum extending over all molecular electrons, and the latter as
\begin{align}
P_{\text{nuc}}^{(0,1)}(p) = &\sum_n \alpha |\langle \varphi_0 | \phi_n \rangle|^2 \left[Z\left(\frac{8W}{\pi p} +\frac{8 p}{3 \pi W}\right) \right. \nonumber \\ 
&\left. +\sum_S Z_S\left(\frac{W}{p^2}+\frac{1}{W}\right)\langle \xi_{000}^0 | \frac{1}{R_S} | \xi_{000}^0 \rangle \right],
\label{eq:coulomb_saenz}
\end{align}
where the sum $n$ extends over all final states, the sum $S$ takes into account all spectator nuclei in the molecule, $|\xi_{000}^0 \rangle$ is the rovibronic ground state of the molecule, and $R_S$ is the distance operator between the decaying atom and the spectator nucleus $S$. The first part in this equation represents the Fermi function to first order, and is larger by a factor $p$ than the effects of the spectator nuclei simply from the prefactor. We could stop here and let the influences of molecular electrons and spectator nuclei be calculated numerically using standard quantum chemical calculations \cite{Yamanouchi2001}. 

In the spirit of this work, however, we intend to arrive at an analytical approximation to the required order of precision. In order to obtain an estimate for the change in screening due to molecular effects, we ignore the energy difference in final atomic states and use closure to perform the sum over $n$ (which is to first order corrected for by using the atomic mismatch correction, see Sec. \ref{sec:atomic_mismatch}). For each atom in the molecule, we can consider it to have an inert atomic inner structure for its electronic configuration, coupled with a shared wave function describing the valence electrons participating in chemical bonds\footnote{We are able to make this approximation because of three restrictions on the participating states, the first two of which depend on the properties of the overlap integral them \cite{Atkins1984}: ($i$) they must have the same rotational symmetry around the internucleus axis; ($ii$) they cannot be too diffuse nor too compact; ($iii$) the energies must be similar.}. The full electronic part of the molecular wave function with $i$ atoms can be written as
\begin{equation}
\psi_{\text{Mol}}^{\text{el}} = | \text{valence} \rangle \times \prod_i | \text{inert} \rangle_i,
\label{eq:molecular_electronic_breakdown}
\end{equation}
where we have implicitly used the Born-Oppenheimer approximation, and $i=0$ corresponds to the decaying nucleus. For the sake of notation, we write $| \text{inert} \rangle \equiv | \text{i} \rangle$ and $| \text{valence} \rangle \equiv | \text{v} \rangle$. The latter can be described in a typical quantum chemical treatment using molecular orbitals \cite{Roothaan1951, Lichten1967, Levine2000, Bransden1983}. Using the clamped-nuclei approximation\footnote{This entails treating the nuclei in a molecular system as infinitely massive compared to the surrounding electrons. Effectively this also means a decoupling of nuclear and electronic motion as in the Born-Oppenheimer approxaimtion. See, e.g., \textcite{Bransden1983}.} for the rovibrational ground state we write $\langle \xi_{000}^0| \frac{1}{R_S} | \xi_{000}^0 \rangle \approx \frac{1}{R_{S, e}}$ with $R_{S, e}$ the equilibrium distance between decaying and spectator nucleus $S$, such that we can write the total Coulombic influence as
\begin{widetext}
\begin{equation}
P_{\text{tot}}^{(0,1)}(p) \approx \alpha \left\{ Z\left(\frac{8W}{\pi p} + \frac{8p}{3 \pi W} \right) +  \left(\frac{W}{p^2} + \frac{1}{W} \right) \left[\sum_S (Z_S-Z_{S,\text{in}})\frac{1}{R_{S,e}} -\,_0\langle \text{i} | \sum_{Z_{\text{in}}}\frac{1}{r} | \text{i} \rangle_0 - \langle \text{v} | \sum_{Z_{\text{val}}}\frac{1}{r} | \text{v} \rangle \right] \right\},
\end{equation}
\end{widetext}
where we have used that $_{i>0}\langle \text{i} | \frac{1}{r} | \text{i} \rangle_{i>0} \approx \frac{1}{R_{S, e}}$. We have now explicitly introduced a screened spectator nucleus with charge $Z_S-Z_{S, \text{in}}$ which can be easily evaluated when $R_{S, e}$ can be estimated to sufficient accuracy. The influence of molecular charge distributions can then be written as
\begin{align}
\Delta S_{\text{Mol}} &= \alpha\left(\frac{W}{p^2} + \frac{1}{W} \right) \left[\sum_S (Z_S - Z_{S, \text{in}})\frac{1}{R_{S,e}} \right. \nonumber \\
&\left. - Z_\text{eff}\langle r^{-1} \rangle_{\text{Val}}\right],
\label{eq:delta_S_mol}
\end{align}
where $Z_\text{eff} = Z_{\text{Val}}-(Z-Z_{\text{in}})$ and $\langle r^{-1} \rangle_{\text{Val}}$ corresponds to the average inverse distance of all valence electrons relative to the decaying nucleus. Then $\Delta S_{\text{Mol}}$ should be added with the regular electronic screening correction of Eq. (\ref{eq:buhring_screening}).

As an example, consider the decay of $^{45}$Ca bound inside CaCl$_2$, with calcium doubly oxidized, Ca(II). The bond length for CaCl was measured to be 2.437 \AA, which we can use as a rough estimate, while $Z_S = 17$ and $Z_{S, \text{in}} = 16$ as only one electron participates in the bonding. In natural units we have then $R_{Cl,e} \sim 600$, such that the first term in Eq. (\ref{eq:delta_S_mol}) is roughly proportional to $3 \cdot 10^{-5}$, i.e. two orders of magnitude smaller than $V_0$. In the evaluation of the second term we have to keep in mind that it constitutes both the valence electrons and the absence of the valence electrons of the decaying atom. We have then $Z_\text{eff} = 2$, while $\langle r^{-1} \rangle_{\text{Val}}$ will be of the same order of magnitude as the bond length since the valence electrons will be located mainly near the Cl atoms. The effect from the second term will thus also be on the order of $2 \cdot 10^{-5}$ with opposite sign to that of the spectator nuclei. The deviation from molecular structure on the electronic screening potential is then of the order of $2\cdot 10^{-5}$. As the screening correction is typically on the (sub)percent level at its maximum, molecular deviations are expected to have an upper limit at the few $10^{-4}$ level.

\subsubsection{Molecular exchange effect}
The atomic exchange process represents the possibility for a direct $\beta$ decay into a bound orbital of the final electronic state. In the case of a molecule, the electronic phase space is greatly enlarged and perturbed relative to the single atomic electronic state. In particular, electrons can now reside in molecular, rather than atomic orbitals, while binding energies for other orbitals can change. As shown explicitly in Eqs. (\ref{eq:exchange_T_s}) and (\ref{eq:exchange_T_p}), the probability for the exchange process into a certain orbital depends on the spatial overlap between continuum and bound state. The continuum state tends to oscillate rapidly for distances large compared to its Compton wavelength. Because of this oscillatory nature, however, we are much more sensitive to the shape of the wave function as compared to the previous section, where we were only interested in the $\langle r^{-1} \rangle$ matrix element. This in turn allowed us to approximate the electronic distribution as in Eq. (\ref{eq:molecular_electronic_breakdown}). At the level of precision we are aiming at, combined with the magnitude of the exchange effect at low energies, this is no longer sufficient. Molecular wave functions for all but the trivial H$_2^+$ system have to be calculated numerically, however, which leaves us with limited possibilities in an analytical description. We will nonetheless introduce the approximation of Eq. (\ref{eq:molecular_electronic_breakdown}), and qualitatively describe the behavior of the valence electrons. An important feature to aid us in this matter is the symmetry group of the molecule.

Molecular orbitals can in a qualitatively enlightening way be constructed from a linear combination of atomic orbitals (LCAO)\footnote{For a quantitative description one has to rely on Hartree-Fock and Kohn-Sham calculations with large basis sets, see, e.g., Refs. \textcite{Jansik2009, Ide2014}. This is beyond the scope of this text, and we must content ourselves with a qualitative outset.} \cite{Bransden1983}. In general we can write all electron wave functions as a LCAO, and minimize the coefficients to provide the lowest energy in a Hartree-Fock scheme. For the sake of the argument, however, we introduce the additional approximation that the internal orbitals for all atoms are the same in the molecular as in the atomic case. The molecular valence orbitals are then a combination of occupied atomic valence orbitals and energetically close atomic excited states\footnote{This is a reasonable approximation, except for when there is degeneracy in the different atomic orbitals. When the same atom is present several times in the molecule and is related through symmetry, different LCAO change the spatial behavior of the wave functions while the energy stays the same.}. We assume then the contributions of the inner orbitals to the exchange effect are unchanged, while that of the final occupied orbital will be perturbed. As the electron density near the decaying atom can both increase and decrease in a molecular bond, the overlap integral in Eqs. (\ref{eq:exchange_T_s}) and (\ref{eq:exchange_T_p}) will do the same. Here we reach the limit of the analytical description, and the molecular bonding possibilities are too extensive to provide parametrisations for the effect. When the valence electron is nearly fully removed in an ionic bond, the exchange effect will be approximately zero, as the internuclear distances are much larger than the $\beta$ Compton wavelength where the wave function rapidly oscillates. In the reverse extreme situation, the valence electron density is doubled. This then naively doubles the exchange correction. Conservatively, then, we treat the contribution of the last orbital with a 100\% error bar. This way, in all but the most extreme ionic bonds, we additionally absorb small errors coming from small binding energy changes in the inner atomic orbitals. Taking Fig. \ref{fig:exchange_45Ca} as an example, this introduces a $< 1 \cdot 10^{-4}$ error from 15\,keV onwards, and grows to a 0.5\% error at 1\,keV. For $^{241}$Pu, the valence orbital participating in exchange is 7$s$, whose contribution drops to below $10^{-4}$ after 3\,keV while its maximum quickly grows to 1\% in the first 0.5\,keV. An estimate for the effect can be obtained from the analytical results first obtained by \textcite{Harston1992} using screened hydrogenic orbitals. The effects of the change of the wave function in a lattice versus that of a gas has been studied by \textcite{Kolos1988} in the case of tritium, and results showed no significant change. Due to our conservative error bar, this effect is completely absorbed and can be neglected.

\section{Overview and crosscheck}
\label{sec:bss_overview}

\begin{table*}
\caption{Overview of the features present in the $\beta$ spectrum shape (Eq. (\ref{full-expression})), and the effects incorporated into the Beta Spectrum Generator Code \cite{HayenTBP}. Here the magnitudes are listed as the maximal typical deviation for medium $Z$ nuclei with a few MeV endpoint energy. Some of these corrections fall off very quickly (e.g., the exchange correction, $X$) but can be sizeable in a small energy region. Varying $Z$ or $W_0$ can obviously allow for some migration within categories for several correction terms.}
\begin{ruledtabular}
{\renewcommand{\arraystretch}{1.2}
\begin{tabular}{c l l r}
Item & Effect & Formula & Magnitude \\
\hline
1 & Phase space factor &  $pW(W_0-W)^2$ & \multirow{2}{*}{Unity or larger} \\
2 & Traditional Fermi function &  $F_0$ (Eq. (\ref{eq:F_0})) &\\
\hline
3 & Finite size of the nucleus &  $L_0$ (Eq. (\ref{L0})) & \multirow{5}{*}{$10^{-1}$-$10^{-2}$} \\
4 & Radiative corrections &  $R$ (Eq. (\ref{eq:radiative})) & \\
5 & Shape factor & $C$ (Eq. (\ref{V-CZW}) and (\ref{A-CZW})) & \\
6 & Atomic exchange &  $X$ (Eq. (\ref{eq:exchange_fit})) & \\
7 & Atomic mismatch & $r$ (Eq. (\ref{eq:atomic_mismatch_full})) & \\
\hline
8 & Atomic screening &  $S$ (Eq. (\ref{eq:buhring_screening})) $^{\text{a}}$ & \multirow{10}{*}{$10^{-3}$-$10^{-4}$}\\
9 & Shake-up & See item 7 \& Eq. (\ref{eq:screening_shakeup}) $^{\text{b}}$ & \\
10 & Shake-off & See item 7 \& Eq. (\ref{eq:change_screening_shake-Off_final}) \& $\chi_{\text{ex}}^{\text{cont}}$ (Eq. (\ref{eq:shakeoff_exchange_integral})) $^{\text{c}}$ & \\
11 & Isovector correction & $C_I$ (Eq. (\ref{eq:C_I_wilkinson})) & \\
12 & Distorted Coulomb potential due to recoil & $Q$ (Eq. (\ref{eq:recoil_coulomb})) & \\
13 & Diffuse nuclear surface & $U$ (Eqs. (\ref{eq:U_analytical}) and (\ref{eq:nuclear_diffuse})) & \\
14 & Nuclear deformation & $D_\text{FS}$ (Eq. (\ref{eq:D_FS})) \& $D_C$ (Eq. (\ref{eq:D_C0})) & \\
15 & Recoiling nucleus &  $R_N$ (Eq. (\ref{R_N})) & \\
16 & Molecular screening & $\Delta S_{\text{Mol}}$ (Eq. (\ref{eq:delta_S_mol})) \\
17 & Molecular exchange & Case by case \\
\hline
18 & Bound state $\beta$ decay & $\Gamma_b/\Gamma_c$ (Eq. (\ref{eq:bound_state_ratio})) $^{\text{d}}$ & \multirow{3}{*}{Smaller than $1\cdot 10^{-4}$} \\
19 & Neutrino mass & Negligible & \\
20 & Forbidden decays & Not incorporated & \\
\end{tabular}
}
\end{ruledtabular}
\label{table:overview_bsgc}
\begin{flushleft}
\footnotesize{$^{\text{a}}$ Here the Salvat potential of Eq. (\ref{eq:salvat_potential}) is used with $X$ (Eq. (\ref{eq:buhring_X})) set to unity.} \\
\footnotesize{$^{\text{b}}$ The effect of shake-up on screening was discussed in Sec. \ref{sec:shake_up} with Eq. (\ref{eq:screening_shakeup}).} \\
\footnotesize{$^{\text{c}}$ Shake-off influences on screening and exchange corrections were discussed separately in Sec. \ref{sec:shake_off}. This has to be evaluated in a case by case scenario.} \\
\footnotesize{$^{\text{d}}$ This does not affect the spectral shape, as discussed in Sec. \ref{sec:bound_state}, but does enter the $\mathcal{F}t$ analysis.}
\end{flushleft}

\end{table*}
Table \ref{table:overview_bsgc} shows all effects included in our description of the $\beta$ spectrum shape with corresponding references to the equations used. It remains then to be considered how well our description stacks up against others results from the literature. There exists limited information on precise shape factor evaluation, other than what has been performed by, e.g., \textcite{Behrens1978}, which is, however limited to the shape factor evaluation in an approximate manner. 

\begin{figure}[h!]
\includegraphics[width=0.48\textwidth]{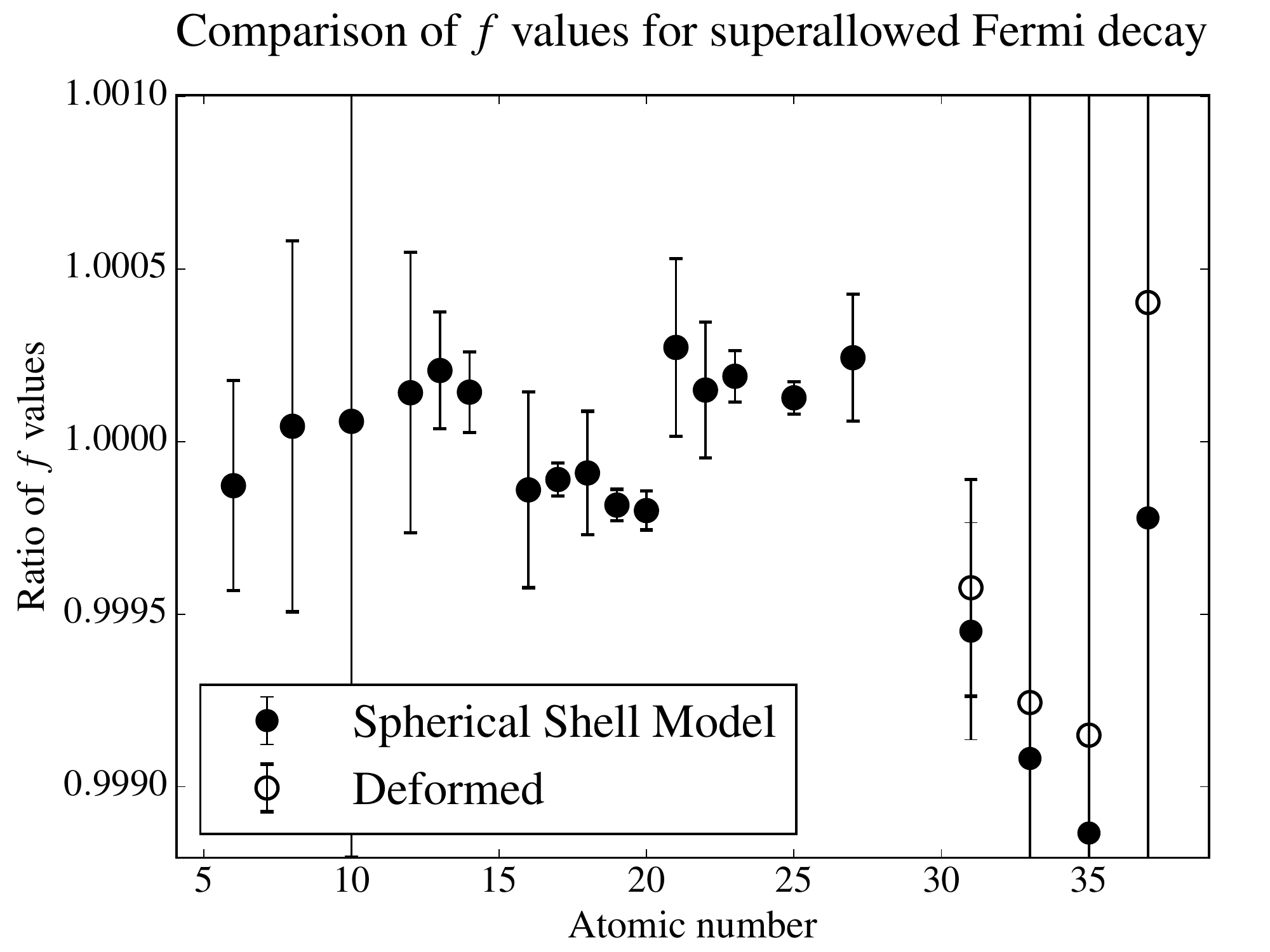}
\caption{Ratio of $f$ values for all superallowed $0^+ \to 0^+$ Fermi decays up to mass 74 included in the analysis by \textcite{Hardy2015}. Uncertainties mainly results from the uncertainties on $Q$ values to illustrate the importance of possible deviations on the $V_{ud}$ analysis. For the heaviest nuclei we show results for both the spherical shell and a deformed shell filling. These last four nuclei ($^{62}$Ga, $^{66}$As, $^{70}$Br, and $^{74}$Rb) all have their valence nucleons outside the $N=Z=28$ shell closure. These nuclei are, however extremely exotic, and show strong deformations and shape coexistence. In the Nilsson model a complex interplay between the $2p_{1/2}$, $1f_{5/2}$ and $1g_{9/2}$ orbitals arises. A reversal between the first two would not influence our isovector correction as $w$ is equal. The same is not true in case the $1g_{9/2}$ becomes filled, as is the case for $^{70}$Br.}
\label{fig:comp_fV_Towner}
\end{figure}

This is an interesting moment to compare what we have done so far to precise $\mathcal{F}t$ values for superallowed Fermi transitions calculated by Towner and Hardy \cite{Hardy2015, Towner2015}. A ratio of $f$ values calculated within our framework relative to the results of \textcite{Hardy2015} is shown in Fig. \ref{fig:comp_fV_Towner}. 

In general we have excellent results, and all residuals are in the few $10^{-4}$ region. This is particularly interesting, as for the heaviest nuclei in question we have moved far away from stability, with $^{74}$Rb being 11 neutrons away from its closest stable isotope and subject to strong deformations and shape coexistence. The daughter nuclei of the last four nuclei investigated, $^{62}$Ga, $^{66}$As, $^{70}$Br and $^{74}$Rb, have deformations with $\beta_2$ equal to 0.195, 0.208, -0.307 and 0.401, respectively\footnote{Clearly this is a very interesting and challenging region, as experimental evidence is often not in agreement with theoretical predictions. Specifically in the case of $^{70}$Se, the daughter nucleus of our largest deviation, the sign of $\beta_2$ is contested \cite{Hurst2007}. The total influence of deformation on the $ft$ value is, however symmetric relative to $\beta_2$ (see Fig. 2 by \textcite{Wilkinson1994}), such that we are not as sensitive to the sign. Either way, the last proton will end up in the $g_{9/2}$ orbital in the Nilsson model.} \cite{Moeller2015}. We conclude that, as we are in the aimed-for range for even these extremely exotic nuclei, and we can trust the validity of our approach for pure Fermi decays in a completely analytical manner without the need for additional computation.

\begin{figure}[h!]
\includegraphics[width=0.48\textwidth]{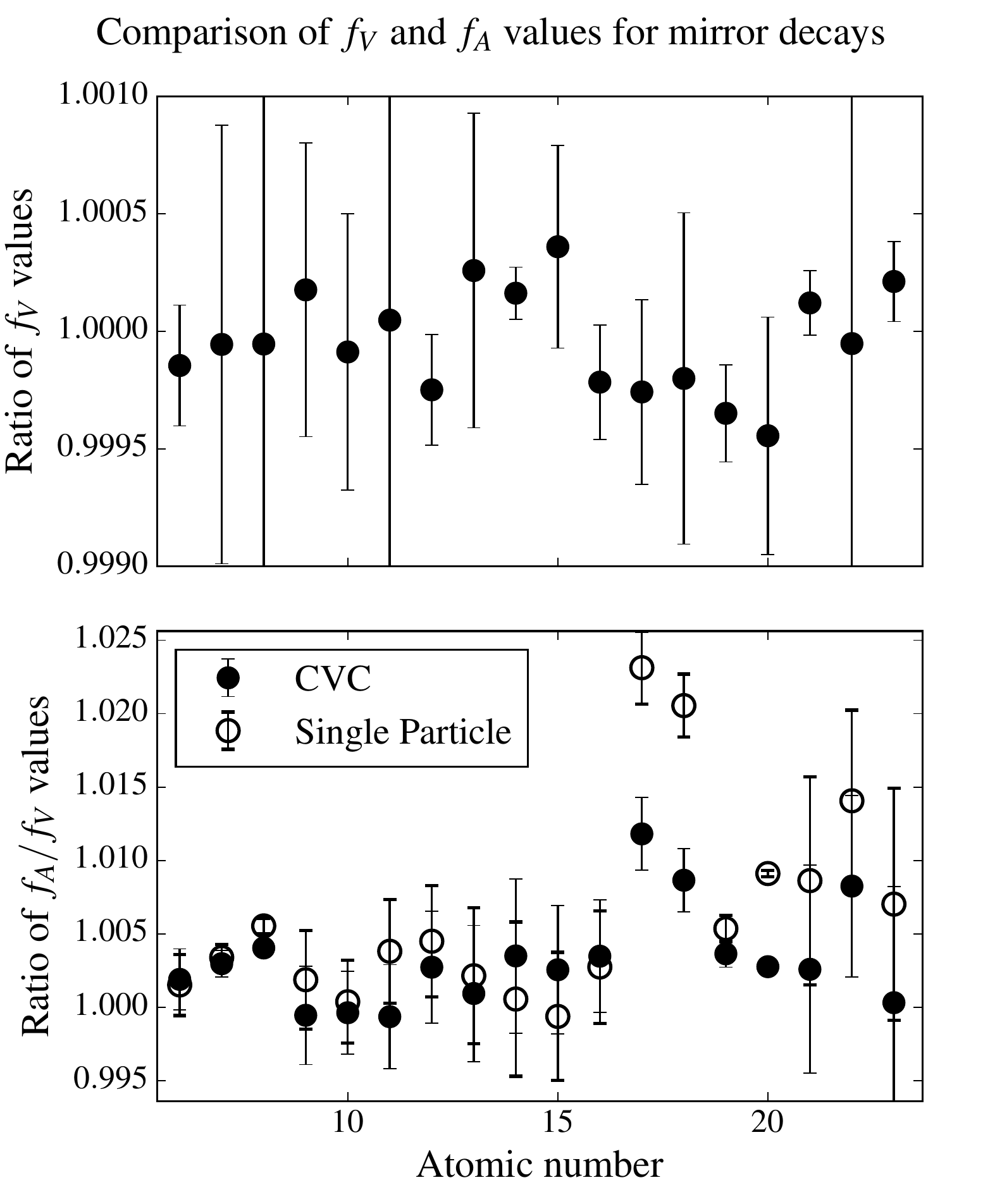}
\caption{Comparison of pure $f_V$ (top) and $f_A/f_V$ (bottom) values for mirror decays as calculated by \textcite{Towner2015} and by using our formalism described here. We see an excellent agreement for the $f_V$ values, with differences being smaller than $4\cdot 10^{-4}$, as we expected from the results in Fig. \ref{fig:comp_fV_Towner}. For the calculation of $f_A/f_V$, the weak magnetism contributions were calculated using CVC in one case and the single-particle estimates discussed in Sec. \ref{sec:single_particle}. Overall a good agreement is found except for special cases where the spherical harmonic oscillator evaluation breaks down, i.e. for $Z=17$ and $Z=18$. The importance of an accurate representation of weak magnetism is underlined by these $^{33}$Cl and $^{35}$Ar isotopes, for which the shell model is also unable to correctly calculate $b/Ac_1$.}
\label{fig:comp_fA_fV_Towner}
\end{figure}

An additional measure that also tests the Gamow-Teller parts can be found in the compilation of mirror decays. Its importance cannot be overstated \cite{Severijns2008}, as experimental precision is ever-improving. Here the final $\mathcal{F}t$ value depends on the mixing ratio $\rho$ and the ratio of vector and axial vector $f$ values
\begin{equation}
\mathcal{F}t^{\text{mirror}} \equiv \frac{2\mathcal{F}t^{0^+ \to 0^+}}{1+\frac{f_A}{f_V}\rho^2}.
\end{equation}
The separate $f$ values have been calculated by \textcite{Towner2015} using the same methods as those used in Fig. \ref{fig:comp_fV_Towner}. We have now two possibilities to compare our results. As these are all mirror nuclei, the Holstein form factor $d$ vanishes identically and CVC allows us to precisely calculate the weak magnetism term from existing experimental data \cite{SeverijnsTBP}. On the other hand, we are able to approximate the latter using single-particle matrix elements as we have discussed in Sec. \ref{sec:single_particle}. Both results are shown in Fig. \ref{fig:comp_fA_fV_Towner}.

\begin{figure}[h!]
\includegraphics[width=0.48\textwidth]{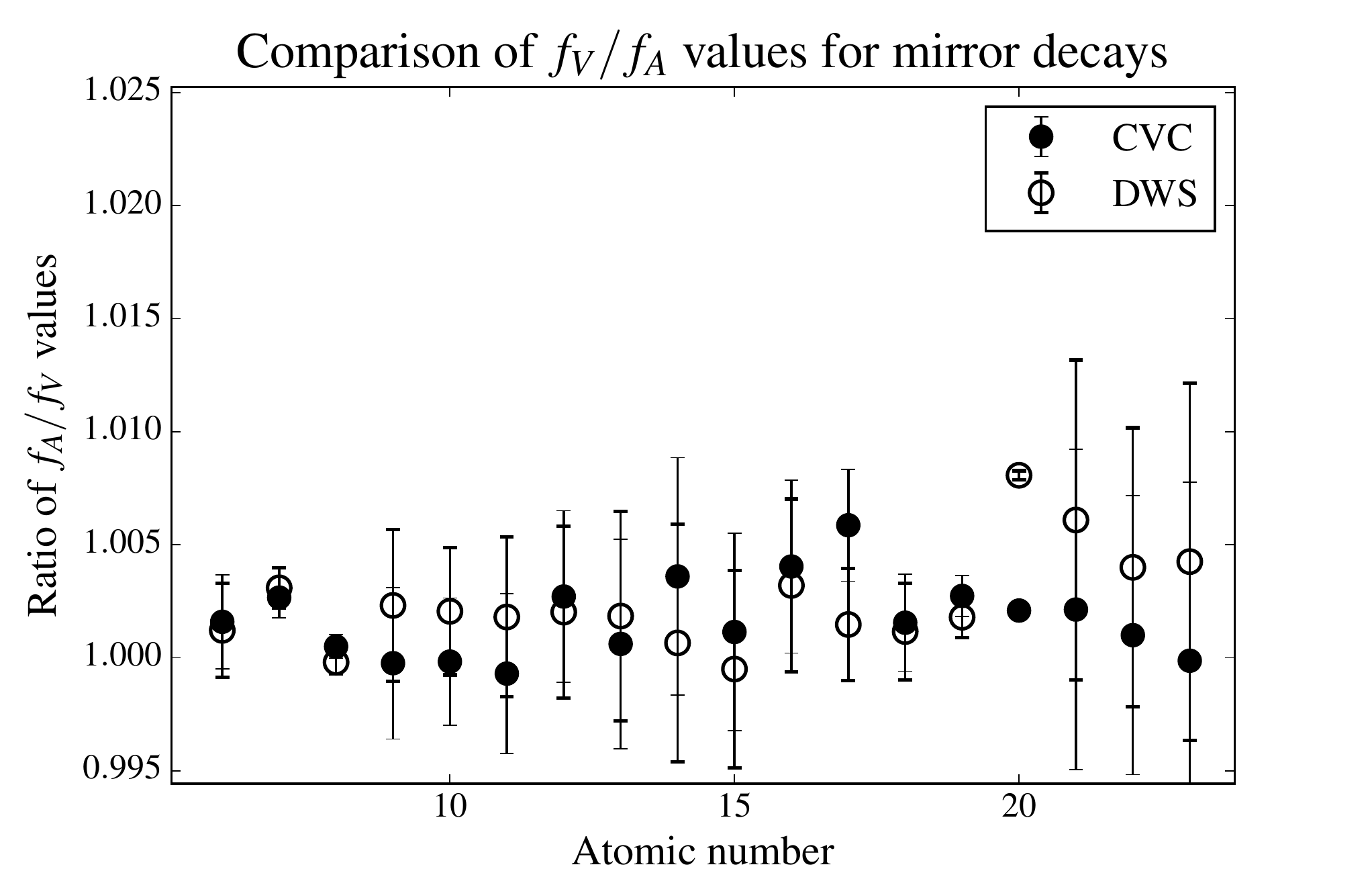}
\caption{Comparison of $f_A/f_V$ values as calculated by \textcite{Towner2015} and those by using our formalism described here, with the addition of deformed Woods-Saxon (DWS) result as described in the text. We have retained the same vertical scale of Fig. \ref{fig:comp_fA_fV_Towner} to show the vast improvement over the spherical harmonic oscillator results. The calculation of the nuclear matrix elements, while still evaluated in an extreme single particle fashion, are now not any more analytically available. The single particle wave function is expanded in a spherical harmonic oscillator basis for which the coefficients have to be calculated by numerical diagonalization of the deformed Hamiltonian. This is performed automatically by the C++ code accompanying this work \cite{HayenTBP}. As shell model calculations reach only a slightly better precision in estimating $b/Ac_1$ (around 10\%) \cite{SeverijnsTBP}, the observed deviations from unity when using CVC results can be attributed to this fact.}
\label{fig:comp_fA_fV_Towner_deformed}
\end{figure}

As expected from the results for superallowed decays, the agreement in the vector sector is exquisite, with all differences smaller than $4 \cdot 10^{-4}$. For the axial vector part the general agreement is good, but there are some distinct features. We see that for cases where the extreme single-particle approach is justified, the deviation is in the $10^{-4}$ range both for CVC and single-particle results. In the cases where this is absolutely not the case, such as the outliers at $^{33}$Cl and $^{35}$Ar, the disagreement reaches $1\%$. A similar failure is then expected in the evaluation of the ratio of matrix elements in $\Lambda$, defined in Eq. (\ref{eq:Lambda_CA}). Important to note, however, is that the shell model has issues pinning down the right values in these cases as well \cite{SeverijnsTBP}. The accuracy with which the shell model is able to explain these values is found to be on the order of 10\% \cite{SeverijnsTBP}, meaning deviations away from unity in our comparison do not necessarily stem from a failure in our evaluation of $\Lambda$, and are at least in part due to the dominant weak magnetism correction. Judging from the comparison between the CVC and single-particle results, one would conclude there to be a complete failure of the extreme single-particle model to accurately predict the weak magnetism contribution. A surprising yet pleasant result is found, however, when one moves from a spherical harmonic oscillator to a deformed Woods-Saxon potential but retains the extreme single particle approximation \cite{SeverijnsTBP}. Excellent agreement is found throughout the entire region for which experimental results are available, and even the outliers can be nicely reproduced when introducing strong oblate deformations based on mean-field results \cite{Moeller2015}. Using this approach to calculate all relevant matrix elements, we show the new results for $f_A/f_V$ in Fig. \ref{fig:comp_fA_fV_Towner_deformed}. While not any more analytically available, it shows a remarkable agreement with much more advanced shell model calculations while using the \emph{extreme single particle} approximation.

Remaining uncertainties can safely be attributed to differences in the exact value of weak magnetism, for which the shell model is shown to perform only marginally better than our deformed single particle approach \cite{SeverijnsTBP}. Combined with the excellent agreement in the vector sector through a comparison of $f_V$, these results show both the potential of the extreme single particle method and the consistency of the developed formalism described here.

On the other hand, this raises important questions on the accuracy of the $f_A/f_V$ calculations when experimental results enter this domain. Seeing as to how differences in weak magnetism predictions can shift these values by several parts in $10^3$, this undermines the claims of reaching 0.01\% in theoretical calculations \cite{Towner2015}. The currently most precisely measured mirror $\mathcal{F}t$ isotope, $^{19}$Ne, is seen to behave rather well under the single-particle approximation and is not so much affected by this concern. Experimental campaigns are under way, however, for precision measurements of the $\beta$-$\nu$ correlation of $^{32}$Ar, $^{19}$Ne $^{35}$Ar \cite{Severijns2017, Lienard2015, Couratin2013} and the $\beta$-asymmetry parameter of $^{35}$Ar \cite{Severijns2017} and $^{37}$K \cite{Fenker2016}. For the extraction of $V_{ud}$ to be valid, a significant amount of attention needs to be given to a precise evaluation of the $\Lambda$ factor in Eq. (\ref{eq:Lambda_CA}). When this can be done reliably, the formalism developed here can be combined with the experimentally determined weak magnetism contribution. Following this, we comment once more on the significance of the induced pseudoscalar component in Eq. (\ref{eq:Lambda_CA}). Its contribution, assuming the free nucleon value $g_P = -229$, is comparable to that of the ratio of matrix elements, meaning strong deviations are expected to occur. It is possible that shifts can occur even on the per mille level. When ignoring this contribution, a reasonable error must be attributed to all $f_A/f_V$ calculations until a way is found to accurately account for this effect.

The results shown when moving to a deformed Woods-Saxon potential were obtained by the custom code that calculates the full $\beta$ and (anti)neutrino spectrum shape. It allows for a calculation of all nuclear relevant matrix $^{V/A}\mathcal{M}_{KLs}$ in an extreme single particle fashion, with several options for customization in order to correctly specify the single particle state. It allows for a coupling with more advanced shell model and mean field codes by expressing the transition matrix elements in a single particle basis. It is properly discussed in a separate publication, and will be publicly available \cite{HayenTBP}. Results concerning weak magnetism including contributions from this custom code are discussed elsewhere \cite{SeverijnsTBP}.

\section{Beta-spectrum shape sensitivity to weak magnetism and Fierz terms}
\label{sensitivity}

In the previous section we have shown our independent correspondence with the currently best numerically calculated $f$ values in the work by \textcite{Towner2015}. We can then reliably use the expressions found in this document, and use it as the basis in looking for new physics results at the per mille level. As mentioned in the introduction, investigation of the Fierz term has the important advantage of looking for deviations linear in new, exotic coupling constants. The $\beta$ spectrum shape is an ideal observable for this as the sensitivity to the Fierz term varies as $1/E$ while the effect of weak magnetism on the $\beta$ spectrum shape varies proportional to $E$.

\subsection{Spectral sensitivity}
\label{sec:spectral_sensitivity}

The weak magnetism and a possible non-zero Fierz interference term both modify the shape of the $\beta$ spectrum in an energy-dependent fashion. Historically, this was investigated experimentally by defining a so-called shape factor after which one defines the slope of the resultant ratio (see, e.g., the work by \textcite{Calaprice1976}). Experimentally an unnormalized shape factor is obtained by dividing the number of counts observed for each $\beta$-particle energy by the quantity $K(Z, W, W_0, M)pW (W-W_0)^2$ (see Eq. (\ref{full-expression})) which is then normalized to unity at some energy $W_\text{norm}$ by dividing these unnormalized values by the value at $W_\text{norm}$ to yield the normalized shape factor\footnote{To remain consistent with our notation, we have written the shape factor as $C$ instead of the spectral function $h_1$ as per Holstein and coauthors.} \cite{Calaprice1976}
\begin{equation}
\label{SE}
S(W) = C(Z, W) ~ / ~ C(Z, W_\text{norm}) ~ .
\end{equation}

\noindent The slope $dS/dW$ then provides the physics information. A similar ratio can be constructed for the Fierz term. A typical value for $b/Ac_1$ is about 5 (see, e.g., the work by \textcite{Wauters2010}). For $b_\text{Fierz}$ a precision of 1\% or better is typically required in view of current constraints on $C_S^{(\prime)}$ or $C_T^{(\prime)}$ coupling constants \cite{Severijns2006, Naviliat-Cuncic2013, Vos2015}. Moreover, as was shown recently, at a precision of about 0.1\% measurements of $b_\text{Fierz}$ in nuclear $\beta$ decay and neutron decay remain competitive with direct searches for new bosons related to scalar and tensor type weak interactions at the LHC collider in the channel $p + p \rightarrow e + MET + X$ (with MET standing for the missing transverse energy) which has an underlying dynamics similar to $\beta$ decay at the parton level \cite{Naviliat-Cuncic2013, Khachatryan2015}.
\begin{figure}[h!]
\centering
\includegraphics[width=0.51\textwidth]{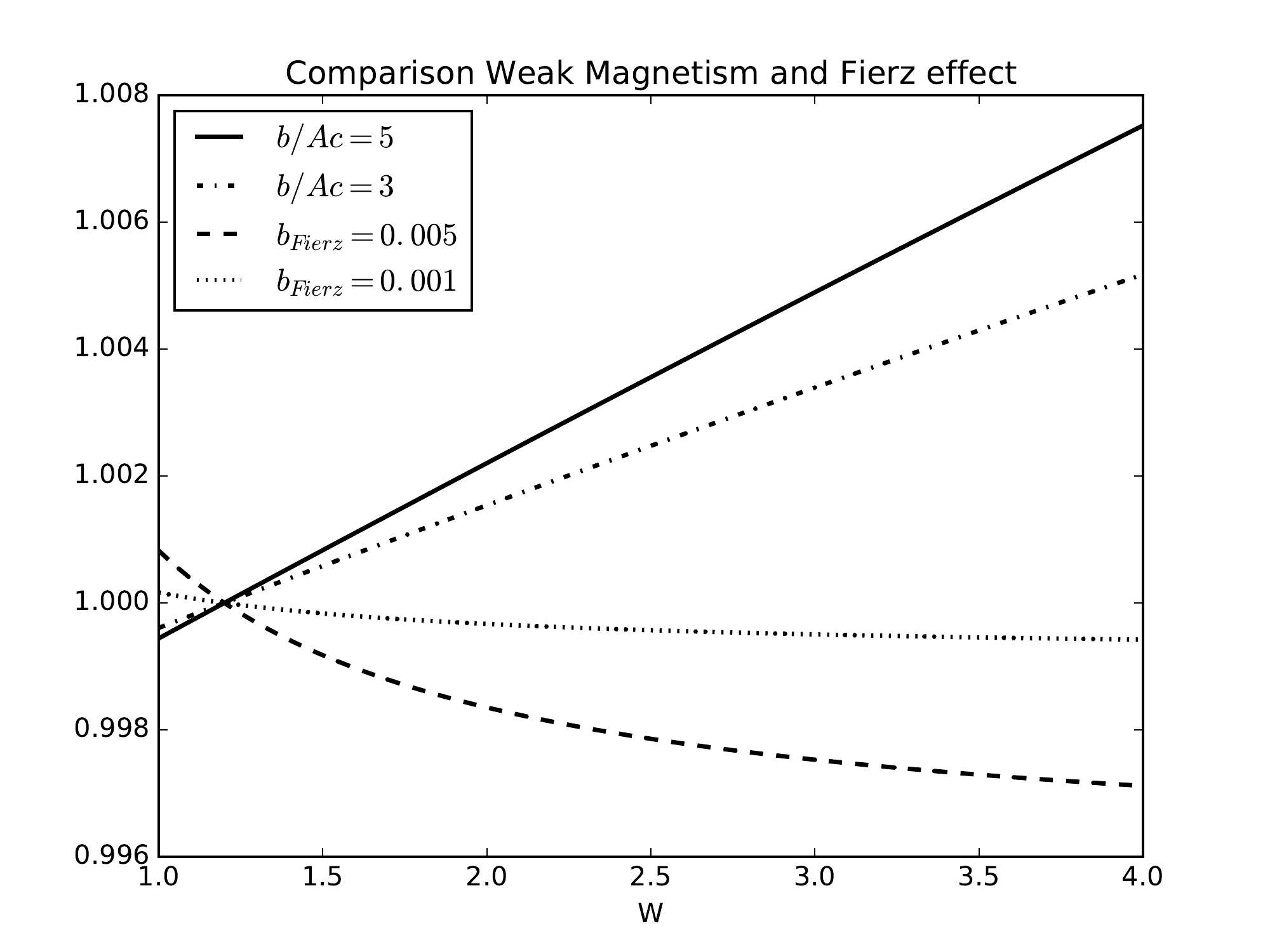}
\caption{Example of the effects of a weak magnetism and Fierz term on the $\beta$ spectrum shape. Both effects are normalized at 100\,keV, i.e. a ratio is constructed as in Eq. (\ref{SE}). This clearly shows the advantage of measuring low energy transitions for Fierz, while high energy endpoint transitions are favourable for weak magnetism studies.}
\label{fig:compare_fierz_wm}
\end{figure}
To generate some perspective on the magnitude of this slope when introducing a Fierz or weak magnetism term, we show in Fig. \ref{fig:compare_fierz_wm} an example shape factor, $S(W)$, for $b/Ac_1 = 5$ and $b/Ac_1 = 3$, and $b_{\text{Fierz}} = 0.005$ and $b_{\text{Fierz}} = 0.001$.
This clearly stresses the need for an accurate value of $b/Ac_1$ when extracting Fierz information. Further, this requires an excellent theoretical description of the beta spectrum shape, as has been presented here. The slope for $b_{\text{Fierz}} = 0.005$ over the first 250\,keV interval is equal to $-0.33\%$\,MeV$^{-1}$. The energy-dependent information of the theoretical spectrum must thus be accurate enough to guarantee all remaining slope artifacts to be $< 0.1\%\,$MeV$^{-1}$. Special attention has been given when describing atomic corrections to avoid any residual energy-dependent slopes.

At the lowest energies, care has to be taken when interpreting experimental results, as the Fierz term spectral modification is approximately linear in this regime. Experiments are then sensitive to $\sim (b_{\text{wm}}-b_{\text{Fierz}})E$, and there is no way to decouple the separate contributions from their energy-dependent behavior. This again underlines the required precision and accuracy in evaluating the weak magnetism contribution when searching for new physics (see also \textcite{Gonzalez-Alonso2016}).

\subsection{Effective field theory approaches}
\label{sec:eft}
The traditional search for exotic interactions in $\beta$ decay is described by introducing additional coupling constants and currents in the $\beta$ decay Hamiltonian. This makes direct comparison with results from high energy colliders like LHC difficult. It is here that effective field theories can bridge the gap. Assuming the mass scale of new physics, $\Lambda_{BSM}$, to be beyond what is probed right now at LHC, both high and low energy experiments can be directly compared in an effective field theory. Here additional coupling constants are introduced that are sensitive to exotic currents in higher order operators. A significant amount of effort has been put into its development, showing that both low and high energy frontiers can be competitive in its current experimental stages \cite{Erler2005, Ramsey-Musolf2008, Bhattacharya2012, Cirigliano2013a, Cirigliano2013, Naviliat-Cuncic2013, Holstein2014, Vos2015}. The classical $\beta$ correlation terms of \textcite{Jackson1957} have been written in terms of these extra coupling terms, allowing both energy frontiers to provide an upper limit on the same parameters.

Care has to be taken to avoid overlap when combining the formalism presented here with, for example, that of \textcite{Bhattacharya2012}. As an example, the Fierz term, $b_{\text{Fierz}}$, for neutron decay presented therein carries a non-zero SM contribution (Eq. (B15) in \textcite{Bhattacharya2012})
\begin{equation}
b^{\text{SM}} = -\frac{m_e}{M_N}\frac{1+2\mu_V\lambda + \lambda^2}{1+3\lambda^2},
\end{equation}
where $\mu_V$ is the difference between proton and neutron magnetic moment, and $\lambda = g_A/g_V$. The full correction term is written as $m_eb^{\text{SM}}/E_e$. The presence of $\mu_V$ points to a weak magnetism contribution, and consequently the term can indeed be retraced to the equivalent terms in Eq. (\ref{A-CZW}).

\section{Conclusions and outlook}

In this review, we have presented a fully analytical description of the $\beta$ spectrum shape, for the first time combining kinematical, electromagnetic, nuclear and atomic corrections to a few parts in $10^{4}$. These corrections are expected to hold from 1\,keV to the endpoint. The importance of the atomic corrections was underlined, showing that although the largest deviations are found at low energy, their influence can be felt at high energy. The theoretical framework for the description of atomic screening was combined with precise Dirac-Hartree-Fock-Slater calculations and showed excellent agreement with numerical results down to the lowest energies. This avoids the ambiguous evaluation of the screening exponent as it was described before. Atomic exchange calculations were performed for the entire atomic number chain, and a fit was provided to enable analytical evaluation with high precision. The oft-neglected contributions stemming from exchange with atomic $p_{1/2}$ orbitals was shown to be significant, and included into our analytical description. Further atomic corrections from different sources were combined with the work done by Wilkinson dealing with electromagnetic finite size corrections.

An overview of the corrections sensitive to nuclear structure was given and a full correction was derived for both Fermi and Gamow-Teller decay. We have presented these results in the transparent notation by Holstein while maintaining the precision of the lepton wave function and nuclear current decomposition as per Behrens and B\"uhring. Besides the weak magnetism contribution, we remain sensitive to nuclear structure information through the explicit evaluation of several matrix elements, and the proposed isovector correction. We have shown that the former can be analytically calculated to a sufficient precision given that the state is properly described in a spherical extreme single-particle manner. When this is not the case, it was shown that moving to a deformed Woods-Saxon potential while retaining the extreme single particle approach provides excellent agreement. If wanted, more extensive methods such as the shell model can provide the relevant matrix elements which can directly be incorporated into the formulae presented here. Further, we note the influence of the induced pseudoscalar contribution to the Gamow-Teller shape factor and point to the importance of its correct evaluation when experimental precision for mirror decays reaches that of the superallowed transitions.

The formalism presented here was compared against the most precise numerical calculations of the total integral of the $\beta$ spectra performed by Towner and coauthors. For both the superallowed and mirror decays, a very good agreement was found throughout the entire investigated mass range. Even for the most exotic cases where deformation effects provide a significant contribution, excellent agreement was found. The calculation of the Gamow-Teller spectrum is more difficult as here we are directly influenced by nuclear structure influences we can circumvent in the vector sector. Using the extreme single-particle approach, very good results nevertheless were obtained for the cases where this is reasonably justified based on electromagnetic moment data. When this is not the case, significant deviations occur, as expected. Moving to a proper deformed potential alleviates these problems for the investigated cases. Important to note is, however, that the shell model calculations which were used to compare against, also cannot reliably predict the weak magnetism contribution to the argued precision. This raises serious questions on the validity and accuracy of the presently available calculations.

We therefore conclude that the work presented here can accurately describe the $\beta$ spectrum shape to the required precision given that the nuclear structure input can be reliably calculated, be that through use of CVC, the extreme single-particle approximation or more advanced methods. Spectral shape measurements have an additional advantage here in that they are not as sensitive to nuclear structure effects outside of the weak magnetism effect, and can thus be the basis for experimental Beyond Standard Model searches exploring the per mille regime.

When looking for Fierz contributions one must remain wary, however, since Standard Model effects such as the weak magnetism contribution and their corresponding uncertainties can interfere destructively with our ability to extract meaningful results. Great control in its evaluation is required in case CVC cannot be invoked and one must rely on alternate methods of calculating the relevant matrix elements. Ideally this value is known to about $10\%$ in order to look for per mille level Fierz contributions, as discussed in the previous section. This is clearly a difficult task, and is currently the largest uncertainty by almost an order of magnitude in the precise determination of the $\beta$ spectrum shape, unless it can be determined through CVC arguments. Even so, one must be careful in the evaluation of the CVC result, as only the isovector component is of relevance.

The full calculation of the work presented here was automated in a custom code, discussed in a separate publication \cite{HayenTBP}. Based on simple configuration files, it aims for flexibility and user-friendliness in the use of this vast size of work.

Nevertheless, the importance of an accurate $\beta$ spectrum shape cannot be understated in the search for Beyond Standard Model physics in the electroweak sector due to the linearity of exotic coupling constants in the observables. Assuming no exotic currents it is conversely also a perfect tool to study the weak magnetism contribution. Particularly for higher masses little is known experimentally about about its magnitude. This also forms an important ingredient in the analysis of the reactor antineutrino anomaly \cite{Mueller2011}. In the current analysis, a constant value is assumed for weak magnetism throughout the entire fission fragment region. Clearly, this is not an optimal method of treating this complex effect, as can already be deduced from the discussion on single-particle matrix elements given in this document. Additionally, a correct treatment of higher order effects in the translation from electron to antineutrino spectra forms the basis of a correct analysis procedure. The formulae presented here serve as the basis for this analysis, where we note the added complexity of the shape factor compared to its usual treatment. Atomic effects, though mainly confined to the lower energy regions, become transported to the end of the antineutrino spectrum and remain highly relevant. This further underlines the importance of a precise description of atomic effects at low energy, and will no doubt be the subject of further research.

\begin{acknowledgements}
We wish to thank Anna C. Hayes for the fruitful discussions regarding the issues mentioned in this manuscript. This work has been partly funded by the Belgian Federal Science Policy Office, under Contract No. IUAP EP/12-c and the Fund for Scientific Research Flanders (FWO).
\end{acknowledgements}

\appendix

\section{General shape factor}
\label{app:general_shape_factor}
In Sec. \ref{sec:coupled_BB} we have given results assuming isospin invariance, in that we replaced the nuclear wave functions with the full charge distribution in the evaluation of the nuclear matrix elements. Doing so, this leaves us with the question of how to accurately treat said charge distribution. The general equations for superallowed Fermi decay are the following
\begin{subequations}
\begin{align}
^{V}C_0 &= -\frac{(W_0R)^2}{3} ^VF1110 \mp \frac{2}{9}\alpha Z W_0R\,^VF1111 \nonumber \\
&~~~-\frac{(\alpha Z)^2}{3}\,^VF1222 \nonumber \\
^{V}C_1 &= \frac{4}{9} W_0 R^2\,^VF1110 \nonumber \\
&~~~ \mp \frac{2}{3}\alpha ZR\left(^VF1221-\frac{1}{3}\,^VF1111\right),  \\
^{V}C_{-1} &= \frac{2}{9} W_0 R^2\,^VF1110 \mp  \frac{\alpha ZR}{3}\,^VF1211 \\
^{V}C_2 &= -\frac{4}{9} R^2 \,^VF1110.
\end{align}
\end{subequations}
using the notation by \textcite{Wilkinson1993b} to write
\begin{equation}
^VFk_emn\rho = \frac{^VF_{000}^{(1)}(k_e,m,n,\rho)}{^VF_{000}^{(0)}}.
\end{equation}
Due to the property that $I(k_e, m, n, 0) \equiv 1$, we have simply 
\begin{equation}
^VF1110 = \frac{^VF_{000}^{(1)}}{^VF_{000}^{(0)}} = \frac{\langle r^2 \rangle}{R^2}.
\end{equation}
The results of Eq. (\ref{V-CZW}) assumed a uniformly charged sphere with radius $R$ such that $\langle r^2 \rangle_{\text{exp}} = 3R^2/5$. We can, however, do better than this. One method uses the modified Gaussian distribution of Eq. (\ref{eq:mod_gauss}), where we now have one fit parameter, $A$. This can be calculated using the method outlined by \textcite{Wilkinson1993c} from which the different ratios can be calculated using
\begin{align}
F1111 &= 0.757 + 0.0069(1-\exp(-A/1.008)) \label{eq:F1111_A} \\
F1221 &= 0.844 - 0.0182(1-\exp(-A/1.974)) \\
F1222 &= 1.219 - 0.0640(1-\exp(-A/1.550)) \label{eq:F1222_A}
\end{align}
such that we are consistent with our isovector correction employing a modified Gaussian charge distribution.

The Gamow-Teller shape factor was evaluated in the same spirit. Here the situation is more complicated, however, as we have more than one type of form factor contributing. The most general expression can be written as 
\begin{subequations}
\begin{align}
^AC_0 &= -\frac{1}{3}(W_0^2-1)R^2\,^AF_{101}^{(1)} \nonumber \\
&~~~\pm \frac{2}{3}\alpha Z \left(\sqrt{\frac{1}{3}}\,^AF_{110}^{(0)}+\sqrt{\frac{2}{3}}\,^VF_{111}^{(0)} \right) \nonumber \\
&~~~\mp \frac{2}{27}\alpha Z W_0R\left(-^AF_{101}^{(1)}(1,1,1,1) \right.\nonumber \\
&~~~\left. + 2\sqrt{2}\,^AF_{121}^{(0)}(1,1,1,1)\right) \nonumber \\
&~~~-\frac{1}{3}(\alpha Z)^2\,^AF_{101}^{(1)}(1,2,2,2) \nonumber \\
&~~~-\frac{2}{3}W_0R\left(-\sqrt{\frac{1}{3}}\,^AF_{110}^{(0)}+\sqrt{\frac{2}{3}}\,^VF_{111}^{(0)}\right) \nonumber \\
&~~~+\frac{1}{9}R^2\left(\frac{11}{3}\,^AF_{101}^{(1)}-\frac{4}{3}\sqrt{2}\,^AF_{121}^{(0)}\right) \\
^AC_1 &= \frac{4}{3}\sqrt{2}R\,^VF_{111}^{(0)} - \frac{4}{27}W_0R^2\left(-5 \,^AF_{101}^{(1)}+\sqrt{2}\,^AF_{121}^{(0)}\right) \nonumber \\
&\mp \frac{2}{3}\alpha Z R \left( \frac{1}{9}\,^AF_{101}^{(1)}(1, 1, 1, 1) - \frac{2}{9}\sqrt{2}\,^AF_{121}^{(0)}(1, 1, 1, 1) \right. \nonumber \\
&\left. + \,^AF_{101}^{(0)}(1,2,2,1) \right) \\
^AC_{-1} &= -\frac{2}{3}R\left(\sqrt{\frac{1}{3}}\,^AF_{110}^{(0)}+\sqrt{\frac{2}{3}}\,^VF_{111}^{(0)}\right) \nonumber \\
&~~~+\frac{2}{27}W_0R^2\left(-\,^AF_{101}^{(1)}+2\sqrt{2}\,^AF_{121}^{(0)}\right) \nonumber \\
&~~~\pm \frac{\alpha Z R}{3}\,^AF_{101}^{(1)}(1,2,1,1) \\
^AC_2 &= -R^2\left(\frac{20}{27}\,^AF_{101}^{(1)}-\frac{4}{27}\sqrt{2}\,^AF_{121}^{(0)}\right)
\label{eq:AC_comps_general}
\end{align}
\end{subequations}
where we have omitted a common division by $^AF_{101}^{(0)}$ in all terms for notational clarity. As was mentioned in Sec. \ref{sec:isospin_breakdown}, the treatment of the Gamow-Teller shape factor is slightly more complex, since different types of form factors are present. There we have split up the nuclear structure dependent parts, i.e. terms containing $^VF_{111}^{(0)}$, $^AF_{110}^{(0)}$ and $^AF_{121}^{(0)}$, from the terms containing $^AF_{101}^{(0)}(1, m, n, \rho)$. For the ratios of the later we can use the same parametrisation as for vector decay in Eqs. (\ref{eq:F1111_A})-(\ref{eq:F1222_A}). The evaluation of the nuclear structure dependent terms can occur in the standard way via single-particle estimates or shell model calculations.

In this treatment we have assumed only a nuclear charge distribution and neglected the influence of atomic electrons on the electrostatic potential. The largest effect was, however, already taken care of when introducing the screening correction in Sec. \ref{screening} as the modification of the Fermi function, Eq. (\ref{eq:F_general}), due to screening. Further, due to the nature of the form factors coefficients all results are written in an expansion of the root mean square radius, which is matched to the experimental value for all chosen charge distributions. In order to estimate the relevance of the spatially averaged screening effect, we consider as an example the two largest contributors
\begin{align}
I(k, 1, 1, 1;r) &= (2k+1)r^{-2k-1}\int_0^rx^{2k}U(x)dx\\
I(k, 2, 2, 1;r) &= 2(2k+1)r^{-2}\int_0^rU(y)y^{-2k}\\
&\times \int_0^yx^{2k}U(x)dxdy
\end{align}
where $U(x)$ is defined by
\begin{equation}
V(x) = -\frac{\alpha Z}{R}U(x).
\end{equation}
Assuming now, for the sake of the argument, that the difference between the screened potential of Eq. (\ref{eq:salvat_potential}) and that of a point charge holds also within the nucleus, the first-order expansion results in a difference $\Delta U  = \sum_i \alpha_i \beta_i \sim \mathcal{O}(\alpha)$. It can be checked numerically that the shape factor of Eq. (\ref{V-CZW}) changes by a few parts in $10^{4}$ up to large $Z$. In these high$Z$ cases, however, the order of perturbation to which all expressions are derived is insufficient to provide the highest level of precision, such that these changes are insignificant. In order to guarantee best performance in this situation, a further expansion is required or replaced instead with a fully numerical approach.

\section{Single-particle matrix elements}
\label{app:single_particle_tables}
The Gamow-Teller shape factor contains several weak form factors which cannot always be transformed to their electromagnetic analogs through the use of CVC. In the notation of Behrens and B\"uhring these are $^VF_{111}^{(0)}$, $^AF_{110}^{(0)}$ and $^AF_{121}^{(0)}$. The former two are known as the weak magnetism and induced tensor contributions, and can only be related to CVC results for specific transitions as discussed in Sec. \ref{sec:symmetries}. The latter needs to be evaluated in our definition of $\Lambda$ in Eq. (\ref{eq:Lambda_CA}). For all these form factors we require their ratio with the main Gamow-Teller form factor, $^AF_{101}^{(0)}$. As we have seen in Sec. \ref{sec:analytical_me}, these form factors can be written as a combination of single-particle matrix elements when employing the impulse approximation. These results were compiled by \textcite{Behrens1971} such that one has
\begin{align}
^A\mathcal{M}_{1L1}^{(0)} &\approx \frac{\sqrt{2}}{\sqrt{2J_i+1}} G_{1L1}(\kappa_f, \kappa_i) \frac{\langle r^L \rangle}{R^L} \label{eq:M_1L1_approx}\\
^{A/V}\mathcal{M}_{11s}^{(0)} &\approx \frac{\sqrt{2}}{\sqrt{2J_i+1}}\left\{ \text{sign}(\kappa_i)G_{11s}(\kappa_f, -\kappa_i) \right.\nonumber \\
&\times \langle  g_f | \left(\frac{r}{R}\right)| f_i \rangle + \text{sign}(\kappa_f)G_{11s}(-\kappa_f, \kappa_i) \nonumber \\
&\left. \langle f_f | \left(\frac{r}{R}\right) | g_i \rangle \right\},
\label{eq:M_KL1_approx}
\end{align}
with $s=1$ for $^V\mathcal{M}_{111}^{(0)}$ and $s=0$ for $^A\mathcal{M}_{110}$. Here, the $G_{KLs}(\kappa_f, \kappa_i)$ capture all spin-angular information of the reduced matrix elements. These coefficients were introduced by \textcite{Weidenmuller1961} and are listed in the work by \textcite{Behrens1982}. For $^AF_{121}^{(0)}$ and $^VF_{111}^{(0)}$, $K$ and $s$ are equal to those of $^AF_{101}^{(0)}$ such that the ratio simplifies considerably. For the former, this means the ratio can be written as
\begin{align}
\frac{G_{121}(\kappa_f, \kappa_i)}{G_{101}(\kappa_f, \kappa_i)} &= -\frac{C(l(\kappa_f)~l(\kappa_i)~2;00)}{C(l(\kappa_f)~l(\kappa_i)~0;00)}\left\{\begin{array}{ccc}
1 & 1 & 2 \\
j_f & \frac{1}{2} & l(\kappa_f) \\
j_i & \frac{1}{2} & l(\kappa_i)
\end{array}\right\} \nonumber \\
&~~~\times \left\{\begin{array}{ccc}
1 & 1 & 0 \\
j_f & \frac{1}{2} & l(\kappa_f) \\
j_i & \frac{1}{2} & l(\kappa_i)
\end{array} \right\}^{-1} \frac{\langle r^L \rangle }{R^L}
\end{align}
where $C(\ldots )$ is a regular Clebsch-Gordan coefficient, and the quantities in brackets are Wigner-$9j$ symbols. Here $l(\kappa) = \kappa$ if $\kappa > 0$ and $l(\kappa) = |\kappa|-1$ if $\kappa < 0$. 

We list the relevant results from Appendix F of \textcite{Behrens1978} for the convenience of the reader. These can be directly deduced from Eqs. (\ref{eq:M_1L1_approx})-(\ref{eq:M_KL1_approx}) using the non-relativistic approximation to express the small radial function $f(r)$ as a function of $g(r)$. These are categorized by the initial and final single particle spins participating in the correction.
\begin{align*}
j_i = j_f = &~ l+\frac{1}{2} \\
\mathcal{M}_{101}^0 &= \frac{\sqrt{2}}{\sqrt{2J_i+1}}\left[ \frac{(l+1)(2l+3)}{2l+1}\right]^{1/2}\cdot I \\
\mathcal{M}_{121}^0 &= -\frac{\sqrt{2}}{\sqrt{2J_i+1}}l\left[\frac{2(l+1)}{(2l+1)(2l+3)} \right]^{1/2} \cdot \frac{\langle r^2 \rangle}{R^2} \\
\mathcal{M}_{111}^0 &= -\frac{\sqrt{2}}{\sqrt{2J_i+1}}(l+1)\left[\frac{6(l+1)(2l+3)}{2l+1} \right]^{1/2} \\
&\times \frac{1}{2M_NR} I \\
\mathcal{M}_{110}^0 &= -\frac{\sqrt{2}}{\sqrt{2J_i+1}} \left[\frac{3(l+1)}{(2l+1)(2l+3)}\right]^{1/2} \xi  \frac{\langle r^2 \rangle}{2R} \\ \\
j_i = j_f = &~ l - \frac{1}{2} \\
\mathcal{M}_{101}^0 &= -\frac{\sqrt{2}}{\sqrt{2J_i+1}}\left[ \frac{l(2l-1)}{2l+1}\right]^{1/2}\cdot I \\
\mathcal{M}_{121}^0 &= \frac{\sqrt{2}}{\sqrt{2J_i+1}}(l+1)\left[\frac{2l}{(2l-1)(2l+1)} \right]^{1/2} \cdot \frac{\langle r^2 \rangle}{R^2} \\
\mathcal{M}_{111}^0 &= -\frac{\sqrt{2}}{\sqrt{2J_i+1}}l\left[\frac{6l(2l-1)}{2l+1} \right]^{1/2} \\
&\times \frac{1}{2M_NR} I \\
\mathcal{M}_{110}^0 &= -\frac{\sqrt{2}}{\sqrt{2J_i+1}} \left[\frac{3l}{(2l+1)(2l-1)}\right]^{1/2} \xi  \frac{\langle r^2 \rangle}{2R} \\ \\
j_f = l ~ \pm ~ &\frac{1}{2} ~~ j_i = l \mp \frac{1}{2} \\
\mathcal{M}_{101}^0 &= -\mp\frac{\sqrt{2}}{\sqrt{2J_i+1}}2\left[ \frac{l(2l+1)}{2l+1}\right]^{1/2}\cdot I \\
\mathcal{M}_{121}^0 &= \mp\frac{\sqrt{2}}{\sqrt{2J_i+1}}\left[\frac{l(l+1)}{2(2l+1)} \right]^{1/2} \cdot \frac{\langle r^2 \rangle}{R^2} \\
\mathcal{M}_{111}^0 &= \pm\frac{\sqrt{2}}{\sqrt{2J_i+1}}\left[\frac{3l(2l+1)}{2(2l+1)} \right]^{1/2} \\
&\times \frac{1}{M_NR} I \\
\mathcal{M}_{110}^0 &= \pm\frac{\sqrt{2}}{\sqrt{2J_i+1}} \left[\frac{3l}{(2l+1)(2l+1)}\right]^{1/2} \\
&\times \left[\pm \frac{2l+1}{2M_NR}I + \xi \frac{\langle r^2 \rangle}{2R} \right]
\end{align*}
with $I=\int g_f(r)g_i(r)r^2dr\approx 1$ in the non-relativistic approximation.
Here $\xi \langle r^2 \rangle / R^2$ is the evaluation of
\begin{equation}
\int_0^\infty g_f\left\{E_i-E_f-(V_i-V_f) \right\}\left(\frac{r}{R} \right)^2g_i r^2dr
\end{equation}
evaluated for a spherical harmonic oscillator wave function to give
\begin{equation}
\xi = \frac{2\nu}{M_N}[2(n_i-n_f)+l_i-l_f]
\end{equation}
with $\nu$ the harmonic oscillator parameter defined as in Eq. (\ref{eq:R_nl_HO}).

Assuming a spherical extreme single-particle approach, we then recover the results of Eq. (\ref{eq:M_ratio_SP}). In the deformed case as discussed in Sec. \ref{sec:deformed_single_particle}, however, the situation is not as straightforward and we require full expressions for the different matrix elements. 

\section{Many-particle matrix elements in $jj$-coupling}
\label{sec:many_particle}
The formulae presented in Sec. \ref{sec:single_particle}, Sec. \ref{sec:deformed_single_particle} and Appendix \ref{sec:single_particle_tables} dealt with odd-$A$ $\beta$ decays, which in the extreme single particle approach we have taken here considers initial and final states consisting of one single nucleon. In Sec. \ref{sec:analytical_me} we have discussed that the many-particle angular momentum couplings reside in a factor $C(K)$, which depends only on the tensor rank $K$ of the operator. This fact is of great assistance, as it drops out completely when taking the ratio of two form factors with identical rank. In the spirit of the extreme single-particle approach taken here, it is worthwhile to discuss odd-$Z$, odd-$N$ (o-o) to even-$Z$, even-$N$ (e-e) $\beta$ decays and vice versa. For this we use the results by \textcite{Rose1954}, written using the isospin formalism \cite{Wilkinson1969}. In line with our previous extreme single particle methods, we consider now two particles in initial and final states, coupled to a core isospin. We then write for o-o to e-e transitions
\begin{align}
&\langle j_2 j_2 J_f M_f T_f T_{3f} | \sum_{n=1,2} \left\{O_{KLs}^M \tau_\pm \right\}_n | j_1 j_2 J_i M_i T_i T_{3i} \rangle \nonumber \\
& = \sqrt{\frac{\hat{J_i}\hat{J_f}\hat{T_i}\hat{T_f}}{1+\delta_{j_1j_2}}}(-)^{J_f-M_f} \left(\begin{array}{ccc}
J_f & K & J_i \\
-M_f & M & M_i
\end{array} \right) (-)^{T_f-T_{3f}} \nonumber \\
&\times \left(\begin{array}{ccc}
T_f & 1 & T_i \\
-T_{3f} & \pm 1 & T_{3i}
\end{array} \right) (-)^K \left\{\begin{array}{ccc}
\frac{1}{2} & T_f & \frac{1}{2}(T_i+T_f) \\
T_i & \frac{1}{2} & 1
\end{array} \right\} \nonumber \\
& \times \left[2 \left\{\begin{array}{ccc}
j_2 & J_f & j_2 \\
J_i & j_1 & K
\end{array} \right\} \langle j_2 || O_{KLs} || j_1 \rangle  + [1 - (-)^{j_1+j_2}]\right. \nonumber \\
&\left. \times \left\{\begin{array}{ccc}
j_2 & J_f & j_1 \\
J_i & j_2 & K
\end{array} \right\} \langle j_2 || O_{KLs} || j_2 \rangle \delta_{j_1j_2} \right]\langle \frac{1}{2}||\bm{t}||\frac{1}{2}\rangle
\label{eq:two_particle_me}
\end{align}
where the quantities in curly brackets are Wigner-6$j$ symbols. We have explicitly extracted the results from the Wigner-Eckart theorem in both spin and isospin spaces. For a final matrix element one has to average over initial spin projections and sum over the final $M_f$. The isospin components can be directly evaluated, using $\langle \frac{1}{2} || \bm{t} || \frac{1}{2} \rangle = \sqrt{3/2}$. An equivalent formula can be written down for e-e to o-o $\beta$ decays.

In the case of deformation the angular momentum, $J$, is not any more a good quantum number, and Eq. (\ref{eq:two_particle_me}) has to be rewritten. Using the results of Ref. \cite{Berthier1966}, the spin-reduced matrix element for even-even to odd-odd decays is found to be
\begin{align}
&\langle \phi(J_fK_f;\Omega_f) || \sum_{n=1,2} \{O_{KLs}\tau^\pm_n\} || \phi(J_iK_i=0;\Omega_i=0) \rangle \nonumber \\
&= \sqrt{\frac{\hat{J_i}\hat{J_f}}{2(1+\delta_{K_f0})}} \left(\begin{array}{ccc}
J_f & K & J_i \\
-K_f & K_f & 0 
\end{array} \right) \times [1+(-1)^{J_i}] \nonumber \\
&\times \sum_{j_2j_1} C_{j_2-\Omega_2}C_{j_1\Omega_1} (-1)^{j_2-\Omega_2} \left(\begin{array}{ccc}
j_2 & K & j_1 \\
-\Omega_2 & K_f & -\Omega_1
\end{array} \right) \nonumber \\
&\times \langle j_2 || O_{KLs} || j_1 \rangle
\label{eq:deformed_matrix_element_even}
\end{align}
while the reverse case can be found in several publications \cite{Berthier1966, Behrens1982}. Here $C_{j\Omega} = (-)^{\frac{1}{2}-j}\pi_{\Omega}C_{j-\Omega}$ with $\pi_{\Omega}$ the parity of the $\Omega$ orbital.

\section{Relativistic Coulomb amplitudes}
\label{app:relativistic_me}

In Sec. \ref{sec:relativistic_terms} we discussed the influence of the relativistic matrix elements, and concluded that for spectral shape measurements these are insignificant on the aimed-for level of precision. We discuss them briefly for completeness. These contributions consist of two parts, the nuclear structure embedded into the ratio of form factors and the electromagnetic influence through the slowly varying Coulomb functions. We specify here for Fermi transitions the energy-dependent functions using the older leptonic wave expansion by \textcite{Behrens1969} as shown in Table \ref{tab:rwf_expansion_bb}.

\begin{table}[h]
\caption{The coefficients $H_{2\sigma}, D_{2\sigma+1}, h_{2\sigma}$, $N_{\sigma}$ and $d_{2\sigma+1}$ used in the expansion of the leptonic radial wave functions.}
\begin{ruledtabular}
{\renewcommand{\arraystretch}{1.8}
\begin{tabular}{l l}
$\overline{W}_e = W_e \pm 3 \alpha Z/(2R)$ & $\bar{p}_e^2 = \bar{W}_e^2-1$ \\
$H_{2\sigma} = D_{2\sigma+1}=0$ for $\sigma < 0$ & $H_{2\sigma+1} = D_{2\sigma}$ for all $\sigma$ \\
\hline
$H_0(k_e) = 1$ & $h_0(k_e) = 0$ \\
$H_2(k_e) = -\frac{(\bar{p}R)^2}{2(2k_e+1)}$ & $h_2(k_e) = 0$ \\
& $h_4(k_e) = \frac{\pm \alpha Z R}{2(2k_e+1)(2k_e+3)}$ \\
\hline

$D_1(k_e) = \frac{\overline{W}_eR}{2k_e+1}$ & $d_1(k_e) = \frac{R}{2k_e+1}$ \\
$D_3(k_e) = -\frac{\overline{W}_eR(\bar{p}_eR)^2}{2(2k_e+1)(2k_e+3)}$ & $d_3(k_e) = -\frac{R(\bar{p}_eR)^2}{2(2k_e+1)(2k_e+3)}$ \\
$~~~~~~~~~~~~~\mp \frac{\alpha Z}{2(2k_e+3)}$ & \\
\hline
$N_0(k_{\nu}) = 1$ & $N_1(k_{\nu}) = \frac{p_{\nu}R}{2k_{\nu}+1}$ \\
$N_2(k_{\nu}) = -\frac{(p_{\nu}R)^2}{2(2k_{\nu}+1)}$ & 
\end{tabular}
}
\end{ruledtabular}
\label{tab:rwf_expansion_bb}
\end{table}

The leading order Coulomb functions for pure Fermi transitions can be written as
\begin{equation}
^VC(Z, W)_{\text{rel}} = \frac{^VF^{(0)}_{011}}{^VF^{(0)}_{000}}\, ^Vf_2(W) + \frac{^VF^{(1)}_{011}}{^VF^{(0)}_{000}}\, ^Vf_3(W).
\end{equation}
The Coulomb functions $f_2$ and $f_3$ are written in terms of the elements in Table \ref{tab:rwf_expansion_bb} as
\begin{subequations}
\begin{align}
^Vf_2(W) &= -2(D_1+N_1)+2\frac{\mu_1\gamma}{W}d_1 \\
^Vf_3(W) &= -2(D_3+N_1H_2-N_2D_1-N_3)\nonumber \\
&~~+ 2\frac{mu_1\gamma}{W}(d_3-N_2d_1)
\end{align}
\end{subequations}
It is worth noting here that $D_1(1)+N_1(1) = \alpha Z/2 + W_0R/3$ is independent of $W$. Typical values for $\overline{f_i(W)}$ for several $0^+\to 0^+$ transitions can, for example, be found in \textcite{Behrens1968}, and are usually on the percent level of smaller. The relativistic matrix elements $^VF_{011}^{(N)}$ obey recursion relations after invoking CVC, specified in Eq. (\ref{eq:F011_CVC}). 

\section{Comparison of finite size effects and electromagnetic corrections in the Behrens-B\"uhring and Holstein formalisms}
\label{app:em_comparison}

As discussed in Sec. \ref{sec:complete_expression}, there is some confusion in the name `finite size effects', as it entails different things for different authors. In this work we have mainly based our approach on the rigorous work of Behrens and B\"uhring, culminating in the standard work by the same authors \cite{Behrens1982}. We have, on the other hand, written some of these results in the more transparent formalism of Holstein. In light of transparency and their increasing importance in other fields not directly related to nuclear physics, we have attempted to elucidate this more quantitatively and convince the reader of the internal consistency of the approach presented in this work.

\subsection{Generalization including electrostatics}
The simplest $\beta$ decay Hamiltonian is written down as a simple current-current interaction
\begin{align}
H_\beta = &\frac{G\cos \theta_C}{\sqrt{2}}\bar{u}(p)\gamma_\mu (1+\gamma_5)v(l)\langle f_{\bf{p}_2} | V^\mu + A^\mu | i_{\bf{0}} \rangle \nonumber \\
& + \text{H.c.}
\end{align}
where plane wave have been used for the leptons, and $\bf{p}_2$ and $\bf{0}$ denote final and initial nuclear momenta. The $S$ matrix is developed to first order and the corresponding spectrum is calculated using standard techniques. Given that the nucleus interacts electromagnetically with the outgoing leptons, one must incorporate the Coulomb interaction. As it is several times stronger than the weak interaction it can, on the other hand, not be treated perturbatively \cite{Halpern1970}. Neglecting the difference between initial and final Coulomb fields, the lepton wave functions are considered solutions of the Dirac equation in the final Coulomb field\footnote{As discussed in Sec. \ref{Fermi function}, neglecting this difference is corrected for in the radiative corrections of Sec. \ref{sec:radiative_corr}.}.

As discussed previously, the Behrens-B\"uhring formalism starts from the results by \textcite{Halpern1970}. The approach taken by Holstein \emph{et al.} \cite{Holstein1974b, Calaprice1976} starts from the generalized matrix element as described by Armstrong and Kim \cite{Armstrong1972, Holstein1979}
\begin{align}
\mathcal{M} = &\frac{G\cos \theta_C}{\sqrt{2}}\int d^3r\bar{\Psi}_e(\mathbf{r}, \mathbf{p})\gamma_{\mu}(1+\gamma_5)v(l) \nonumber \\
&\times \int \frac{d^3k}{(2\pi^3)}e^{i\mathbf{r}\cdot \mathbf{k}}\langle f_{\mathbf{p_2}+\mathbf{p}-\mathbf{k}}|V^{\mu}+A^{\mu}|i_{\mathbf{0}}\rangle,
\label{eq:matrix_element_armstrong}
\end{align}
where $\bar{\Psi}_e$ is the electron wave function in the presence of a nuclear Coulomb potential. Both results are identical up to this point after neglecting the difference in initial and final Coulomb potentials. In both the Holstein and Behrens-B\"uhring formalisms the nuclear current is now expanded using a series of form factors, replacing the impulse-approximation results of Eqs. (\ref{eq:currents_BB_HS}). Those of the former are useful for allowed transitions whereas those of the latter make no inherent distinction between the order of the transitions. Translation tables between both descriptions can be found in several publications \cite{Behrens1978, Behrens1982} and will not be repeated here. Terms relevant for the discussion below can be easily deduced from Tables \ref{table:matrix_elements_BB} and \ref{table:form factors}.

Both approaches use as a starting point the so-called Behrens-J\"anecke Fermi function, which corresponds to Eq. (\ref{eq:F_general}). In works by Holstein and derivatives thereof it is typically expressed as the point charge Fermi function, $F_0$, times some correction factor as discussed in Sec. \ref{Fermi function}. The crucial difference between both approaches, however, now lies in two points, both in favour of the Behrens-B\"uhring formulation. The first is the expansion of the electron wave function, and the corresponding electrostatic finite size corrections discussed in Sec. \ref{size-and-mass}. The second concerns the combination of the expansion of the nuclear current with that of the lepton wave function.

\subsection{Approximations by Holstein and notes for the wary}

We concern ourselves first with the expansion of the electron wave function. As was briefly mentioned in Sec. \ref{Fermi function} and Sec. \ref{sec:coupled_BB}, the Behrens-B\"uhring approach expands these as a function of $(\alpha Z)^\rho$, $(WR)^{\nu-\rho}$ and $(m_eR)^{\mu-\nu}$ with the coefficients of these expansions encoded in the functions $I(|\kappa|, \mu, \nu, \rho; r)$. The latter are sensitive to the nuclear potential and examples can be found in the works by \textcite{Behrens1970}. In the \textcite{Holstein1974b} approach one expands $\bar{\Psi}_e$ keeping only leading $j=1/2$ and $j=3/2$ terms, such that it becomes a function of $f_{\kappa}, g_{\kappa}$ with $\kappa \in \{-2, -1, 1, 2\}$. Specific results of the expansion can be found for instance in \textcite{Armstrong1972, Holstein1974a, Holstein1974b}. Here the first difference occurs, as the expansion of the latter is performed only for a uniformly charged sphere. Our discussion in Sec. \ref{size-and-mass} shows that this is not sufficient, and effects of moving to a diffuse charge distribution cannot be ignored.

The calculation proceeds analogously to that of \textcite{Armstrong1972} where a weak charge density is defined through the Fourier transform of the leading order expansion of the nuclear current. We have then
\begin{equation}
M = \frac{G\cos \theta_C}{\sqrt{2}}\int d^3r \bar{\Psi}_e(\mathbf{r}, \mathbf{p})\gamma_\mu (1+\gamma_5)v(l)\rho^\mu(\mathbf{r})
\label{eq:matrix_element_Armstrong_weak_charge}
\end{equation}
where $\rho^\mu(\mathbf{r})$ is simply the Fourier transform of the involved form factors
\begin{equation}
\rho^\mu(\mathbf{r}) = \int \frac{d^3s}{(2\pi)^2}e^{i\bf{r}\cdot \bf{s}}\langle f_{\mathbf{p_2}+\mathbf{p}-\mathbf{k}}|V^{\mu}+A^{\mu}|i_{\mathbf{0}}\rangle.
\label{eq:weak_charge_definition}
\end{equation}
The second difference now appears in the expansion of the nuclear current, and the consequent artificial split of nuclear structure terms and the Coulomb corrections in the Holstein approach. This is in part because Coulomb corrections to induced corrections were neglected \cite{Holstein1974b}, such that for Gamow-Teller decay it is approximated as
\begin{equation}
\langle f_{\mathbf{p}_2}|V^{\mu} + A^{\mu}|i_{\mathbf{0}}\rangle \approx - g^{\mu \kappa}c(W_0^2-p_2^2)C^{M' \kappa;M}_{J'1;J},
\label{eq:nuclear_current_calaprice}
\end{equation}
with $C$ a regular Clebsch-Gordan coefficient. The weak charge density is then typically normalized by extracting a factor $c(0)$. 

Here once again great care must be taken when comparing to the Holstein formalism, as the corrections are written differently in different papers \cite{Holstein1974b, Holstein1974a, Calaprice1976, Holstein1979} by the same authors. In particular the used Fermi function differs with factor $(1+\gamma)/2$ depending on the publication, resulting in changes of order $(\alpha Z)^2$ when naively comparing formulas. Further, the redefinition of the so-called spectral functions taking into account Coulomb interactions has been written in two different ways. The first \cite{Holstein1974a, Holstein1974a} replaces
\begin{equation}
h_i^{(0)}(W) \to F(Z, W)[h_i^{(0)}(W) + \Delta h_i(W)]
\label{eq:holstein_old_coulomb_replacement}
\end{equation}
where $F(Z, W) = (1+\gamma)/2F_0$, with $\Delta h_i(W)$ defined in those works. The later works \cite{Calaprice1976, Holstein1979}, on the other hand, artificially split the transition matrix element of Eq. (\ref{eq:matrix_element_Armstrong_weak_charge}) into a nuclear structure part and their Coulomb corrections
\begin{widetext}
\begin{align}
N(W)dW &\propto \widetilde{h}_1(W) ~ \Big{[} |A|^2 + |B|^2 +|C|^2 +|D|^2  + \frac{2}{3} \text{Re} \left( A^* D + B^* C \right)   \nonumber \\
& ~~~~ + 2 \frac{m_e}{W} \text{Re} \left( A^* B + C^* D + \frac{1}{3} A^* C + \frac{1}{3} B^* D \right) - \frac{2}{9} \frac{p}{W} \text{Re} \left( A^* F - B^* G + 3 F^* D - 3 G^* C \right) \Big{]}
\label{em-complete-bis}
\end{align}
\end{widetext}
with $\widetilde{h}_1(W)$
\begin{align}
\widetilde{h}_1(W) =  &c_1^2 - \frac{2}{3} \frac{W_0}{M} c_1 \left( c_1 + d \pm b \right)\nonumber \\
& + \frac{2}{3} \frac{W}{M}\left[ c_1 \left( 5 c_1 \pm 2 b \right)\right] - \frac{m_e^2}{3 M W} \Big{[}2 c_1^2 \nonumber \\
&\left. + c_1 \left( d \pm 2 b \right) - c_1 h \frac{W_0-W}{2 M} \right]
\label{h-tilde-1}
\end{align}
and $A$-$G$ denoting integrals of the different parts of the leptonic expansion with the weak charge distribution of Eq. (\ref{eq:weak_charge_definition}). These are given, for example, by \textcite{Holstein1974a} and will not be repeated here for brevity. Here the Fermi function consistently refers the so-called Behrens-J\"anecke Fermi function, discussed in more detail in the following section. It is Eqs. (\ref{em-complete-bis}) and (\ref{h-tilde-1}) that should be compared against the combination $F_0L_0UC$ in the current work. For clarity, we will continue with the later results of Eqs. (\ref{em-complete-bis}) and (\ref{h-tilde-1}) and refer only to the older works when necessary in Appendix \ref{app:weak_charge_CI}.

Expanding Eq. (\ref{eq:nuclear_current_calaprice}) to zeroth order will introduce the Fermi function, while the first order introduces the leptonic convolution. These will be discussed separately. Finally, the inclusion of the Coulomb corrections to induced currents will be briefly touched.

\subsection{Fermi function and finite size correction}

The Coulomb corrections of Eq. (\ref{em-complete-bis}) depend on the precise evaluation of the integrals $A$-$G$, which are in turn dependent on the weak charge distribution of Eq. (\ref{eq:weak_charge_definition}) \cite{Calaprice1976}. In a first approximation we can approximate the Gamow-Teller form factor $c(q^2)$ as a constant, in which case $\rho(r)$ becomes a simple Dirac delta function. Now only $A$ and $B$ survive such that
\begin{align}
|A|^2 + |B|^2 &= \frac{1}{2}N^2\frac{2m_e}{W}\int d^3r \delta^3(\mathbf{r}) \nonumber \\
&\times \left[g_{-1}^2(r)+\left(\frac{W}{p}\right)^2f_{1}^2(r) \right] \\
&=F_0L_0 \nonumber
\end{align}
when taking into account the differing normalization definitions of the wave functions in the Behrens-B\"uhring and Holstein formalisms. As $f_1$ and $g_{-1}$ have only been approximated for a uniformly charged sphere for low $Z$ in the Holstein approach, the expression obtained, for example, by \textcite{Calaprice1976, Huffaker1967} 
\begin{equation}
|A|^2+|B|^2 \approx F_0\left[1\mp \frac{13}{15}\alpha Z W R\right]
\end{equation}
is clearly less precise than our Eq. (\ref{L0}) as was discussed already in Sec. \ref{Fermi function}. The term can easily be recognized as a simplified version of that found in Eq. (\ref{L0}). Further, it contains no corrections stemming from a diffuse nuclear charge, for which we have derived explicit expressions and analytical parametrisations in Sec. \ref{sec:U}.


\subsection{Nuclear and leptonic convolution}

The finite nuclear size does not only affect the Coulomb potential as felt by the charged lepton and the behavior of its wave function near the origin. The interaction volume becomes a sphere with radius $R$, and thus the matrix element requires an average for all the nucleon positions. This is clearly visible in Eq. (\ref{eq:matrix_element_Armstrong_weak_charge}), and is inherently present in our definition of the $C$ factor of Sec. \ref{sec:coupled_BB}. In the initial results by \textcite{Huffaker1967} the shape factor contains an additional term
\begin{equation}
C_0(W) = \frac{\langle r^2 \rangle}{R^2}\alpha Z W R,
\label{eq:convulation_huffaker}
\end{equation}
as a remnant of explicit averaging of $f_1$ and $g_{-1}$ inside the nuclear volume. For a constant nuclear wave function, i.e. a rectangle of width $R$, one finds $\langle r^2 \rangle = \frac{3}{5}R^2$ so that $C_0(W) = \frac{3}{5}\alpha Z WR$. The approach of Holstein and co-authors builds on this result and provides a more consistent discussion by introducing the elementary particle treatment. We move beyond the approximation of the previous section and now expand Eq. (\ref{eq:nuclear_current_calaprice}) to first order in $q^2$, $c(q^2) \approx c_1 + c_2q^2$ to find
\begin{align}
\rho &\approx \int \frac{d^3k}{(2\pi)^3}e^{i\mathbf{r}\cdot \mathbf{k}}\left(1+\frac{c_2}{c_1}q^2\right) \nonumber \\
&= \left(1+\frac{c_2}{c_1}(W_0^2+\mathbf{\nabla}^2) \right)\delta^3(\mathbf{r}),
\label{eq:weak_charge_density_second_order}
\end{align}
where $\mathbf{\nabla}$ is the gradient operator. In this manner, all further expansions in $q^2$ will always result in a constant factor times $\delta^3(\mathbf{r})$ as $\mathcal{F}[q^n] \propto \nabla^n\delta^3(\mathbf{r})$. This entails that all leptonic radial wave functions will still be evaluated at the nuclear center, be it multiplied with some constant factor depending on nuclear structure. As such, all results will always be proportional to the Fermi function defined at the origin, such that it can always be meaningfully extracted as we have always done so far. Again, however, results presented by Holstein and colleagues have only been listed for uniformly charged spheres. Using Eq. (\ref{eq:weak_charge_density_second_order}) we find only only $A$, $D$ and $F$ integrals survive, all of which are proportional to the square root of the Fermi function as expected \cite{Calaprice1976}. Equations (13) and (A.16) in \textcite{Calaprice1976} are then to be compared against Eq. (\ref{eq:convulation_huffaker}) and (\ref{A-C0-C1-C2}). 
A critical ingredient in the quantitative connection between the different formalisms lies in the evaluation of $c_2/c_1$. In the impulse approximation it can be written as \cite{Calaprice1976}
\begin{align}
\frac{c_2}{c_1} &\approx \frac{1}{6}\frac{\langle \beta || \tau^{\pm}\sigma r^2 || \alpha\rangle}{\langle \beta || \tau^{\pm}\sigma || \alpha \rangle} \nonumber \\
&+\frac{1}{6\sqrt{10}}\frac{\langle \beta || \tau^{\pm}[\sigma\times\sqrt{\frac{16}{5}\pi}Y_2(r)]||\alpha \rangle}{\langle \beta || \tau^{\pm}\sigma||\alpha\rangle},
\label{eq:GT_HS_c2c1}
\end{align}
and is related to the shape of nuclear wave functions. In the initial isospin invariant formalism, these were always treated as rectangles with only non-zero values for $r < R$, i.e. a uniformly charged sphere with radius $R$. Using this, one arrives at $c_2/c_1 = \frac{1}{10}R^2$. The correction factor to the Behrens-J\"anecke Fermi function (i.e. $F_0L_0$ in our formalism) in the Holstein formalism is then written as
\begin{align}
_{HS}C(Z, W)_0 &\approx 1 + R^2/5 - (W_0R)^2/5 - \frac{9}{20}(\alpha Z)^2 \nonumber \\
&\pm \frac{2}{15}\alpha Z W_0 R \mp \frac{11}{15}\alpha Z WR \nonumber \\
& + \frac{2}{5}W_0R^2W - \frac{2}{5}R^2W^2.
\label{eq:C_LC_HS}
\end{align}

If we use the older expansion of $\bar{\Psi}_e$ \cite{Behrens1969, Wilkinson1990}, we have for Gamow-Teller decay
\begin{equation}
_{\text{old}}C(Z, W)_0 = 1 + ~^AC_0 + ~^AC_1W + ~^AC_2W^2,
\end{equation}
where
\begin{align}
^AC_0 &= -\frac{9}{20}(\alpha Z)^2-(W_0R)^2/5\pm\alpha ZW_0R/15 + R^2/5, \nonumber \\
^AC_1 &= \mp 2\alpha ZR/3+4W_0R^2/9, \nonumber \\
^AC_2 &= -4R^2/9.
\end{align}
It is clear every term can be identified with a similar one in Eq. (\ref{eq:C_LC_HS}), except for a factor 2 difference in the $\alpha ZW_0R$ term. All of these results are again valid only for a uniformly charged sphere, both in the expansion of the lepton wave function and in the calculation of the weak charge density. In Sec. \ref{app:general_shape_factor} we have looked at the deviation of the former due to a more realistic charge distribution. Further, to correct for the breakdown of the latter we introduced the isovector correction $C_I$ in Sec. \ref{sec:isospin_breakdown}. Due to its large contribution this can for sure not be neglected. In combination with a more precise expansion of the lepton wave functions, it is clear the approach presented in this work is superior to that of Holstein.

\subsection{Induced terms}
Finally, the Holstein formalism requires an adjustment for the approximation introduced in Eq. (\ref{eq:nuclear_current_calaprice}). Initially, Coulomb corrections to the induced terms in the nuclear current were neglected as the latter are already suppressed by a factor $q/M$. As discussed by \textcite{Bottino1973, Bottino1974}, however, this approximation is not strictly valid. The reasoning starts from the result by \textcite{Armstrong1972} showing that $\langle f(\mathbf{p}_f)|$ should be replaced by $\langle f(\mathbf{p}_f+\mathbf{p}_e-\mathbf{p})|$ where $\mathbf{p}$ is an internal momentum of the Fourier transform. This has the consequence that induced terms are now a function of $q^{\prime} = (\mathbf{p}_f+\mathbf{p}_e-\mathbf{p})-\mathbf{p}_i$ rather than $q = \mathbf{p}_f-\mathbf{p}_i$, thereby introducing Coulomb corrections to the induced terms. This can clearly be seen from the fact that the additional term
\begin{align}
\frac{1}{(2\pi)^3}\int& d\vec{p}\left[\int d\vec{r} \left(\frac{\bar{\Psi}_e(\mathbf{r}, \mathbf{p}_e)}{\bar{u}_e(\mathbf{p}_e)}\right) e^{i\mathbf{r}\cdot\mathbf{p}} \right] \nonumber \\
&\times \frac{F_M(q^{\prime})}{2m_p} \times \sigma^{\alpha \beta}(\mathbf{p}_e-\mathbf{p})_{\beta}
\end{align}
becomes identically zero when $\bar{\Psi}_e$ reduces to $e^{-i\mathbf{r}\cdot \mathbf{p}_e}\bar{u}_e(\mathbf{p}_e)$ in the absence of Coulomb interactions. The correction factor given by \textcite{Calaprice1976, Holstein1974c}
\begin{equation}
\delta h_1(Z) \approx \frac{\sqrt{10}}{6}\frac{\alpha Z}{MR}c_1(2b+d^{II}\pm d^I \pm c_1)
\label{eq:coulomb_induced}
\end{equation}
can be obtained from the results of Eqs. (9) and (13) in \textcite{Bottino1974} after proper conversion from $F_M$ ($F_A$) to $b$ ($c$) using the definition of the nuclear currents (Eq. (5) in \textcite{Bottino1974} and, e.g., Eq. (1) in \textcite{Holstein1974b}), with an exception of the $d^I, d^{II}$ terms. The results from \textcite{Bottino1974} can, however, easily be extended to include this result, and a quick glance on the behavior of $b$ and $d$ terms leads directly to Eq. (\ref{eq:coulomb_induced}). A discussion on the signs for electron and positron decay can for instance be found in \textcite{Holstein1974b}. As there is no artificial separation of nuclear structure and Coulombic terms and all terms of the nuclear current are retained, these corrections occur naturally in the $C$ factor of the Behrens-B\"uhring formalism followed in this work.

\subsection{Summary}
While the Holstein formalism initially profits from a more transparent description of the different factors and their origins participating in the $\beta$ spectrum shape, several approximations introduced along the way limit the usefulness of the quoted final formulae when requiring a high precision description. Several terms have to be added, such as Coulomb corrections to induced currents, thereby introducing confusion as to the origin of these terms and how they play with the other results. This is further hindered by a difference in definitions throughout several papers about the correct manner of introducing Coulomb corrections. We have looked at individual parts of the calculation and for each of them discussed the superiority of the formalism presented in this work. To facilitate the interpretation of experimental results and cut back on notational clutter, the formulae in Sec. \ref{sec:coupled_BB}, while calculated in the Behrens-B\"uhring formalism, are presented using the well-known Holstein form factors.

\section{Validity of harmonic oscillator functions}
\label{app:weak_charge_CI}
Recently the issue of finite size corrections was addressed by \textcite{Wang2016}, incorporating density functional theory (DFT) results to calculate the weak charge density. The analytical formulation to first order in $\alpha Z$ and $R$ by \textcite{Holstein1974b} was used, and rewritten in terms of nuclear Zemach moments \cite{Zemach1956}
\begin{align}
\delta_\text{FS}^\text{Wang} &= -\frac{\alpha Z}{3}\left(4W\langle r \rangle_{(2)} + W\langle r \rangle_{(2)}^r - \frac{W_{\nu}\langle r \rangle_{(2)}^r}{3}\right. \nonumber \\
&~~~\left.+\frac{1}{W}(2\langle r \rangle_{(2)}-\langle r \rangle_{(2)}^r) \right),
\label{eq:FS_Wang}
\end{align}
where $\langle r \rangle_{(2)}$ and $\langle r \rangle_{(2)}^r$ are defined as
\begin{align}
\langle r \rangle_{(2)} &= \int d^3s\, s \int d^3r\, \rho_\text{w}(r)\rho_\text{ch}(|\vec{r}-\vec{s}|), \\
\langle r \rangle_{(2)}^r &= \int d^3s\,s \int d^3r\, \rho_\text{w}(r)\, r\, \frac{\partial}{\partial r} \rho_\text{ch}(|\vec{r}-\vec{s}|).
\end{align}
Here $\rho_\text{w}$ and $\rho_\text{ch}$ are the weak and regular charge densities, respectively. As discussed in the previous section, care must be taken when looking to compare Eq. (\ref{eq:FS_Wang}) to those found in this manuscript. \textcite{Holstein1974b} defines the Coulomb correction to the spectral function through Eq. (\ref{eq:holstein_old_coulomb_replacement}), where $\Delta h_1(W)$ is given by a simplified version of the Coulomb terms of Eq. (\ref{em-complete-bis}) without the last term, and subtracting the Fermi function $(1+\gamma)/2F_0$. This is then developed to only first order in $\alpha Z$ to yield Eq. (25) in \cite{Holstein1974b}. It is thus a combination of electrostatic finite size effects contained in $L_0$ and the leptonic convolution. The previous section has discussed in great detail the improvements made by both formalisms, and we will not spend any more time on explicitly showing the correspondence to the formulae presented there. We will instead treat the evaluation of the Zemach moments in a purely spherical harmonic oscillator fashion and compare the results with those of \textcite{Wang2016} obtained through DFT. This will serve as a benchmark for the validity of using harmonic oscillator wave functions in the evaluation of our more carefully expanded results.

As we discussed previously, a uniform density is not appropriate for the weak charge density. As a result, we added the $C_I$ correction discussed in Sec. \ref{sec:isospin_breakdown}. Here we introduced the rms radius of the weak charge density, $\langle r^2 \rangle_\text{w}$, as the essential parameter in our correction. \textcite{Wang2016} calculated this quantity for a series of nuclei ranging from $A=14$ to $A=120$ using both DFT and uniformly charged radius results. We presented the way to analytically calculate these radii using a charge-insensitive harmonic oscillator distribution model. We use the charge radii listed by \textcite{Wang2016} to obtain an optimal comparison of our method. This is plugged into Eq. (\ref{eq:R_nl_HO}) to evaluate $\langle r^2 \rangle_\text{w}$ for the last occupied transforming nucleon. We use the approximation dubbed as `Behrens-B\"uhring', which involves approximating the form factor $c(q^2) \approx c_1(1+\langle r^2 \rangle_{\text{w}}/6)$. The Zemach moments are then reduced to
\begin{align}
\langle r \rangle_{(2)} &\approx \langle r \rangle_\text{ch} + \frac{\langle r^2 \rangle_\text{w}\langle r^{-1} \rangle_\text{ch}}{3}, \label{eq:approx_wang_r2}\\
\langle r \rangle_{(2)}^r &\approx \frac{2\langle r^2 \rangle_\text{w}\langle r^{-1} \rangle_\text{ch}}{3},
\label{eq:approx_wang_r2r}
\end{align}
The last of these is, however, a poor approximation, as can be seen from numerical results listed by \textcite{Wang2016}. If we attempt to extend this approximation by expanding $c$ to $\mathcal{O}(q^4)$, dimensional analysis tells us this factor will be proportional to $-\langle r^4 \rangle_\text{w}\langle r^{-3} \rangle_\text{ch}$, which is, however, not convergent. We thus keep in mind that Eq. (\ref{eq:approx_wang_r2r}) tends to overestimate the integral by about 30\%.
Table \ref{tab:comp_wang} shows a comparison of our analytical methods with DFT results for the relevant parameters in our corrections.

\begin{table}[h]
\caption{Comparison of DFT and analytical results using both uniformly charged sphere and harmonic oscillator (HO) functions as described in the text. We compare the weak charge radius $\langle r^2 \rangle_\text{w}$ used in our our isovector correction (Eq. (\ref{eq:C_I_wilkinson}); column 3), together with the Zemach moments discussed by \textcite{Wang2016} for several $\beta^-$ transitions. Here $BB$ stands for the approximations listed in Eqs. (\ref{eq:approx_wang_r2}) and (\ref{eq:approx_wang_r2r}). Units are in fm and fm$^2$. These were used together with a uniform and harmonic oscillator charge distribution.}
\begin{ruledtabular}
{\renewcommand{\arraystretch}{1.2}
\begin{tabular}{c c c c c c}
& $\langle r^2 \rangle_\text{w}^\text{Wang}$ & $\langle r^2 \rangle_\text{w}^\text{HO}$ & $\langle r \rangle_{(2)}^{UniBB}$ & $\langle r \rangle_{(2)}^{HO-BB}$ & $\langle r \rangle_{(2)}^\text{Wang}$  \\
\hline
$^{14}$C/$^{14}$N & 8.78 & 7.77 & 3.52 & 3.67 & 3.66\\
$^{25}$Na/$^{25}$Mg & 11.63 & 11.82 & 4.01 & 4.41 & 4.22\\
$^{35}$S/$^{35}$Cl & 13.77 & 12.97 & 4.43 & 4.70 & 4.62\\
$^{45}$Ca/$^{45}$Sc & 17.12 & 18.09 & 4.71 & 5.38 & 5.07\\
$^{61}$Cr/$^{61}$Mn & 18.88 & 19.02 & 5.02 & 5.58 & 5.36\\
$^{64}$Co/$^{64}$Ni & 18.87 & 18.97 & 5.12 & 5.60 & 5.41\\
$^{100}$Nb/$^{100}$Mo & 25.80 & 27.17 & 5.86 & 6.62 & 6.28\\
$^{104}$Nb/$^{104}$Mo & 26.14 & 27.77 & 6.07 & 6.73 & 6.33\\
$^{121}$Sn/$^{121}$Sb & 28.11 & 33.81 & 6.26 & 7.31 & 6.56
\end{tabular}
}
\end{ruledtabular}
\label{tab:comp_wang}
\end{table}
While the agreement is fair for lower masses, differences appear for the highest masses that were investigated. The reason for this is twofold
\begin{enumerate}
\item The approximation we have thus far used by considering only the final decaying neutron is not any more valid for the higher mass, $A > 100$, regions. Here the difference between quantum numbers for proton and neutron orbitals becomes too large and Eq. (\ref{eq:R_nl_HO}) is an unsustainable estimate. A more correct approach is then to calculate the actual overlap integral $\int dr\, r^2 R_{nl}^{\alpha}R_{n'l'}^{\beta}/\int dr\, R_{nl}^{\alpha}R_{n'l'}^{\beta}$ where $\alpha$ and $\beta$ are the neutron and proton harmonic oscillator wave functions.
\item Our harmonic oscillator functions are charge-insensitive, while the DFT results are not. We expect then an additional effect arising from the breakdown of isospin invariance, introducing the so-called nuclear mismatch $\delta_C$ \cite{Towner2010}. This has a strong $Z$ dependent behavior as is to be expected, and is a worthy subject in its own right. This is, however, beyond the scope of the work here.
\end{enumerate}

Application of number 1 yields results accurate to within 10\% of the DFT results. The remaining discrepancy we attribute to inherent limitations of the harmonic oscillator treatment, and to a lesser extent the nuclear mismatch problem of number 2. The latter can be assumed to drop out to first order when taking ratios of matrix elements suffering from the same issue. This precision is, however, sufficient for our current purposes, and we conclude the harmonic oscillator wave functions yield appropriate results. An extension of the latter to deformed nuclei is straightforward, as discussed in several places throughout this text. Results obtained in a deformed Woods-Saxon potential using harmonic oscillator basis functions show excellent agreement with experimental weak magnetism data, extensively discussed by \textcite{SeverijnsTBP}. This allows us to put our trust into the approach presented here.

\section{Tabulation of fit parameters for exchange correction}
\label{app:exchange_fit_coeff}

\newpage

\begin{widetext}
\csvautobooklongtable[ table head=\caption{Fit coefficients as defined in Eq. (\ref{eq:exchange_fit}) for $Z=2$ to $Z=120$.}\label{tab:exchange_fit_coeff}\\\hline
               \csvlinetotablerow\\\hline
               \endfirsthead\hline
               \csvlinetotablerow\\\hline
               \endhead\hline
               \endfoot]{FitData.csv}
\end{widetext}

\bibliography{library}

\end{document}